\documentclass{llncs}

\usepackage[T1]{fontenc}
\usepackage[misc]{ifsym}

\usepackage{graphicx}
\usepackage{array,multirow}
\usepackage{amsmath}
\usepackage{amsfonts}
\usepackage{subcaption}
\usepackage{comment}
\usepackage{hyperref}
\usepackage[color=cyan!20]{todonotes}
\usepackage{siunitx}
\usepackage{booktabs}
\usepackage{cleveref}
\usepackage{algorithm}
\usepackage{algpseudocode}
\usepackage{enumitem}

\usepackage{tikz}
\usetikzlibrary{positioning}
\usetikzlibrary{shapes.geometric}

\usepackage{orcidlink}

\usepackage{xspace}

\newcommand{\etc}{\textit{etc.}\xspace}

\newcommand{\eg}{\textit{e.g.,}\xspace}

\newcommand{\Stp}{\mathsf{Setup}}
\newcommand{\Eval}{\mathsf{Eval}}
\newcommand{\Vf}{\mathsf{Verify}}

\newcommand{\view}{\mathrm{view}}
\newcommand{\Hinf}{\widetilde{H}_\infty}

\newcommand{\NHC}{\mathsf{NHC}}
\newcommand{\Ext}{\mathsf{Ext}}

\newcommand{\EntRate}{\mathsf{EntRate}}
\newcommand{\Ecal}{\mathcal{E}}

\newtheorem{assumption}{Assumption}

\newtheorem{goal}{Goal}

\crefname{assumption}{Assumption}{Assumptions}
\Crefname{assumption}{Assumption}{Assumptions}
\crefname{goal}{Goal}{Goals}
\Crefname{goal}{Goal}{Goals}
\crefname{definition}{Definition}{Definitions}
\Crefname{definition}{Definition}{Definitions}
\crefname{proposition}{Proposition}{Propositions}
\Crefname{proposition}{Proposition}{Propositions}
\crefname{remark}{Remark}{Remarks}
\Crefname{remark}{Remark}{Remarks}

\usepackage{fullpage}

\begin{document}

\title{High-Frequency Financial Data for Public Randomness}

\author{
	Silvia Onofri\inst{1}\orcidlink{0000-0001-5093-3456}(\Letter) \and 
    Andrey Shternshis\inst{2}\orcidlink{0000-0002-9299-369X} \and
    Stefano Marmi\inst{1}\orcidlink{0000-0001-5105-8876} \and
	Jeremy Clark\inst{3}\orcidlink{0000-0002-3533-5965}
	}

\institute{
    Scuola Normale Superiore, Pisa, Italy\\ \email{\{silvia.onofri,stefano.marmi\}@sns.it} \and
    Lund University, Lund, Sweden\\ \email{andrey.shternshis@scilifelab.se} \and
    Concordia University, Montreal, Canada\\ \email{j.clark@concordia.ca}
}

\maketitle

\begin{abstract}

A randomness beacon provides cryptographic protocols with a publicly verifiable source of unbiased randomness for tasks such as selection, challenge generation, and auditing. Financial data is an attractive entropy source for such beacons because it is publicly observable, difficult to predict, and, in some markets, generated continuously. Prior work has characterized the entropy in daily stock closing prices, limiting beacon production to once per trading day. We introduce a beacon that draws entropy from trading data across a configurable window of time, and uses an entropy-based circuit breaker to fail when entropy is insufficient. Each trade-to-trade price change contributes one bit, its sign, and we quantify the entropy of each window with a conservative lower bound, keeping a margin against manipulation of the market data. We demonstrate that Binance cryptocurrency trading markets could have enabled a beacon every 5 minutes without failing once across the entire year 2025. Finally, we adapt the security analysis of an existing delay-based beacon protocol to our use case.
\end{abstract}



\section{Introduction}

Cryptographic protocols use randomness extensively. In many settings, the randomness is generated privately and does not need to be verified by any other party. However, for occasions that require \textit{verifiable} randomness (such as leader selection in consensus mechanisms, audits in voting systems, or challenge generation in proof-of-storage protocols~\cite{CMB23,KWJ24}), what exactly is being verified about the randomness? If a protocol $\Pi$ produces a random value $B$ from distribution $\mathcal{D}$ (\eg a uniform random binary string of length 256) at time $t$, a verifier could be interested in learning any of the following:

\begin{enumerate}[label=(\alph*)]
    \item $B$ is the output of correctly executing $\Pi$,
    \item $B$ was sampled from $\mathcal{D}$ (or from a distribution indistinguishable from $\mathcal{D}$), even in the presence of an active adversary,
    \item $B$ could not be predicted (beyond what is implied by $\mathcal{D}$) before time $t$, and
    \item $B$ is the unique value for time $t$ that all verifiers can obtain.
\end{enumerate}

A physical randomness protocol (\eg dice, lottery balls, \etc) only provides (b) if it is evident that it cannot be adversarially controlled (\eg through loaded dice) and (c)--(d) if the process is reliably broadcast. A pseudorandom generator (PRG) requires a seed as input, pushing (a--d) onto the seed. A verifiable random function (VRF) solves (a) and (b) but does not itself solve (c)--(d). If the input to the VRF has properties (c)--(d), then the output of the VRF will as well. A verifiable delay function (VDF) gives (a) and (d) and delays (c), since its output cannot be computed sooner than a fixed time after its input is known, but it adds no randomness, so (b) must come from its input.

A coin-tossing protocol (such as Ethereum's RANDAO or dozens of recent proposals~\cite{CMB23}) run among several participants is designed to provide (a)--(d), but it relies not only on the usual honesty assumption of protocol compliance, but also on at least one honest participant contributing sufficiently high-quality randomness. Drawing randomness from financial data is similarly designed to provide (a)--(d). It streamlines the protocol by not requiring multiple participants or multiple rounds (as commit/reveal-based coin tossing does), and, rather than taking randomness as an input supplied by participants, it derives randomness directly from an observed external process. In the literature, this is called a randomness beacon. However one must now justify that the entropy source is sufficiently random and resistant to manipulation.

Beacon outputs based on financial data were first studied in depth by Clark and Hengartner~\cite{CH10}. Closing prices of stocks are analyzed because closing prices are widely available online and in print, however they only provide one value a day, and only on trading days ($\approx$252 in a year). In this paper, we propose drawing randomness from financial trades on markets that operate continuously 24/7. This enables beacon outputs to keep up with protocols that require verifiable randomness more often than once a day, such as consensus algorithms like Ethereum where validators are randomly assigned roles every 6.4 minutes. 

\subsection{Contributions}

In this paper, we answer the following research questions about a financial-data beacon:

\begin{itemize}
    \item \textbf{RQ1:} How quickly can high-frequency cryptocurrency market data generate enough entropy to produce a 256-bit beacon output? 
    \item \textbf{RQ2:} To what extent does combining data from multiple cryptocurrencies increase the rate at which entropy accumulates?
    \item \textbf{RQ3:} How should a beacon respond if a time window produces insufficient entropy (\eg low trading activity which is outside of the beacon's control)? 
    \item \textbf{RQ4:} Does an existing delay-based beacon protocol remain verifiably unbiased when it reads high-frequency windows, despite bounded adversarial perturbations to its underlying financial data?
\end{itemize}

Briefly, using 2025 Binance data, we find, with regards to RQ1, that five minute trading windows are sufficient for producing 256-bit strings, while reporting on other window lengths. Pooling assets was necessary to achieve this (re:RQ2). For example, under our conservative criterion, approximately 20\% of five-minute BTCUSDT windows in 2025 failed to meet our entropy target, whereas none of the five-minute windows for the mixed basket failed. In regards to RQ3, we propose a `circuit breaker' that will produce a `fail' if the conditions for a sufficiently random beacon output cannot be met. Finally, in regards to RQ4, we use the protocol in~\cite{LB25}, and we redo its security proof for our setting.

To what extent are our answers to RQ1--4 already known (or easily extrapolated) from the existing literature? We argue that the high-frequency setting significantly changes how we answer them. For RQ1, the literature measures entropy from one closing price to the next, whereas our high-frequency methodology harvests the randomness in the intraday price path. The difference is substantial: a naive extrapolation of prior daily closing-price estimates would suggest substantially less than one bit per five-minute period, whereas we observe hundreds of bits over the same duration. Our methodology is model-free, avoiding the geometric Brownian motion assumption made in prior work~\cite{CH10}. For baskets of assets (RQ2), we also avoid having to specify a model of inter-asset correlations and instead measure the entropy rate directly from the merged price series. 

The way we model entropy also enables us to track the entropy available to each beacon output and trigger a circuit breaker whenever it falls below a specified threshold (RQ3). This distinguishes our design from existing beacons based on financial data, which produce outputs regardless of the observed entropy. Such a safeguard becomes particularly important at shorter timescales, where unusual trading patterns or service outages can have a disproportionate effect on a five-minute window compared with an entire trading day.

For RQ4, like the literature, we consider a beacon protocol that draws market data into an entropy pool and derives a beacon output from the pool. As the entropy pool populates, the adversary can monitor it closely, determine the beacon output implied by the current state, and evaluate how that output would change under many possible market interventions. 
This can bias the output of the beacon, and against an adversary with unlimited computing power no deterministic function of the pool can prevent it~\cite{AORSV20}. Landis and Bonneau address this issueb with a verifiable delay function~\cite{LB25}, which makes it computationally infeasible for the adversary to learn the beacon output quickly enough to use it when choosing how to intervene.

However there is a second related issue. A verifier might check that the price history (\eg from Binance's historical prices, available the following day) is bit-for-bit identical to the data used to populate the entropy pool (\eg from Binance's live WebSocket stream). If there is a discrepancy, the verifier cannot distinguish whether it arose from harmless API behaviour, such as data reordering, race conditions, or timing adjustments, or from adversarial manipulation designed to remain close to the ground truth while giving the adversary many candidate price histories over which to grind and select the most favourable beacon output. This broader issue has already been observed in the wild: when the 2009 Takoma Park municipal election~\cite{CCC+10} used financial data to audit its cryptographic proof of results and the independent auditor obtained slightly different prices~\cite{Adi09}. Unlike market manipulation, which is costly, this attack is essentially costless but requires control over the beacon producer. We adapt VDF-based mitigation~\cite{LB25} to address this related, but distinct, attack.

\section{Entropy generation [RQ1 \& RQ2]}
\label{sec:entropy}

A beacon's output entropy is bounded by its source market window. To determine the necessary window size and entropy capacity, this section proceeds in three steps: defining the trade-to-binary conversion, establishing an entropy estimator, and evaluating a year of market data.

Answering RQ1 and RQ2, we find that five minutes of trades across eleven tokens yield several thousand bits of entropy, comfortably exceeding the 512 bits required for a 256-bit output. We give this only as an example, since users can choose themselves the tokens and the frequency to use: indeed, our empirical analysis looks for a tradeoff between quality of the entropy generation and frequency of the outputs.

\subsection{Related work}
\label{subsec:rwentropy}

\paragraph{Randomness beacons.}
A beacon, in Rabin's original sense~\cite{Rab83}, is a service that emits a fresh value at a known rate, in public, so that every party reads the same value and nobody could have known it in advance. Beacons based on physical processes have been used in election auditing~\cite{Hal08,CHF08,Res09}. Current applications and designs are surveyed in~\cite{CMB23,KWJ24} and a reporting
standard is given in~\cite{KBPB19}.

\paragraph{Beacons from a public source.}
Beacons divide into those that draw randomness from their participants, by
commitment and reveal or by threshold cryptography, and those that read a process
nobody controls. The first kind~\cite{CMB23,KWJ24} needs at least one participant to be honest
and to hold good randomness, which no transcript can show. The second
replaces that hypothesis by one about the source. Financial markets were studied in depth by Clark and Hengartner~\cite{CH10}, who fit a geometric Brownian motion to historical closing
prices and bound by simulation the min-entropy of the next day's outcome. The other public source that has
been tried is a blockchain. Bonneau, Clark and Goldfeder read a beacon off Bitcoin
block headers and price the bias a miner buys by discarding a block it dislikes~\cite{BCG15}; Pierrot and Wesolowski quantify that
manipulation~\cite{PW18}; and B\"unz, Goldfeder and Bonneau answer it with a
proof of delay~\cite{BGB17}, the countermeasure this work inherits
through~\cite{LW15}. Landis and Bonneau bring the delay back to market data and analyze it against an adversary that trades in order to move the output~\cite{LB25}; their construction is the basic version we use in \Cref{sec:v1}.
 
\paragraph{Our work.}
Our approach diverges from prior work in three independent ways. The first is the density of the data: by encoding every strict
intraday price move rather than a daily summary, we capture orders of magnitude
more entropy per unit of time. The second is continuous operation: cryptocurrency
markets have no weekends and no overnight gap, so the period of the beacon
becomes a parameter of the protocol, free to match a consumer such as Ethereum's
$6.4$-minute epoch, instead of being fixed by the trading hours of a venue. The
third is the method: previous estimates model the price process, and a basket
then needs a model of how its constituents move together, whereas we apply a
block estimator directly to the realized binary string. Being model-free, it
assumes no distribution, returns its own standard error, and makes pooling
tokens a matter of applying the same estimator to a longer string.

\paragraph{Entropy of financial time series.}
Entropy serves as a primary metric for capturing the uncertainty, unpredictability, and structural complexity of price return series \cite{Olbrys26,Molgedey00}. For instance, empirical studies demonstrate that the entropy of daily closing spot prices depends on geographical region \cite{Mensi12,Risso09}. The entropy of weekly changes in US market indices is statistically indistinguishable from maximum entropy at the $99\%$ confidence
level \cite{Alvarez21}. According to \cite{Fernandes25}, the entropy rate is
higher in cryptocurrency prices than in traditional asset prices; at daily
frequencies, most cryptocurrencies exhibit entropy rates ranging from $0.96$ to
$0.99$ \cite{Pinto25}. While most studies estimate entropy using daily time
series, others explore high-frequency intraday data
\cite{Calcagnile20,Chen24,Shternshis25}. 

\subsection{From trades to bits}
\label{subsec:bits}
A beacon has two parts: it collects bits from a source, and it passes them through a cryptographic construction that turns them into an output with properties (a)--(d) of the introduction. Only the first part is particular to financial data, and it is the subject of this section. What the second part needs from the first is a quantity of \emph{entropy}, measured in bits: a window carries $\eta$ bits of entropy against a given observer if, from that observer's point of view, no single outcome of the window is more likely than $2^{-\eta}$. The cryptographic construction adds no statistical entropy, since its output is a deterministic function of the window and of public values. What it adds is computational: the delay function keeps the adversary from evaluating the output while it can still act on the window, so the output stays unpredictable until the delay elapses, provided the window was unpredictable when it opened. The entropy of the market is therefore what the design relies on. We ask for $512$ bits to emit $\lambda=256$, and \Cref{rem:lhl} in \Cref{sec:primitives} explains the factor of two.

The quantity just described is the min-entropy of the window conditioned on the observer's view. When the observer is the adversary, together with its own trading strategy, it measures the randomness of the window that the adversary can neither predict nor steer: whatever the adversary does, no reading of the window has probability above $2^{-\eta}$. This is the part of the market's entropy that is not under the adversary's control, and it is the quantity the security analysis assumes (\Cref{ass:source}). Our empirical analysis measures instead the Shannon entropy rate of the realized strings, which is what the literature on financial time series reports and which can be estimated without a model of prices. The two are related in one direction: Shannon entropy bounds min-entropy from above, so a window whose Shannon entropy falls short of the target cannot meet it against any observer. The Shannon test is therefore a necessary condition, and it is the one the circuit breaker enforces. In the other direction we rely on a bounded adversary, one that controls a limited share of the trading in a window rather than the market as a whole, and on the margin by which the measured entropy exceeds the target. In the configuration of \Cref{subsec:bound}, the target is met as long as the min-entropy outside the adversary's control is at least $73\%$ of the conservative estimate in the least favourable five-minute window of 2025, and at least $12\%$ of it in the median window. \Cref{app:cbcertifies} states what this margin covers and what it does not.

To feed a high-quality seed, we collect binary strings from ultra-high-frequency (UHF) transaction data on a selection of
cryptocurrencies. A similar methodology was used in~\cite{OSM25}. We fix a data provider and a set of tokens $T_1,\dots,T_J$; each token $T_j$ carries a stream of transaction prices $p_{j,1},p_{j,2},\dots$,
indexed in the order in which trades occur. Every time a new transaction is
recorded on one of our observed tokens, we compute the ratio
$r=\frac{p_{j,t}}{p_{j,t-1}}$. We then append a bit to the binary string
$\overline{b}\in\{0,1\}^*$ according to
\begin{equation}
\label{eq:binarization}
\overline{b} \;\rightarrow\;
\begin{cases}
	\overline{b}\,1 & \text{if } r>1,\\
	\overline{b} & \text{if } r=1 \text{ (no bit is appended)\,,}\\
	\overline{b}\,0 & \text{if } r<1\,.\\
\end{cases}
\end{equation}

Each comparison is always between two consecutive prices of the \emph{same}
token, but a single string $\overline{b}$ can interleave bits coming from
different tokens, ordered by transaction timestamp. Three properties of the rule
matter later. It is public, so anybody holding the trades of the window can
recompute the string, and anybody can compute what an honest party reading a
different window would have produced. It discards the magnitude of a move and
keeps only its sign, which is what makes the string insensitive to the price
scale and comparable across tokens. Finally, unchanged prices contribute no bits. This ensures that inactive windows yield shorter bit strings rather than biased, predictable ones. As a result, the entropy accounting below can be stated in bits of the string.

\paragraph{Entropy accounting.}
The literature \cite{Calcagnile20,Mensi12} measures how efficient a
market is, and reports entropy rates, that is entropy per symbol. A beacon needs
entropy per unit of time, which is the rate multiplied by the number of symbols
the market produces, and the two are not maximized by the same choices; the
distinction is what \Cref{app:aggregation,subsec:bound} turn on.

To determine the entropy contained in a string $\overline{b}$ of length $N$, we
use the estimator we introduce in \Cref{app:algoentrrate}. It returns a pair
\begin{equation}
  \label{eq:estimator}
  (\hat h,\hat\sigma_h)\;=\;\EntRate(\overline{b}),
  \qquad \hat h\in[0,1],\qquad \hat\sigma_h\ge 0,
\end{equation}
the estimated entropy rate and its standard error, both normalized per bit so
that they are directly comparable. We write
\begin{equation}
  \label{eq:Ecal}
  n_E\;=\;\Ecal(\overline{b})\;=\;N\hat h,
  \qquad
  \sigma_{n_E}\;=\;N\hat\sigma_h,
  \qquad
  \Ecal^{-}(\overline{b})\;=\;n_E-3\sigma_{n_E}
\end{equation}
for the estimated number of entropy bits carried by $\overline{b}$, its standard error and its $3\sigma$ lower confidence bound, and $\eta$ for the target that a window must clear.

Two thresholds have to be considered. The \emph{gate threshold} $\eta$, used by the circuit breaker of \Cref{sec:circuitbreaker} (the \emph{gate}), is a bound on a on a statistic, $\Ecal^{-}$, computed from one realized string. The \emph{entropy floor} $\eta^{*}$ (that will be used in \Cref{ass:source}) is a bound on the conditional min-entropy of that string from the adversary's point of view, and it is what enters every security bound. \Cref{rem:proxy} lists what separates the two thresholds, and we take $\eta^{*}=\eta=512$ and $\lambda=256$ throughout.

\subsection{The estimator}
\label{app:algoentrrate}

For each binary string $\overline{b}=b_1 b_2 \dots b_N$ of length $N$ produced as
above, we estimate its entropy rate $h \in [0,1]$ with a block estimator: we choose
a block length $k$ that grows slowly with $N$, tabulate the empirical distribution
of all overlapping $k$-blocks in the string, compute the resulting Shannon entropy
and normalize it by $k$.
Since ${\hat{h}}$ is a statistic computed on a finite sample, we also return its
standard error $\hat\sigma_h$: for a multinomial sample of size $c=N-k+1$ over the alphabet of $k$-blocks, with empirical frequencies $p_w$ and block entropy $H_k=-\sum_w p_w\log p_w$,
\begin{equation}
  \label{eq:basharin}
  \operatorname{Var}\!\left(H_k\right)
  \;=\;
  \frac{1}{c}\left(\sum_{w} p_w \log^2 p_w \;-\; H_k^2\right) + O\!\left(c^{-2}\right),
  \qquad
  \hat\sigma_h \;=\; \frac{\sqrt{\operatorname{Var}(H_k)}}{k \log 2},
\end{equation}
so that $\hat\sigma_h$ is normalized exactly like $\hat h=H_k/(k\log 2)$ and the two are directly comparable. \Cref{alg:entropyrate} details the procedure.
 
\begin{algorithm}[htbp]
	\caption{Empirical block entropy-rate estimator, with standard error}
	\label{alg:entropyrate}
	\begin{algorithmic}[1]
		\Require binary string $b_1,\dots,b_N \in \{0,1\}^N$, alphabet size $A=2$
		\Ensure estimated entropy rate $h$ and its standard error $\sigma_h$
		\If{$N < 4$}
		\State \Return \textsc{undefined} \Comment{string too short}
		\EndIf
		\State $k \gets \left\lfloor \tfrac{1}{2}\log_A N \right\rfloor$
		\Comment{block length grows slowly with $N$}
		\If{$k < 1$}
		\State \Return \textsc{undefined}
		\EndIf
		\State $c \gets N-k+1$ \Comment{number of overlapping $k$-blocks}
		\State $\mathrm{Count}[w] \gets 0$ for every $w \in \{0,1\}^k$
		\For{$i = 1,\dots,c$}
		\State $w \gets b_i b_{i+1}\dots b_{i+k-1}$
		\Comment{$k$-bit block starting at position $i$}
		\State $\mathrm{Count}[w] \gets \mathrm{Count}[w] + 1$
		\EndFor
		\State $H \gets 0$; \; $S \gets 0$ \Comment{$S$ accumulates $\sum_w p_w \log^2 p_w$}
		\For{each block $w$ with $\mathrm{Count}[w] > 0$}
		\State $p \gets \mathrm{Count}[w] / c$
		\State $H \gets H - p \log p$
		\State $S \gets S + p \log^2 p$
		\EndFor
		\State $V \gets \max\!\left\{ (S - H^2)/c,\; 0 \right\}$
		\Comment{delta-method variance, clipped for round-off}
		\State $h \gets H / (k \log A)$ \Comment{normalize by block length and alphabet}
		\State $\sigma_h \gets \sqrt{V} / (k \log A)$ \Comment{same normalization}
		\State \Return $(h, \sigma_h)$
	\end{algorithmic}
\end{algorithm}

\subsection{Data}
\label{app:empirical}

Our data are the trades executed on the spot market of Binance~\cite{binance} during 2025 for the $11$ tickers ADAUSDT, BCHUSDT, BNBUSDT, BTCUSDT, DOGEUSDT, ETHUSDT, LINKUSDT, SOLUSDT, TRXUSDT, XLMUSDT and XRPUSDT. We take them from Binance's public archive of market data, which publishes one compressed file per ticker and per day (UTC), listing every trade of that day as returned by Binance's REST endpoint for historical trades: for each trade, its identifier, price, quantity and timestamp, in microseconds for 2025. We use the price and the timestamp. 
Throughout, \textsc{Mixed} denotes the
series obtained by pooling all eleven tickers, interleaving their bits by transaction timestamp as in \Cref{subsec:bits}.

\subsection{Aggregation is not the right choice}
\label{app:aggregation}

Our first attempt was the methodology of~\cite{OSM25}, where consecutive price
observations are aggregated: instead of comparing every pair of consecutive
trades, the tick sequence is subsampled with a step $a>1$ and only prices $a$
trades apart are compared, on the premise that coarser sampling averages out
microstructure noise and raises the entropy rate. That premise is about the
quality of the bits. Our objective is the rate at which they arrive, so what
matters here is the product of the two, and the trade-off has to be measured
rather than assumed.

\begin{table}
\caption[Entropy and bit-generation statistics by ticker and aggregation level]{
Statistics over the daily strings of 2025 for each ticker and aggregation level $a\leq 5$. For $a>1$ each day gives $a$ strings, one for each starting trade of the subsampling, and the statistics are taken over all of them. $\min\hat h$ is the smallest entropy rate among the daily strings. The time per bit of a string is $86400/N$ seconds, $N$ being its number of bits; P95, P99, Max, Min and Mean sec.\ are the $95$th and $99$th percentiles, maximum, minimum and mean of this quantity over the days. $T_{512}^{x}=512\,x/\min\hat h$, with $x$ the Min, Mean or Max sec.\ value, is the time needed to collect $512$ bits of entropy at that time per bit and at the lowest entropy rate of the year, which need not come from the same day.}
\label{tab:stats1}
\begin{tabular}{lr|r|rrrrr|rrr}
\toprule
Ticker & $a$ & $\min\hat h$ & P95 & P99 & Max sec. & Min sec. & Mean sec. & $T_{512}^{min}$ & $T_{512}^{mean}$ & $T_{512}^{max}$\\
\midrule
ADAUSDT & 1 & 0.89 & 5.15 & 6.83 & 10.57 & 0.07 & 2.16 & 37.73 & 1236.24 & 6050.11 \\
ADAUSDT & 2 & 0.87 & 6.18 & 8.60 & 13.79 & 0.08 & 2.51 & 49.94 & 1479.01 & 8125.72 \\
ADAUSDT & 3 & 0.85 & 6.92 & 9.85 & 15.76 & 0.10 & 2.76 & 61.63 & 1650.15 & 9437.57 \\
ADAUSDT & 4 & 0.84 & 7.51 & 10.60 & 17.28 & 0.12 & 2.97 & 74.14 & 1809.03 & 10529.19 \\
ADAUSDT & 5 & 0.83 & 8.17 & 11.22 & 18.60 & 0.14 & 3.18 & 86.92 & 1957.97 & 11466.49 \\
\midrule
BCHUSDT & 1 & 0.90 & 12.35 & 17.22 & 23.07 & 0.68 & 5.88 & 385.32 & 3326.04 & 13060.77 \\
BCHUSDT & 2 & 0.89 & 13.98 & 20.60 & 26.09 & 0.88 & 6.64 & 503.88 & 3798.45 & 14929.83 \\
BCHUSDT & 3 & 0.89 & 15.08 & 22.80 & 28.21 & 0.98 & 7.21 & 565.03 & 4156.22 & 16256.57 \\
BCHUSDT & 4 & 0.89 & 16.05 & 24.78 & 29.96 & 1.09 & 7.75 & 623.54 & 4452.32 & 17205.15 \\
BCHUSDT & 5 & 0.90 & 17.34 & 27.27 & 31.12 & 1.20 & 8.28 & 686.70 & 4723.28 & 17759.78 \\
\midrule
BNBUSDT & 1 & 0.47 & 1.12 & 1.45 & 2.02 & 0.05 & 0.50 & 56.90 & 540.07 & 2185.41 \\
BNBUSDT & 2 & 0.50 & 1.47 & 2.04 & 2.90 & 0.07 & 0.64 & 73.15 & 657.51 & 2993.73 \\
BNBUSDT & 3 & 0.49 & 1.80 & 2.48 & 3.58 & 0.09 & 0.75 & 89.89 & 785.57 & 3727.65 \\
BNBUSDT & 4 & 0.50 & 2.07 & 2.88 & 4.18 & 0.10 & 0.86 & 104.61 & 889.82 & 4306.39 \\
BNBUSDT & 5 & 0.50 & 2.34 & 3.23 & 4.69 & 0.12 & 0.97 & 117.63 & 987.36 & 4769.40 \\
\midrule
BTCUSDT & 1 & 0.32 & 0.35 & 0.46 & 0.54 & 0.02 & 0.15 & 38.00 & 238.29 & 861.02 \\
BTCUSDT & 2 & 0.32 & 0.53 & 0.71 & 0.86 & 0.03 & 0.21 & 47.85 & 337.94 & 1385.76 \\
BTCUSDT & 3 & 0.31 & 0.65 & 0.97 & 1.11 & 0.04 & 0.26 & 59.65 & 422.46 & 1808.31 \\
BTCUSDT & 4 & 0.32 & 0.78 & 1.21 & 1.36 & 0.04 & 0.31 & 71.35 & 497.58 & 2176.94 \\
BTCUSDT & 5 & 0.32 & 0.91 & 1.40 & 1.59 & 0.05 & 0.36 & 82.32 & 564.33 & 2502.05 \\
\midrule
DOGEUSDT & 1 & 0.66 & 1.86 & 2.90 & 3.69 & 0.04 & 0.75 & 33.29 & 580.95 & 2873.63 \\
DOGEUSDT & 2 & 0.63 & 2.12 & 3.70 & 4.29 & 0.05 & 0.83 & 40.00 & 676.31 & 3509.66 \\
DOGEUSDT & 3 & 0.61 & 2.27 & 4.44 & 4.89 & 0.05 & 0.88 & 45.59 & 734.33 & 4090.63 \\
DOGEUSDT & 4 & 0.61 & 2.47 & 4.67 & 5.32 & 0.06 & 0.92 & 51.08 & 776.86 & 4481.56 \\
DOGEUSDT & 5 & 0.61 & 2.63 & 4.85 & 5.67 & 0.07 & 0.97 & 56.92 & 810.81 & 4756.03 \\
\midrule
ETHUSDT & 1 & 0.35 & 0.31 & 0.44 & 0.66 & 0.02 & 0.13 & 29.05 & 187.96 & 968.91 \\
ETHUSDT & 2 & 0.35 & 0.37 & 0.57 & 0.83 & 0.03 & 0.16 & 40.32 & 236.79 & 1234.19 \\
ETHUSDT & 3 & 0.34 & 0.44 & 0.68 & 0.98 & 0.04 & 0.19 & 52.02 & 285.41 & 1458.74 \\
ETHUSDT & 4 & 0.35 & 0.51 & 0.77 & 1.12 & 0.04 & 0.22 & 63.34 & 329.09 & 1648.66 \\
ETHUSDT & 5 & 0.35 & 0.60 & 0.87 & 1.26 & 0.05 & 0.26 & 75.16 & 374.54 & 1844.24 \\
\bottomrule
\end{tabular}
\end{table}

\begin{table}
\caption[Entropy and bit-generation statistics (continued)]{
\Cref{tab:stats1}, continued.}
\label{tab:stats2}
\begin{tabular}{lr|r|rrrrr|rrr}
\toprule
Ticker & $a$ & $\min\hat h$ & P95 & P99 & Max sec. & Min sec. & Mean sec. & $T_{512}^{min}$ & $T_{512}^{mean}$ & $T_{512}^{max}$\\
\midrule
LINKUSDT & 1 & 0.70 & 15.00 & 21.22 & 27.26 & 0.34 & 6.75 & 245.87 & 4928.06 & 19914.45 \\
LINKUSDT & 2 & 0.79 & 21.92 & 32.09 & 41.70 & 0.44 & 9.46 & 285.37 & 6093.51 & 26872.86 \\
LINKUSDT & 3 & 0.85 & 27.43 & 40.68 & 52.17 & 0.54 & 11.46 & 322.74 & 6902.42 & 31423.95 \\
LINKUSDT & 4 & 0.88 & 32.18 & 47.69 & 62.93 & 0.63 & 13.10 & 363.86 & 7600.28 & 36498.03 \\
LINKUSDT & 5 & 0.91 & 35.96 & 54.09 & 72.00 & 0.71 & 14.47 & 403.70 & 8176.90 & 40698.83 \\
\midrule
SOLUSDT & 1 & 0.73 & 1.26 & 1.72 & 2.08 & 0.04 & 0.65 & 28.16 & 456.16 & 1463.15 \\
SOLUSDT & 2 & 0.76 & 1.62 & 2.31 & 2.90 & 0.05 & 0.78 & 31.32 & 525.77 & 1947.27 \\
SOLUSDT & 3 & 0.73 & 1.88 & 2.72 & 3.47 & 0.05 & 0.87 & 35.94 & 606.62 & 2426.85 \\
SOLUSDT & 4 & 0.71 & 2.06 & 3.03 & 3.94 & 0.06 & 0.93 & 39.94 & 668.01 & 2836.40 \\
SOLUSDT & 5 & 0.70 & 2.20 & 3.28 & 4.36 & 0.06 & 0.98 & 43.45 & 715.98 & 3194.74 \\
\midrule
TRXUSDT & 1 & 0.21 & 5.02 & 5.35 & 5.57 & 1.14 & 3.62 & 2747.58 & 8755.15 & 13489.59 \\
TRXUSDT & 2 & 0.29 & 9.31 & 10.07 & 10.37 & 1.68 & 6.50 & 2971.90 & 11500.98 & 18338.79 \\
TRXUSDT & 3 & 0.33 & 13.21 & 14.34 & 14.70 & 2.09 & 9.07 & 3273.34 & 14183.13 & 22991.94 \\
TRXUSDT & 4 & 0.36 & 16.86 & 18.52 & 19.35 & 2.43 & 11.42 & 3437.70 & 16125.13 & 27327.64 \\
TRXUSDT & 5 & 0.39 & 20.40 & 22.52 & 23.56 & 2.73 & 13.59 & 3547.85 & 17656.85 & 30622.60 \\
\midrule
XLMUSDT & 1 & 0.86 & 17.57 & 24.06 & 26.37 & 0.45 & 7.60 & 266.40 & 4507.39 & 15644.01 \\
XLMUSDT & 2 & 0.94 & 21.92 & 32.34 & 36.73 & 0.50 & 9.08 & 272.91 & 4958.55 & 20066.34 \\
XLMUSDT & 3 & 0.95 & 25.40 & 37.88 & 44.74 & 0.53 & 10.09 & 288.65 & 5442.29 & 24144.13 \\
XLMUSDT & 4 & 0.94 & 27.87 & 42.61 & 50.41 & 0.57 & 10.88 & 306.97 & 5905.29 & 27364.36 \\
XLMUSDT & 5 & 0.94 & 29.77 & 46.81 & 56.32 & 0.59 & 11.55 & 325.16 & 6315.71 & 30785.79 \\
\midrule
XRPUSDT & 1 & 0.72 & 1.49 & 2.00 & 2.33 & 0.03 & 0.64 & 20.49 & 453.01 & 1658.28 \\
XRPUSDT & 2 & 0.68 & 1.83 & 2.60 & 3.12 & 0.03 & 0.74 & 25.76 & 557.93 & 2362.48 \\
XRPUSDT & 3 & 0.65 & 2.05 & 2.96 & 3.79 & 0.04 & 0.80 & 30.62 & 632.55 & 2990.56 \\
XRPUSDT & 4 & 0.63 & 2.18 & 3.22 & 4.28 & 0.04 & 0.85 & 35.84 & 687.70 & 3461.49 \\
XRPUSDT & 5 & 0.62 & 2.29 & 3.44 & 4.69 & 0.05 & 0.89 & 41.60 & 731.25 & 3850.04 \\
\bottomrule
\end{tabular}
\end{table}

\Cref{tab:stats1,tab:stats2} report, for each ticker and each aggregation level
$a\le5$, the minimum entropy rate $\min\hat h$ over the daily strings of 2025, statistics of the average time per bit across days, and the implied time to accumulate
$\eta=512$ bits of entropy. The last quantity is what the beacon cares about: it
combines the entropy per bit with the rate at which bits arrive, and it is
reported at the minimum, mean and maximum over the days of the average time per bit.

The tables show that the two effects do not merely trade off, they trade off unfavourably. Where aggregation raises $\min\hat h$ at all, it raises the time per bit by at least as much, so $T_{512}$ never improves, for any ticker. LINKUSDT is the clearest case: $\min\hat h$ climbs from $0.70$ to $0.91$ between $a=1$ and $a=5$, a gain of $30\%$, while $T^{mean}_{512}$ climbs from $4928$ to $8177$ seconds, a loss of $66\%$. The gain in rate is not systematic, and it does not depend on how high the rate is to begin with: aggregation raises it for TRXUSDT, the ticker with the lowest rate, leaves it essentially unchanged for BTCUSDT and ETHUSDT, and there are tickers, such as ADAUSDT and XRPUSDT, where it does not even improve $\min\hat h$ and lowers it instead. In these cases aggregation costs frequency and gives nothing in return. We therefore fix $a=1$ for the rest of the paper.

\subsection{Pooling tickers is the right choice}
\label{app:mixed}

The other way to raise the bit rate is to widen the basket. \Cref{tab:512bits_stats_mixed}
repeats the level-$1$ statistics with the pooled \textsc{Mixed} series added.
Pooling lowers the entropy rate with respect to most tickers: $\min\hat h$ for \textsc{Mixed} is $0.44$, below eight of the eleven individual tickers. This is what the composition of the basket predicts: BTCUSDT and ETHUSDT, the tickers with the shortest time per bit, supply about two thirds of the pooled bits at rates of $0.32$ and $0.35$, and the average of the eleven rates weighted by the bits each ticker supplies is $0.44$ as well. Pooling buys frequency by a far larger factor: the mean time per bit falls to $0.04$ seconds, against $0.13$ for the fastest single ticker, and $T^{mean}_{512}$ falls to $48$ seconds against $188$. The worst case is what decides the design, and there the margin is wider still: $T^{max}_{512}=173$ seconds for \textsc{Mixed} against $861$ for BTCUSDT. A pooled basket therefore collects $\eta$ bits of entropy within five minutes even when the lowest daily rate of the year is combined with the average time per bit of the slowest day, which no single ticker does. These figures are daily averages; \Cref{subsec:bound} checks the five-minute claim window by window.

\begin{table}
\caption[Statistics at aggregation level 1, including the pooled \textsc{Mixed}
series]{%
Statistics over the daily strings of 2025 at aggregation level $a=1$: one string per ticker and day (UTC), and for \textsc{Mixed} one string per day merging the bits of all eleven tickers by timestamp. $\min\hat h$ is the smallest entropy rate among the daily strings. The time per bit of a string is $86400/N$ seconds, $N$ being its number of bits; P95, P99, Max, Min and Mean sec.\ are the $95$th and $99$th percentiles, maximum, minimum and mean of this quantity over the days. $T_{512}^{x}=512\,x/\min\hat h$, with $x$ the Min, Mean or Max sec.\ value, is the time needed to collect $512$ bits of entropy at that time per bit and at the lowest entropy rate of the year, which need not come from the same day.}
\label{tab:512bits_stats_mixed}
\begin{tabular}{lr|r|rrrrr|rrr}
\toprule
Ticker & $a$ & $\min\hat h$ & P95 & P99 & Max sec. & Min sec. & Mean sec. & $T_{512}^{min}$ & $T_{512}^{mean}$ & $T_{512}^{max}$\\
\midrule
ADAUSDT & 1 & 0.89 & 5.15 & 6.83 & 10.57 & 0.07 & 2.16 & 37.73 & 1236.24 & 6050.11 \\
BCHUSDT & 1 & 0.90 & 12.35 & 17.22 & 23.07 & 0.68 & 5.88 & 385.32 & 3326.04 & 13060.77 \\
BNBUSDT & 1 & 0.47 & 1.12 & 1.45 & 2.02 & 0.05 & 0.50 & 56.90 & 540.07 & 2185.41 \\
BTCUSDT & 1 & 0.32 & 0.35 & 0.46 & 0.54 & 0.02 & 0.15 & 38.00 & 238.29 & 861.02 \\
DOGEUSDT & 1 & 0.66 & 1.86 & 2.90 & 3.69 & 0.04 & 0.75 & 33.29 & 580.95 & 2873.63 \\
ETHUSDT & 1 & 0.35 & 0.31 & 0.44 & 0.66 & 0.02 & 0.13 & 29.05 & 187.96 & 968.91 \\
LINKUSDT & 1 & 0.70 & 15.00 & 21.22 & 27.26 & 0.34 & 6.75 & 245.87 & 4928.06 & 19914.45 \\
\textsc{Mixed} & 1 & 0.44 & 0.09 & 0.12 & 0.15 & 0.01 & 0.04 & 6.09 & 47.91 & 173.29 \\
SOLUSDT & 1 & 0.73 & 1.26 & 1.72 & 2.08 & 0.04 & 0.65 & 28.16 & 456.16 & 1463.15 \\
TRXUSDT & 1 & 0.21 & 5.02 & 5.35 & 5.57 & 1.14 & 3.62 & 2747.58 & 8755.15 & 13489.59 \\
XLMUSDT & 1 & 0.86 & 17.57 & 24.06 & 26.37 & 0.45 & 7.60 & 266.40 & 4507.39 & 15644.01 \\
XRPUSDT & 1 & 0.72 & 1.49 & 2.00 & 2.33 & 0.03 & 0.64 & 20.49 & 453.01 & 1658.28 \\
\bottomrule
\end{tabular}
\end{table}

\subsection{Entropy per window}
\label{subsec:bound}
The configuration we propose is \textsc{Mixed} at aggregation level $a=1$, selected in \Cref{app:aggregation,app:mixed}. Here we state the bound it attains. We partition the year 2025 into non-overlapping intervals of $5$, $15$ and $30$ minutes, build one string per interval and evaluate the number of entropy bits on each.

\begin{figure}[!t]
	\centering
	\includegraphics[width=0.8\linewidth]{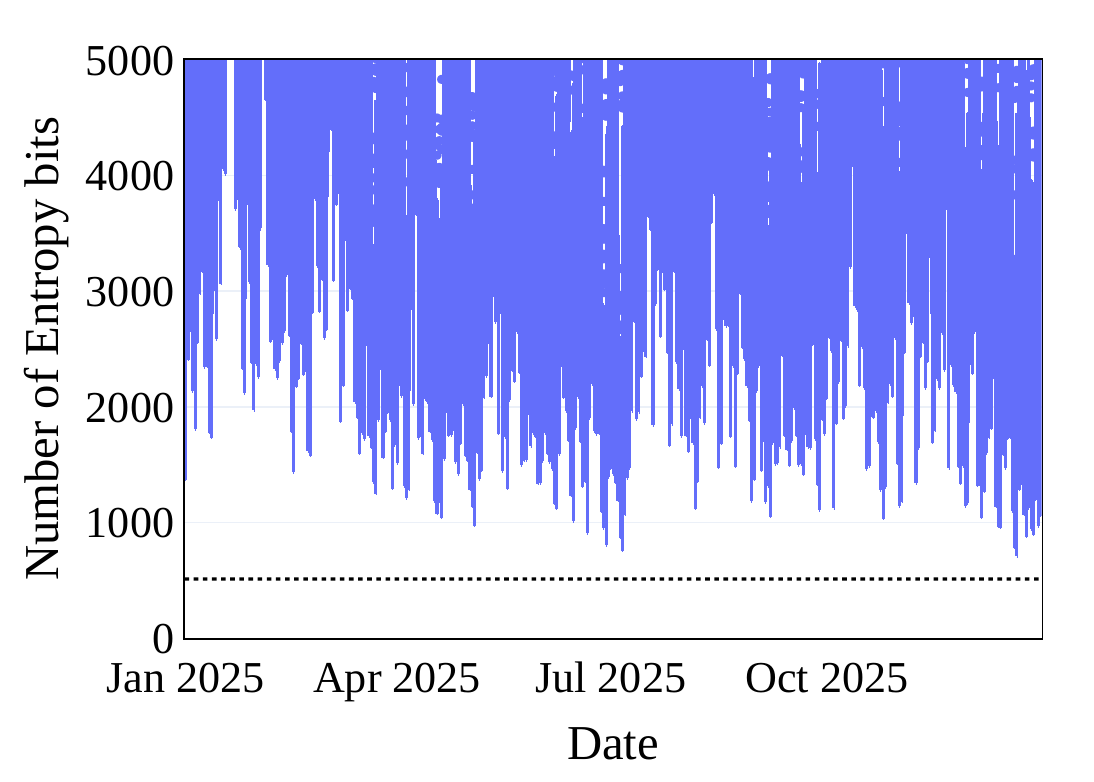}
	\caption{Number of entropy bits produced in 5-minute slots across the year 2025 by strings generated by 11 different cryptocurrencies. The black dotted line identifies our threshold $\eta=512$.}
	\label{fig:MIXED5min}
\end{figure}

\Cref{tab:mixedbound} gives the result. At five minutes the conservative bound $\Ecal^{-}$ falls below the target $\eta=512$ in none of the $105\,120$ intervals of the year; its median over them is $4232$ bits, and its smallest value is $701$ bits, which clears the target by $189$. At fifteen and thirty minutes the smallest
value is five and ten times the target. \Cref{fig:MIXED5min} shows the distribution over the year at five minutes.

The margin that the rule asks for is small, and it shrinks as the window grows: the
correction $3\sigma_{n_E}$ is worth $2.34\%$ of the median $n_E$ at five minutes
and $0.96\%$ at thirty. This is what \eqref{eq:basharin} predicts: the relative correction $3\sigma_{n_E}/n_E=3\sigma_h/\hat h$ decreases as $N^{-1/2}$ in the number $N$ of bits of the window, and the median window has about six times as many bits at thirty minutes as at five, hence a ratio of about $\sqrt6\approx2.4$.

\begin{table}[!t]
\caption{The pooled \textsc{Mixed} series against the target $\eta=512$, with the conservative bound $\Ecal^{-}$ of \eqref{eq:Ecal}, aggregation level $a=1$, year 2025.
$n_{\text{int}}$ is the number of non-overlapping intervals of the given length;
the next two columns are the median and the smallest value of $\Ecal^{-}$ over
them, in bits; the margin is that smallest value less $\eta$; the last column
counts the intervals
on which the circuit breaker fires.}
\label{tab:mixedbound}
\centering\footnotesize
\begin{tabular}{r|rrrrr}
\toprule
Window (min) & $n_{\text{int}}$ & median $\Ecal^{-}$ & smallest $\Ecal^{-}$ & margin & refusals\\
\midrule
5  & 105\,120 & 4232  & 701  & $+189$  & 0\\
15 & 35\,040  & 12\,751 & 2551 & $+2039$ & 0\\
30 & 17\,520  & 25\,957 & 5118 & $+4606$ & 0\\
\bottomrule
\end{tabular}
\end{table}


\section{The circuit breaker [RQ3]}
\label{sec:circuitbreaker}
Thin hours, venue outages and pinned prices produce windows from which no defensible value can be drawn. The design answers them by refusing to publish: for a string $\overline{b}$, with the quantities of \eqref{eq:Ecal}, the \emph{circuit breaker}, or \emph{gate}, fires whenever
\begin{equation}
  \label{eq:cb}
  \Ecal^{-}(\overline{b})\;=\;n_E-3\sigma_{n_E}\;<\;\eta ,
\end{equation}
and we take $\eta=512$. In the protocol of \Cref{sec:protocol} the gate reads the string that the publisher encodes from the window by \eqref{eq:binarization}, so that a window which does not clear the target yields a signed refusal rather than a value the protocol cannot stand behind.

\begin{remark}
\label{rem:onecriterion}
\Cref{eq:cb} is the conjunction of a \textit{plain} and a \textit{conservative}
criterion. Since $\hat\sigma_h\ge0$ unconditionally,
\[
  n_E-3\sigma_{n_E}\;\ge\;\eta
  \quad\Longrightarrow\quad
  n_E\;\ge\;\eta ,
\]
so the two are nested and the conservative failure rate dominates the plain one
by construction. We keep the conservative form as the rule and use the plain
criterion only as the naive baseline against which the cost of the margin is
measured in \Cref{app:circuitbreaker}.
\end{remark}

\subsection{How often the gate fires}
\label{app:circuitbreaker}

We now measure how often the gate \eqref{eq:cb} would have fired. We partition
2025 into non-overlapping intervals of $5$, $15$ and $30$ minutes, build one
string per interval, and evaluate on each the two quantities of
\Cref{sec:circuitbreaker}: the estimated entropy content $n_E=\Ecal(\overline{b})=N\hat h$
and its $3\sigma$ lower bound $\Ecal^{-}(\overline{b})=n_E-3\sigma_{n_E}$, with
$\sigma_{n_E}=N\hat\sigma_h$.

Two criteria are reported. The \emph{conservative} criterion is the rule of the
protocol, $\Ecal^{-}<\eta$; the \emph{plain} criterion, $n_E<\eta$, is the naive
baseline. As \Cref{rem:onecriterion} notes the two are nested, so the conservative
failure rate dominates the plain one by construction and their difference, the
column $\Delta$ of \Cref{tab:nentrbits_std_summary}, is exactly the price of the
margin. The purpose of reporting both is to price it, not to compare two
candidate rules.

The plain criterion is shown in
\Cref{fig:BTC_entropy_4x2_combined,fig:ETH_entropy_4x2_combined,fig:MIXED_entropy_4x2_combined};
the conservative one, on the same tickers and window lengths, in
\Cref{fig:BTC_lb3,fig:ETH_lb3,fig:MIXED_lb3}. \Cref{tab:nentrbits_std_summary}
summarizes both.

\paragraph{The margin is small and shrinks with the window.}
The correction $3\sigma_{n_E}$ is worth $5.08\%$ of the median $n_E$ for BTCUSDT
and $5.35\%$ for ETHUSDT at five minutes, and falls to $2.07\%$ and $2.13\%$ at
thirty, in agreement with the $M^{-1/2}$ scaling of \eqref{eq:basharin}. For
\textsc{Mixed}, whose strings are an order of magnitude longer, it is worth
$2.34\%$ at five minutes and $0.96\%$ at thirty.

\paragraph{The cost in liveness is concentrated at high frequency.}
Over the $105\,120$ five-minute intervals of the year the share of refusals rises
from $17.25\%$ to $20.11\%$ for BTCUSDT, about $58$ of the $288$ daily intervals,
and from $12.11\%$ to $15.04\%$ for ETHUSDT. At fifteen minutes it rises from
$0.63\%$ to $0.82\%$ and from $0.15\%$ to $0.21\%$ respectively. At thirty minutes
BTCUSDT refuses on one or two of the $17\,520$ intervals in the whole year and
ETHUSDT on none. \textsc{Mixed} refuses on no interval at any of the three window
lengths, under either criterion. Requiring the target with a $3\sigma$ margin
therefore does not change which configurations are viable; it quantifies how thin
the margin is in the ones that are marginal.

\begin{figure}[!t]
    \centering
    \begin{subfigure}{0.32\textwidth}
        \centering
        \includegraphics[width=\linewidth]{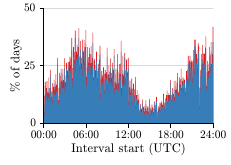}
        \caption{BTCUSDT}
        \label{fig:fri_BTC_5}
    \end{subfigure}
    \hfill
    \begin{subfigure}{0.32\textwidth}
        \centering
        \includegraphics[width=\linewidth]{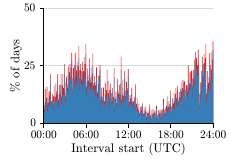}
        \caption{ETHUSDT}
        \label{fig:fri_ETH_5}
    \end{subfigure}
    \hfill
    \begin{subfigure}{0.32\textwidth}
        \centering
        \includegraphics[width=\linewidth]{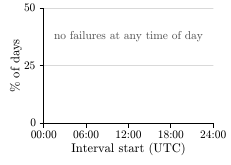}
        \caption{\textsc{Mixed}}
        \label{fig:fri_MIXED_5}
    \end{subfigure}
    \caption{Circuit-breaker rate by time of day, $5$-minute windows, 2025. For each
    of the $288$ interval starts, the share of the $365$ days whose string fails the
    $512$-bit target under the plain criterion (blue) and the conservative one (red).
    Failures for BTCUSDT and ETHUSDT concentrate in the low-activity hours, so the
    circuit breaker is predictable from the clock rather than idiosyncratic;
    \textsc{Mixed} is flat at zero throughout.}
    \label{fig:failrate_byinterval_5}
\end{figure}

\paragraph{Refusals are predictable from the clock.}
\Cref{fig:failrate_byinterval_5} resolves the five-minute failure rate by time of
day. For BTCUSDT and ETHUSDT the refusals concentrate in the low-activity hours
rather than scattering across the day, so a deployment on a single ticker could
anticipate them from the calendar instead of discovering them at the deadline.
\textsc{Mixed} is flat at zero throughout, which is the property that makes the
pooled basket the configuration we propose: not that it fails rarely, but that it
has no hour of the day at which it is close to failing.

\paragraph{Consequence for the parameters.}
A deployment on a single high-capitalization ticker must run at thirty-minute
windows to avoid refusals. A deployment on the pooled basket runs at
five minutes with the margin intact, and by \Cref{prop:liveness} the refusals it
does not experience are the only ones the protocol could have suffered, since no
behaviour of the reporters can close the gate. This is the configuration for which
the claim of \Cref{subsec:bound} holds.

\begin{table}[!t]
\caption{Plain versus conservative circuit-breaker criterion, aggregation level
$a=1$, 2025. $n_{\text{int}}$ is the number of intervals in the year;
$\widetilde{n}_E$ and $\widetilde{3\sigma}$ are the medians of $n_E$ and
$3\sigma_{n_E}$ over those intervals; the last three columns give the share of
intervals failing the plain criterion, the share failing the conservative one,
and their difference in percentage points. The difference is non-negative by \Cref{rem:onecriterion}.}
\label{tab:nentrbits_std_summary}
\centering
\begin{tabular}{lr|rrr|rrr}
\toprule
Ticker & Window & $n_{\text{int}}$ & $\widetilde{n}_E$ & $\widetilde{3\sigma}$ ($\%$ of $\widetilde{n}_E$) & $n_E<512$ & $n_E-3\sigma<512$ & $\Delta$ \\
 & (min) & & (bits) & (bits) & ($\%$) & ($\%$) & (p.p.) \\
\midrule
BTCUSDT & 5  & 105\,120 & 948.38   & 49.08\ \ (5.08) & 17.25 & 20.11 & $+2.86$ \\
BTCUSDT & 15 & 35\,040  & 2720.19  & 80.95\ \ (2.94) & 0.63  & 0.82  & $+0.19$ \\
BTCUSDT & 30 & 17\,520  & 5573.98  & 115.77\ (2.07)  & 0.00  & 0.01  & $+0.01$ \\
\midrule
ETHUSDT & 5  & 105\,120 & 1106.09  & 59.96\ \ (5.35) & 12.11 & 15.04 & $+2.94$ \\
ETHUSDT & 15 & 35\,040  & 3247.51  & 99.47\ \ (3.04) & 0.15  & 0.21  & $+0.06$ \\
ETHUSDT & 30 & 17\,520  & 6636.40  & 142.70\ (2.13)  & 0.00  & 0.00  & $0.00$ \\
\midrule
\textsc{Mixed} & 5  & 105\,120 & 4339.06  & 106.71\ (2.34) & 0.00 & 0.00 & $0.00$ \\
\textsc{Mixed} & 15 & 35\,040  & 12930.36 & 182.93\ (1.36) & 0.00 & 0.00 & $0.00$ \\
\textsc{Mixed} & 30 & 17\,520  & 26238.06 & 260.46\ (0.96) & 0.00 & 0.00 & $0.00$ \\
\bottomrule
\end{tabular}
\end{table}

\begin{figure}[!t]
    \centering
    \begin{subfigure}{0.48\textwidth}
        \centering
        \includegraphics[width=\linewidth]{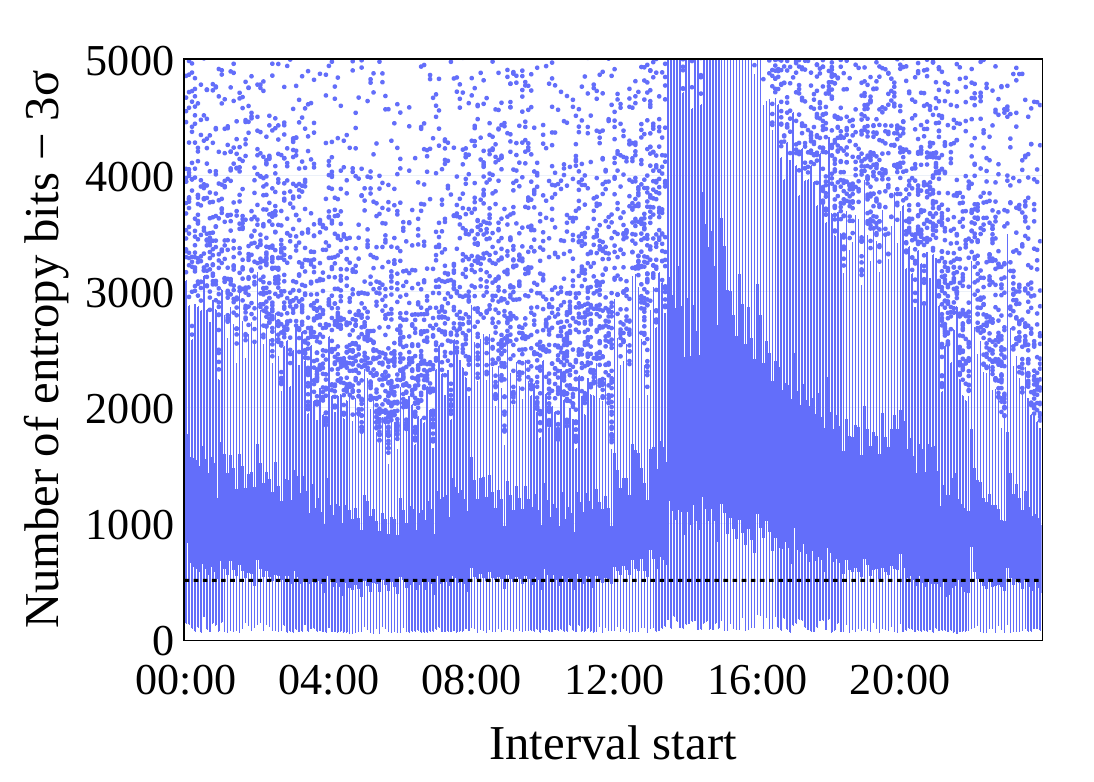}
        \caption{$n_E-3\sigma_{n_E}$ by interval start}
    \end{subfigure}
    \hfill
    \begin{subfigure}{0.48\textwidth}
        \centering
        \includegraphics[width=\linewidth]{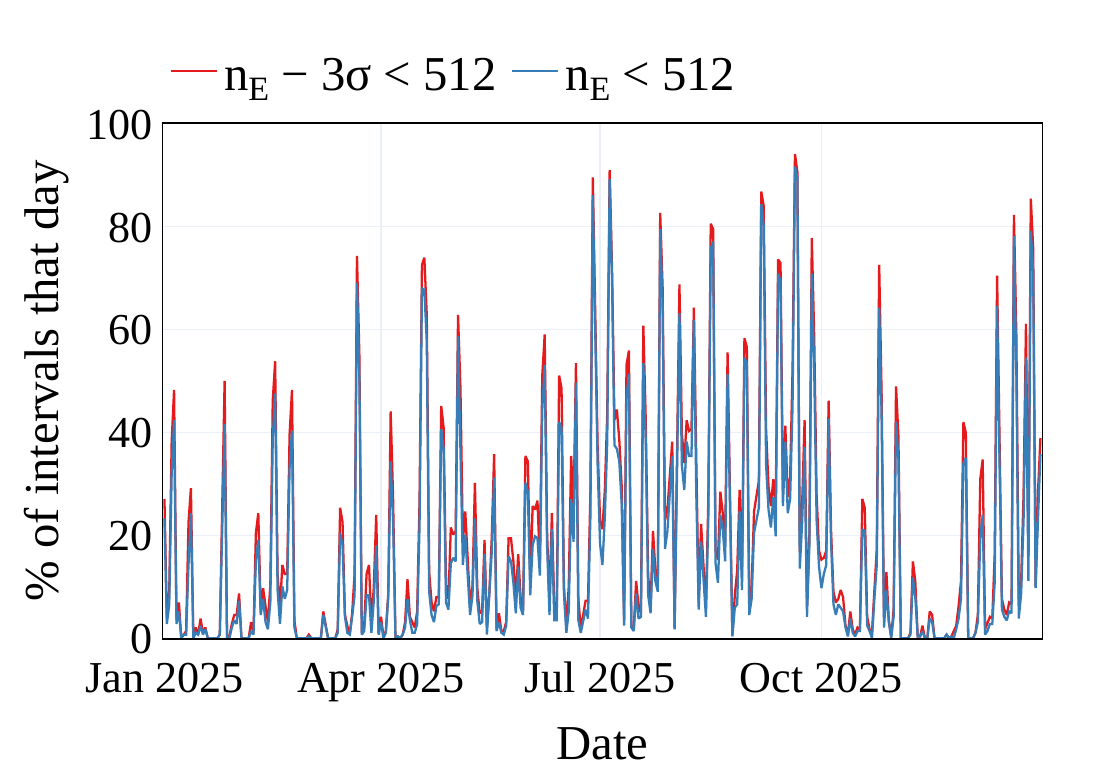}
        \caption{per-day failure rate}
    \end{subfigure}
    \caption{BTCUSDT alone, $5$-minute windows, 2025. Left: distribution of $n_E-3\sigma_{n_E}$ by interval start, against the $512$-bit target (dotted). Right: share of each day's intervals falling below the target, under the plain criterion (blue) and the conservative one (red). The pooled basket never falls below the target (\Cref{tab:mixedbound}).}
    \label{fig:BTC5_body}
\end{figure}

Combining multiple assets eliminates refusals entirely across all 24 hours. We choose this multi-asset approach not merely for a low average failure rate, but because it promises no low-activity hours where it comes close to failing.

\subsection{What the gate certifies}
\label{app:cbcertifies}
The gate and the security analysis look at the window from two sides. \Cref{ass:source} bounds the min-entropy of the window given the adversary's view when the window opens, which includes its trading strategy; the gate sees one realized string and estimates its Shannon entropy rate, which is what the literature on financial time series measures~\cite{Calcagnile20,Mensi12}, so that our figures can be compared with it. \Cref{rem:proxy} explains how the two fit together.

\begin{remark}[The gate as an empirical proxy]
\label{rem:proxy}
\Cref{ass:source} asks that the reading of the window have min-entropy at least $\eta^{*}$ given the adversary's view when the window opens; the gate \eqref{eq:cb} checks a conservative estimate of the Shannon entropy of the realized string against $\eta$. With $\eta^{*}=\eta$, the gate is the measurable counterpart of the assumption. The two quantities differ in three respects, and each is covered by a different part of the design.
\begin{enumerate}
\item \emph{Estimation.} $\hat h$ is computed on one finite string. The correction $3\sigma_{n_E}$ accounts for its dispersion, and the margin by which the pooled basket clears the target leaves ample room for the limitations of the estimator noted in \Cref{app:cbcertifies}: no five-minute window of the pooled basket would have been refused in 2025 even if $\Ecal^{-}$ were $25\%$ lower (\Cref{subsec:bound}).
\item \emph{Shannon against min-entropy.} Shannon entropy bounds min-entropy from above, so the test the gate applies is a necessary condition for \Cref{ass:source}: a window whose Shannon entropy falls short of the target cannot satisfy the assumption against any observer. Above the threshold, the margin of the estimate over the target is what allows for the difference between the two, as quantified in \Cref{subsec:bits}. A deployment that prefers to measure min-entropy directly can run \Cref{alg:entropyrate} on $-\log_2\max_w\hat p_w$ instead of on the Shannon functional, with a more conservative and noisier statistic.
\item \emph{Public against conditional.} The estimator sees the string as the public sees it, while \Cref{ass:source} speaks of what remains uncertain to a party that trades in the window. No statistic computed on a string can separate the two, since the part of the string a party writes can look as random to the public as the rest while holding no uncertainty for that party. This is true of any source of randomness, and it is why standards for entropy sources ask for a justification of the source in addition to statistical estimates~\cite{SP800-90B}. Our justification is the bounded adversary of \Cref{subsec:bits}, which writes a limited share of a window drawn from eleven markets, together with the delay function of \Cref{sec:protocol}, which keeps such a party from turning what it learns during the window into a choice of value (\Cref{ass:delay1}, \Cref{sec:protocol}).
\end{enumerate}
The gate thus does what a measurement can do: it refuses the windows that are thin or pinned for everyone, and it measures, with a margin, how much entropy the market supplied. \Cref{ass:source} states that enough of it lies outside the adversary's control.
\end{remark}


\section{Protocol [RQ4]}
\label{sec:protocol}

\subsection{Related work}
\label{subsec:rwprotocol}

We now turn to the concrete protocol for drawing entropy from prices and producing a verifiable random output. We follow a basic paradigm, originating in \texttt{unicorn}~\cite{LW15}, of hashing the source, evaluating a slow function, and hashing again. This paradigm can be seen in a line of research with varying instantiations of the slow function~\cite{LW15,RSW96,BN00,BBF18,Wes19,Pie19,BBF24,BGB17,CCB24}. Landis and Bonneau instantiate this pattern on financial data and analyze an adversary that trades to influence the beacon~\cite{LB25}; their construction is the basis of our basic beacon in \Cref{sec:v1}.

A competing paradigm for verifiable randomness is distributed coin-tossing, which uses commit-reveal, punishment, secret sharing, or threshold cryptography~\cite{Blu83,ethspec,SJK+17,CD17,CD20,SJSW20,drand,Cle86,CMB23}. Ethereum's RANDAO is a prominent deployment of this approach. Here, multiple participants use one or more rounds of communication to contribute and combine randomness that they themselves generate. By contrast, an external-source beacon can, in its simplest form, draw randomness directly from a single public source without such communication, although our broadcast variant gives up some of this efficiency to improve robustness (see \Cref{subsec:cointoss} for a detailed comparison). Our beacon could also serve as one participant’s randomness source in a distributed coin-tossing protocol, helping satisfy the usual assumption that at least one participant contributes honest, high-quality randomness.

Neither paradigm addresses what a verifier should do when its own reading of the source disagrees with the operator's. Landis and Bonneau speak of \emph{the} price vector, which is reasonable for daily closing prices (although not perfect~\cite{Adi09}). One might consider using a fuzzy extractor~\cite{DORS08}, which turns a noisy reading into a stable key. It does not apply here for two independent reasons: the errors are edit errors rather than Hamming errors, a metric for which no robust construction is known, and whoever computes the public helper string makes its own reading canonical, which is the problem rather than its solution. \Cref{sec:v2} answers with one round of messages over a reliable broadcast~\cite{Bra87,LSP82,DS83,CGR11} instead.

\subsection{The protocol}
The value of each epoch is derived from the bits of \Cref{sec:entropy} by the construction of Landis and Bonneau~\cite{LB25}, which instantiates on market data the paradigm of \texttt{unicorn}~\cite{LW15}: hash the source, evaluate a verifiable delay function, hash again. What we add is a new run of its security proof for our setting, so that the guarantees are known to hold for high-frequency windows, with the circuit breaker in the loop and with data feeds that may disagree. This section states the construction and the results; first we show the setting, the goals and the assumptions in full, and then analyze the primitives.

\subsubsection{The setting}
\label{subsec:setting}

\begin{figure}[!t]
\centering
\begin{tikzpicture}[x=1mm,y=1mm]
  \draw (0,0)  rectangle (34,7);  \node[font=\small] at (17,3.5) {observation};
  \draw (34,0) rectangle (56,7);  \node[font=\small] at (45,3.5) {consensus};
  \draw (56,0) rectangle (112,7); \node[font=\small] at (84,3.5) {delay};
  \node[font=\scriptsize] at (17,10.5) {read the market};
  \node[font=\scriptsize] at (45,10.5) {commit, vote};
  \node[font=\scriptsize] at (84,10.5) {$\Eval(pp,s_e)$};
  \node[font=\small] at (17,-4.5) {$\tau$};
  \node[font=\small] at (45,-4.5) {$\rho$};
  \node[font=\small] at (84,-4.5) {$\Delta>A(\tau+\rho)$};
  \foreach \x in {0,34,56,112} \draw (\x,0) -- (\x,-2);
  \node[font=\scriptsize,anchor=north west] at (-1.5,-8)  {$t_e$};
  \node[font=\scriptsize,anchor=north]      at (34,-8)    {window closes};
  \node[font=\scriptsize,anchor=north]      at (56,-8)    {seed fixed};
  \node[font=\scriptsize,anchor=north east] at (113.5,-8) {$B_e$ published};
\end{tikzpicture}
\caption{One epoch. Everything that determines $B_e$ is fixed at the deadline
$T^{*}_e$; the delay phase adds no freedom, and no evaluation of the delay
function started at any point of the epoch can finish before that deadline. The
consensus phase is absent from the basic version, which reads $\rho=0$.}
\label{fig:timeline}
\end{figure}
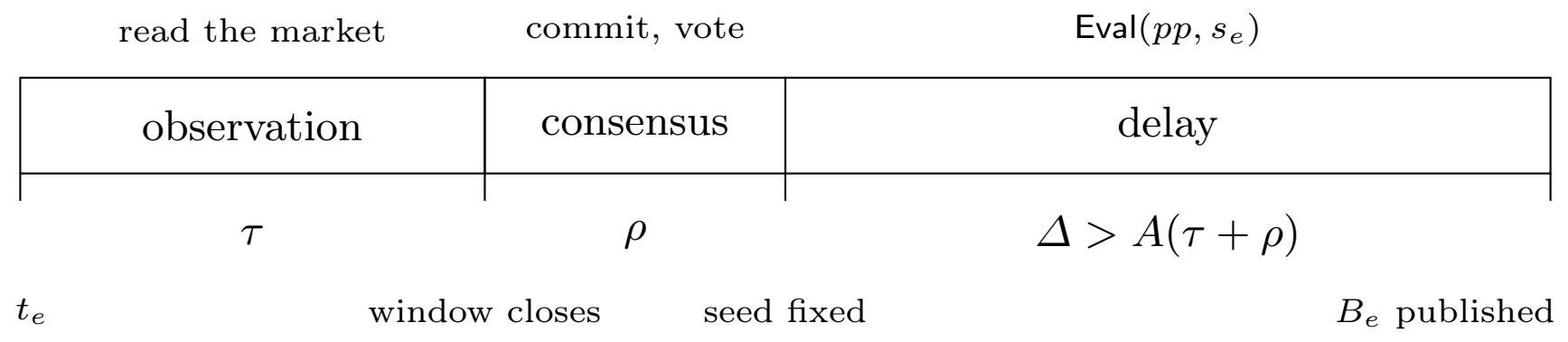

The beacon runs in \emph{epochs}. Epoch $e$ opens at time $t_e$ and observes the
window $[t_e,t_e+\tau)$ of trades; the consensus version of \Cref{sec:v2} adds a
phase of length $\rho$ in which the readings are exchanged, ending at the
deadline $T^{*}_e=t_e+\tau+\rho$; a delay of length $\Delta$ follows, after which
the value of the epoch, $B_e\in\{0,1\}^{\lambda}$, is published.
\Cref{fig:timeline} shows one epoch. Epochs are pipelined, so a value is emitted
every $\tau+\rho$ and reaches its consumers $\Delta$ later.

\paragraph{Parties and strings.}
The basic version has a single \emph{publisher}; the consensus version has a
registry $\mathcal{P}=\{P_1,\dots,P_n\}$ of \emph{reporters}, and we write
$\theta=\lfloor n/2\rfloor+1$ for a majority of them. Either way a
\emph{verifier} may read the window afterwards, through a data provider of its
own choosing. Four strings matter and they are all images of \eqref{eq:binarization}
applied to somebody's record of the window: $b^{*}$, the \emph{reading of the
window}, is what the venue's own record gives, and is what the others are trying
to report; $b_i$ is what reporter $P_i$ read; $b_V$ is what a verifier read; and
$\hat b$ is the \emph{reference string}, the one the protocol actually uses, which
is the publisher's string in the basic version and the majority answer in the
consensus version. They may not coincide, and this is addressed in
\Cref{sec:v2}.

\paragraph{What is computed.}
From $\hat b$ the protocol derives three things in order: a \emph{seed} $s_e$, by
hashing; a delay-function output $y_e$ with its proof $\pi_e$, by evaluating; and
the value $B_e$, by hashing again. Both hashings use one function $H$ into $\{0,1\}^{\lambda}$, modelled as a random oracle, at points that begin with distinct domain separators $\mathsf{dom}_1$ and $\mathsf{dom}_2$, so that no query made for one purpose can be a query made for the other; the delay function is a triple $(\Stp,\Eval,\Vf)$ with public parameters $pp$ and delay
$\Delta$. \Cref{sec:primitives} states both with the properties the analysis uses, and fixes the separators.
The \emph{sequentiality error} $\epsilon_{\mathrm{seq}}$ of the delay function is the probability that an adversary produces $\Eval(pp,s)$ in appreciably less than the delay. It is assumed as a property of the primitive.
Standing between the string and the seed is the \emph{gate} $g_e\in\{0,1\}$, which is the circuit breaker of \Cref{sec:circuitbreaker} evaluated on $\hat b$ against the threshold $\eta$: the epoch publishes a value when $g_e=1$ and a signed refusal when $g_e=0$. We write $\|$ for concatenation and $\langle e\rangle$ for a fixed-length encoding of the epoch index, which binds a message to one epoch and keeps the argument of an oracle unambiguous.

\Cref{tab:notation} in \Cref{app:spec} collects the notation, and
\Cref{app:spec} writes the consensus version out in full.

\subsection{Primitives}
\label{sec:primitives}
The construction uses a hash function and a verifiable delay function; the
consensus version adds a broadcast, a majority vote and a non-hiding commitment.
All are used as black boxes and all are standard.

\paragraph{The hash function.}
Hashing is by a cryptographic function $H$ that maps a string of any length to one of a fixed length $\lambda$, modelled as a random oracle~\cite{BR93}: $H$ is drawn uniformly among all such functions and is evaluated only by querying it, so that its value at a point no party has queried is uniform and independent of every party's view.

The construction hashes in three places, and we keep them apart by prefixing the argument with one of three pairwise distinct \emph{domain separators} fixed in the public parameters: $\mathsf{dom}_1$ where a reading of the window becomes a seed, $\mathsf{dom}_0$ where the seed is advanced because an epoch was skipped, and $\mathsf{dom}_2$ where a seed and a delay output become the value of the epoch. The separators have a fixed length, so the three sets of query points are disjoint and no query made for one purpose is a query made for another. This is all the proofs need, and it is why the analysis speaks of one oracle rather than three independent ones: on disjoint sets of points a random oracle is already independent, so nothing has to be assumed about the relation between the three uses. We write $q_H$ for the total number of queries the adversary makes to $H$.

\paragraph{The delay function.}
A verifiable delay function is a triple $(\Stp,\Eval,\Vf)$ in which
$\Stp(1^{\kappa},\Delta)$ outputs public parameters $pp$, $\Eval(pp,s)\to(y,\pi)$
takes $\Delta$ sequential steps, and $\Vf(pp,s,y,\pi)\in\{0,1\}$ runs in polylogarithmic time~\cite{BBF18}. We require three properties of it, in the form of~\cite[\S2]{BBF18}. It is \emph{correct}: for every $s$, the output of $\Eval(pp,s)$ is accepted by $\Vf$. It is \emph{sound} with error
$\epsilon_{\mathrm{uniq}}$: no efficient party exhibits a $y\neq\Eval(pp,s)$ carrying an accepting proof, so for each $s$ at most one output is accepted, an evaluator who dislikes the output cannot substitute another, and anybody can recompute it. And it is
$(\Delta,A,K,\epsilon_{\mathrm{seq}})$-\emph{sequential}: an adversary holding $K$ parallel processors, each faster than the honest evaluator by a factor $A$,
does not output $\Eval(pp,s)$ in fewer than $\Delta/A$ steps counted from the
instant it learns $s$, except with probability $\epsilon_{\mathrm{seq}}$. We use
sequentiality in the form that forbids a shortlist as well: within the same time
the adversary produces no short list of candidates containing $\Eval(pp,s)$. This
follows from the single-output form by a union bound over the list, and it is the
form a beacon needs, since an adversary that narrows the value to a few
candidates is already dangerous. The analysis also needs sequentiality for many challenges at once. The seed of an epoch is the value of $H$ at a point that depends on the window, and an adversary that trades in the window can query $H$ at the points of all the continuations it can still realise, so it may learn up to $q_H+1$ candidate seeds, any of which may become the seed. We therefore let $\epsilon_{\mathrm{seq}}$ bound the probability that the adversary produces such a list, within $\Delta/A$ steps of learning the challenge, for \emph{any} of at most $q_H+1$ challenges given as oracle values. A reduction that guesses which challenge becomes the seed obtains this multi-instance form from the single-instance one at a cost of a factor of at most $q_H+1$, so the error stays negligible when the single-instance error is negligible and $q_H$ is polynomial. All three properties hold for the constructions
of~\cite{Wes19,Pie19}, surveyed in~\cite{BBF24}, whose ancestor is the time-lock
puzzle of~\cite{RSW96}.

We write $q_E$ for the number of evaluations the adversary completes. The two
budgets are of very different sizes and we keep them apart throughout: $q_H$ is
limited only by the adversary's computation, whereas $q_E$ is limited by its
hardware and by the length of the delay.

\paragraph{Extraction.}
One consequence of the random oracle is used often enough to record. If a string
$X$ is unpredictable to the adversary, in the sense that
$\Hinf(X\mid Z)\ge\eta^{*}$ for its view $Z$, where $\Hinf$ is the average
conditional min-entropy, then it has not queried $H$ at $\pi\|X$, for any fixed
prefix $\pi$, except with probability $q_H2^{-\eta^{*}}$: conditionally on $Z=z$
the query list is determined and contains $\pi\|X$ with probability at most
$q_H\max_x\Pr[X=x\mid Z=z]$, and averaging over $z$ gives $q_H2^{-\Hinf(X\mid Z)}$.
On the complement $H(\pi\|X)$ is uniform, so $H$ turns $\eta^{*}$ bits of
unpredictability into $\lambda$ bits that are within $q_H2^{-\eta^{*}}$ of
uniform.

\begin{remark}[Why the target is twice the output length]
\label{rem:lhl}
In the analysis of this paper the window's entropy enters only through the term $q_H2^{-\eta^{*}}$ of \Cref{lem:ext1}, so against $q_H\le2^{128}$ oracle queries a target of $\eta^{*}=256$ already keeps that term at most $2^{-128}$. The target $\eta^{*}=2\lambda=512$ doubles this, and the extra $256$ bits are a margin. It is also the target that the standard model asks for. Replacing $H$ by a universal hash function with a public seed, the leftover hash lemma~\cite{ILL89,HILL99,BDK+11} emits $\lambda=256$ bits within statistical distance $\tfrac12\sqrt{2^{\lambda-\eta^{*}}}=2^{-129}$ of uniform when $\eta^{*}=512$, and the same accounting applied to a public financial source of low entropy is in~\cite{BCG15}. That lemma requires the source to be independent of the seed, which a party that trades after the seed is published need not respect; with $H$ modelled as a random oracle the requirement does not arise, and the target keeps its margin.
\end{remark}

\paragraph{The broadcast.}
All messages of the consensus version are sent over a broadcast with four
properties. \emph{Delivery}: a message sent when the window closes reaches every
party before the consensus phase ends. \emph{Agreement}: every party holds the
same messages at the commitment deadline. \emph{Attribution}: a message carries
the identity of its sender, and one ascribed to an honest party was indeed sent
by it. \emph{Persistence}: a message once received can be neither retracted nor
altered. These are the properties of reliable broadcast~\cite{Bra87},
\cite[\S3.10]{CGR11}, in the authenticated form that a signature scheme
supplies~\cite{LSP82,DS83}. The resilience that the primitive itself needs is stated here, since it is a property of the channel: the authenticated construction tolerates any number of deviating senders given a public-key infrastructure, and the unauthenticated construction of~\cite{Bra87} requires $n>3f$.

\paragraph{The vote.}
The rule that selects the reference string is a majority vote, in its simplest
form: it counts the strings committed during the epoch and returns one of them if
it has a majority. We write $n$ for the number of reporters and
$\theta=\lfloor n/2\rfloor+1$ for a majority of them.

\begin{definition}[Majority vote]
\label{def:vote}
A \emph{majority vote} takes the strings committed in an epoch and returns either
one of them or $\bot$, subject to two requirements:
\emph{agreement}, that every party which receives the same broadcast messages
obtains the same answer; and \emph{validity}, that the answer is $b\neq\bot$ only
if at least $\theta=\lfloor n/2\rfloor+1$ reporters committed to $b$, and is not
$\bot$ whenever some string was committed by at least $\theta$ reporters.
\end{definition}

\paragraph{The commitment.}
A reporter fixes its string by a commitment, and a non-hiding one suffices: the
market is public, and a reporter conceals nothing that the other parties have not
just watched for themselves. What the message has to do is fix a string at a
deadline, so we ask only that the string can be read off it and that it cannot
afterwards be changed.

\begin{definition}[Non-hiding commitment]
\label{def:commit}
A \emph{non-hiding commitment} of $P_i$ to a string $b\in\mathcal{B}$ for epoch
$e$ is a broadcast message $c=\NHC(e,i,b)$ which is \emph{recoverable}, in that
there is an efficient map $\Ext$ with $\Ext(c)=(e,i,b)$; and \emph{binding}, in
that once $c$ has been received, no efficient party, $P_i$ included, can cause a
string other than $b$ to be attributed to the pair $(e,i)$.
\end{definition}

That a non-hiding commitment is enough is a consequence of the delay and not an
independent design choice, and \Cref{app:primitives} makes the point precisely:
without the delay, the reporter that speaks last reads the others, enumerates the
reference strings its own message can still produce, and computes the value of
each.

\paragraph{Adversary.}

$\mathcal{A}$ is an \emph{admissible adversary} if it holds $K$ processors of speedup $A$, makes at most $q_H$ queries to $H$ and completes at most $q_E$ evaluations of $\Eval$, the two budgets of \Cref{sec:primitives}. It may also trade in the observed tokens. One evaluation costs $\Delta/A$
units of time on one processor, so an adversary preparing over a wall-clock
interval of length $W$ before the window opens has
\begin{equation}
  \label{eq:qE}
  q_E \;\le\; \frac{KWA}{\Delta} \;<\; \frac{KW}{\tau+\rho},
\end{equation}
the second inequality by \Cref{ass:delay}, and by \Cref{ass:delay1} with
$\rho=0$ when the bound is applied to the basic version.
We write $\view_{\mathcal{A}}(t)$ for its view at time $t$, which contains $pp$, the messages delivered by then, the trades it has observed or placed, its own state
and randomness, and its trading strategy for the current window. The view only
grows with $t$, and every action the adversary takes at time $t$ is a function of
$\view_{\mathcal{A}}(t)$, which is the only property of it the analysis uses.

\subsection{The basic beacon}
\label{sec:v1}
We give the protocol in two versions. The basic version of this section is the
construction standard for a public source read at a deadline: hash the source,
delay, hash again. We restate it in our notation and prove what goals it achieves.

A \emph{publisher} reads the window $[t_e,t_e+\tau)$ through a data provider and
encodes it by \eqref{eq:binarization} into a string $\overline{b}$. It computes
\begin{equation}
  \label{eq:seed1}
    s_e \;=\; H\bigl(\mathsf{dom}_1\,\|\,\langle e\rangle\,\|\,s_{e-1}\,\|\,\overline{b}\bigr)
  \qquad
  (y_e,\pi_e)\;=\;\Eval(pp,s_e),
\end{equation}
and
\begin{equation}
  \label{eq:value1}
  B_e \;=\; H\bigl(\mathsf{dom}_2\,\|\,\langle e\rangle\,\|\,s_e\,\|\,y_e\bigr)
  \;\in\;\{0,1\}^{\lambda}.
\end{equation}
It broadcasts $\overline{b}$ when the window closes, at $t_e+\tau$, and $(B_e,y_e,\pi_e)$ when the evaluation completes. The two applications of
$H$ do different work: the first extracts from the window, and the second is
needed because sequentiality makes $y_e$ unpredictable rather than uniform, so
$B_e=y_e$ would not meet the required properties, introduced in the next subsection. 
The second consumes no source entropy, its input being the seed and the delay output.

The delay is what answers an adversary that trades in the window: it cannot learn the value that a candidate window implies before the window closes, so it cannot choose its trades by their effect on the output. Without a delay, an adversary that places the last trades of the window can try each continuation and keep the one whose value it prefers (``Hashing alone'' in \Cref{app:primitives}). This is an instance of a general impossibility: no deterministic function extracts uniform randomness from a source whose adversarial parts may depend on the parts that precede them~\cite{AORSV20}, and the delay function sidesteps it computationally.

\subsubsection{The goals}
\label{subsec:v1goals}
\label{subsec:goals}
The three goals below specify what properties a beacon should provide. The first two are the headline properties
of~\cite[\S II.D]{CMB23}, whose third, liveness, we treat in \Cref{app:v2}; the
third is the verifiability of~\cite[\S5.2]{CH10}, and it is the one that
separates the two versions.

\begin{goal}[$\beta$-unbiasability]
\label{goal:unbias}
$\bigl|\Pr[B_e\in E \mid g_e=1]-p_E\bigr|\le\beta$ for every
$E\subseteq\{0,1\}^{\lambda}$ and every admissible adversary, where
$p_E=|E|/2^{\lambda}$.
\end{goal}
\begin{goal}[$(T,\delta)$-unpredictability]
\label{goal:unpred}
No adversary can output $B_e$ before time $T$ except with probability $\delta$.
\end{goal}
\begin{goal}[Verifiability]
\label{goal:verif}
A party that reads the window through a data provider of its own choosing,
obtaining $b_V$, and runs the audit on the published record obtains
$\mathsf{Audit}=\mathsf{yes}$ whenever that record was produced faithfully from
the market, and $\mathsf{Audit}=\mathsf{no}$ whenever it was not.
\end{goal}
We write $\mathsf{Audit}$ for the protocol's own verification procedure, to keep it apart from the delay function's $\Vf$, which it calls. The first half of the goal is what a deployment needs: a user that reads the market honestly must not be told no. The second is what keeps the first from being met by an audit that answers yes to everything. It is the first half that the basic version loses.
An \emph{admissible adversary} is the adversary described in \Cref{sec:primitives}.

\subsubsection{Assumptions of the basic version}
\label{subsec:v1ass}
Four assumptions serve both versions, and we state them in the form the basic
version needs. Two name a party and a duration and are restated for the registry
of reporters in \Cref{sec:v2}; those are tagged 1a and 2a here, 1b and 2b there,
so that an assumption keeps its number across the two versions.

\renewcommand{\theassumption}{1a}
\begin{assumption}[Honest publisher]
\label{ass:pub}
The publisher obtains its string from the window by \eqref{eq:binarization} and
derives $(B_e,y_e,\pi_e)$ from it by \eqref{eq:seed1}--\eqref{eq:value1}.
\end{assumption}
\renewcommand{\theassumption}{2a}
\begin{assumption}[Delay dominance]
\label{ass:delay1}
$\Delta>A\tau$.
\end{assumption}
\renewcommand{\theassumption}{3}
\begin{assumption}[Source entropy]
\label{ass:source}
For every admissible adversary and every epoch,
$\Hinf\bigl(b^{*}\mid\view_{\mathcal{A}}(t_e)\bigr)\ge\eta^{*}$, where $b^{*}$ is
the reading of the window.
\end{assumption}
\renewcommand{\theassumption}{4}
\begin{assumption}[Reading accuracy]
\label{ass:eps}
There are a radius $\varepsilon\ge0$ and a lag $\rho_{\mathrm{fin}}$, after which
the venue's record no longer changes, such that an honest party reading the window
later than that obtains a string within $\varepsilon$ of $b^{*}$; in particular a
verifier obtains $b_V$ with $d(b_V,b^{*})\le\varepsilon$. Neither parameter is
public: both belong to the verification step alone.
\end{assumption}

\Cref{ass:delay1} asks the delay to outlast the window, so that no candidate
value can be evaluated while the adversary can still trade. \Cref{ass:source} is
the only assumption about the market;
it conditions on the adversary's view at the start of the window, hence on
its own trading strategy, and asserts that a party that trades cannot make any
single reading more likely than $2^{-\eta^{*}}$ (\Cref{rem:proxy} says what the
circuit breaker does and does not certify about it). \Cref{ass:eps} is the only
one whose parameter the protocol never uses, and the whole difference between the
two versions lies in the value it may take: the basic version reaches
\Cref{goal:verif} only at $\varepsilon=0$ and the consensus version at any
$\varepsilon$.

\subsubsection{What the basic version achieves}
\label{subsec:v1thm}
\begin{theorem}
\label{thm:v1}
Let $H$ be a random oracle and the delay function be as in
\Cref{sec:primitives}, and let $p_{\min}>0$ be a lower bound on $\Pr[g_e=1]$.
Under Assumptions \ref{ass:pub}, \ref{ass:delay1} and \ref{ass:source}, the basic version is
$\beta_0/p_{\min}$-unbiasable in the sense of \Cref{goal:unbias} and
$(t_e+\Delta/A,\;\beta_0+q_H2^{-\lambda})$-unpredictable in the sense of
\Cref{goal:unpred}, with
\begin{equation}
  \label{eq:beta0}
  \beta_0 \;=\; q_H\,2^{-\eta^{*}}\;+\;(q_H+1)\,q_E\,2^{-\lambda}\;+\;\epsilon_{\mathrm{seq}} .
\end{equation}
\end{theorem}

The oracle is fixed before the window opens, and the window's entropy enters through the first term: it bounds the probability that the adversary queried, and so prepared for, the point at which the seed is computed.

The proof is in \Cref{app:v1}. Both conclusions rest on \Cref{ass:pub} through a
single step: the string the oracle is queried at
carries $\eta^{*}$ bits given the adversary's view at $t_e$, and that follows for $\overline{b}$ only because \Cref{ass:pub} makes $\overline{b}$ a faithful reading and \Cref{ass:source} gives the window its entropy. 
\Cref{goal:verif} is the goal this version reaches only if we suppose that two honest users always download the same string.

\begin{proposition}
\label{prop:v1audit}
Let $\mathsf{Audit}$ form $s_V=H(\mathsf{dom}_1\|\langle e\rangle\|s_{e-1}\|b_V)$ from the
verifier's own reading and answer yes iff
\[
  \Vf(pp,s_V,y_e,\pi_e)=1
  \quad\text{and}\quad
  B_e = H\bigl(\mathsf{dom}_2\|\langle e\rangle\,\|\,s_V\,\|\,y_e\bigr).
\]
Then $\mathsf{Audit}=\mathsf{yes}$ if and only if the publisher derived the
record from $b_V$, except with probability $\epsilon_{\mathrm{uniq}}$, the
soundness error of the delay function, and the probability of a collision in $H$.
\end{proposition}
\begin{proof}
If $\Vf$ accepts then $y_e=\Eval(pp,s_V)$ up to the soundness error, so the second equality says that the published value is the one $s_V$ determines; and
$s_V$ determines $\overline{b}$ except on a collision. \qed
\end{proof}

The verifier runs $\Vf$ and not $\Eval$, so the audit costs polylogarithmic. But it decides whether the
publisher used \emph{the verifier's own} reading, which is the same question as
faithfulness only when $b_V=b^{*}$. If both feeds are exact, both halves of
\Cref{goal:verif} hold and the basic version attains it.
They stop holding as soon as the feeds are not exact, since two honest parties reading the same window through different providers may not obtain the same string, because sometimes feeds drop
records, deliver them late, and revise them for a short interval after the window closes. Hashes do not compare approximately, so a verifier whose feed dropped a single record is told no by an honest publisher, and the first half of the goal fails. 

\begin{proposition}
\label{prop:noaudit}
Suppose $\varepsilon>0$, so that an honest reading may differ from $b^{*}$. Then
no audit whose inputs are the published record and the verifier's own reading
answers yes to every faithful record and no to every unfaithful one. The basic
version does not attain \Cref{goal:verif}.
\end{proposition}
\begin{proof}
Take $b_V\neq b^{*}$ with $d(b_V,b^{*})\le\varepsilon$ and let each in turn be
the venue's record: either the record is $b^{*}$ and the publisher is honest, or
it is $b_V$ and the publisher substitutes $b^{*}$. The audit sees the same triple
and the same $b_V$ in both, so its verdict is the same, yet only the first is
faithful. \qed
\end{proof}

Then, our basic version cannot achieve \Cref{goal:verif} if
$\varepsilon>0$. If $\varepsilon=0$ the second execution cannot arise, the test
of \Cref{prop:v1audit} decides faithfulness, and the basic version attains all
three goals with no consensus phase. But, in the general approach with $\varepsilon>0$, we should use the consensus version of our protocol, which we introduce in the next subsection.

\subsection{The beacon with consensus}
\label{sec:v2}
The second version replaces the single publisher by the registry $\mathcal{P}$ of
\Cref{subsec:setting}, each reporter reading the window through a provider of its
own. The delay function is unchanged; we add one round of messages and
a public rule for reading them. At the close of the window each reporter broadcasts a non-hiding commitment
(\Cref{def:commit}) to the string it read. The commitments held at the deadline $T^{*}_e$ fix the \emph{admitted set} $S$, a reporter that broadcasts two strings for one epoch being excluded publicly and without adjudication, and the reference string is the majority answer of \Cref{def:vote},
\begin{equation}
  \label{eq:vote}
  \hat b \;=\;
  \begin{cases}
    b    & \text{if } \nu(b)=\bigl|\{i\in S: b_i=b\}\bigr|\ge\theta,\\
    \bot & \text{if no string reaches } \theta,
  \end{cases}
\end{equation}
which is well defined because $\theta>n/2$. From $\hat b$ the epoch proceeds as in
\Cref{sec:v1}: the gate of \Cref{sec:circuitbreaker}, then \eqref{eq:seed1} and
\eqref{eq:value1} with $\overline{b}$ replaced by $\hat b$.

The audit changes with it. A verifier recomputes the value from the broadcast messages, then fixes a tolerance $t_{\mathrm{aud}}$, admits the reporters whose strings lie within $t_{\mathrm{aud}}$ of its own reading, and accepts when a majority are admitted. Neither check asks its feed to be exact, so the tolerance that defeated the basic version is what this one runs on, and that is why the commitments are not hiding: the audit needs the strings in the clear.
\Cref{app:spec} gives the protocol, the audit and the notation in full.

\subsubsection{Assumptions of the consensus version}
\label{subsec:v2ass}
\Cref{ass:source,ass:eps} are carried over word for word, and this time the
radius $\varepsilon$ of \Cref{ass:eps} may be positive. The two assumptions that
name a party and a duration are restated for the registry, and nothing else is
added.

\renewcommand{\theassumption}{1b}
\begin{assumption}[Reporting majority]
\label{ass:agree}
Of the $n$ registered reporters, at least $\theta=\lfloor n/2\rfloor+1$ are
admitted at $T^{*}_e$ and commit the reading of the window:
$\bigl|\{i\in S: b_i=b^*\}\bigr|\ge\theta$. Equivalently, at most $n-\theta$
reporters either fail to be admitted or commit some other string, whether
because they deviated or because their provider delivered a different window.
\end{assumption}
\renewcommand{\theassumption}{2b}
\begin{assumption}[Delay dominance]
\label{ass:delay}
$\Delta>A(\tau+\rho)$.
\end{assumption}

\Cref{ass:agree} speaks of behaviour in one epoch rather than of who is corrupt:
a reporter fails it by deviating or because its provider dropped a record, and
the assumption does not distinguish the two. That indifference is what makes it
testable, the audit returning the reporters outside the tolerance without
deciding why, and it absorbs the turnout condition of the gate.

\subsubsection{What the consensus version achieves}
\label{subsec:v2thm}
\begin{theorem}
\label{thm:v2}
Let $H$ be a random oracle and the delay function be as in
\Cref{sec:primitives}, and let $p_{\min}>0$ be a lower bound on $\Pr[g_e=1]$.
Under Assumptions \ref{ass:agree}, \ref{ass:delay}, \ref{ass:source} and \ref{ass:eps}, the
consensus version attains
the three goals: \Cref{goal:unbias} with $\beta=\beta_0/p_{\min}$ and $\beta_0$
as in \eqref{eq:beta0}; \Cref{goal:unpred} with $T=T^{\min}_e=t_e+\Delta/A$ and
$\delta=\beta_0+q_H2^{-\lambda}$; and \Cref{goal:verif} for any
$t_{\mathrm{aud}}\ge2\varepsilon$, together with a list of the reporters that
deviated by more than that tolerance. It is also live, in the sense made precise
in \Cref{app:v2}.
\end{theorem}
Proved in \Cref{app:v2}, the bounds follow \Cref{thm:v1}, with the consensus phase adding two steps. The first is that
the admitted set, and with it $\hat b$, $g_e$ and $s_e$, is final at $T^{*}_e$
and identical for every reader. The second is that under \Cref{ass:agree} the
reference string is the reading of the window, $\hat b=b^{*}$, and this does the
work twice over: it turns the hypothesis of \Cref{thm:v1} about the published
string into a hypothesis about the market, and it is what puts \Cref{goal:verif}
out of the reach of \Cref{prop:noaudit}, a verifier now seeing the whole set of
readings rather than a single string.
Without \Cref{ass:agree}, a coalition that controls the vote must have evaluated its candidates before the window opened, which leaves it the best of $q_E+1$ of them: the bound degrades to the multiplicative form of \Cref{prop:freedom} rather than collapsing. This is \Cref{prop:degrade}, stated and proved in \Cref{app:v2}.

\subsection{The two versions}
\label{subsec:vcompare}

\begin{table}[!t]
\centering\footnotesize
\caption{Goals attained by each combination of primitives (\S\ref{sec:primitives}).}
\label{tab:primitives}
\begin{tabular}{@{}llll@{}}
\toprule
 & unbiasability & unpredictability & verifiability\\
\midrule
$H$              & no                 & no             & only if $\varepsilon=0$\\
$H$, delay (\S\ref{sec:v1})      & $\beta_0/p_{\min}$ & $t_e+\Delta/A$ & only if $\varepsilon=0$\\
$H$, vote        & no                 & no             & yes, at any $\varepsilon$\\
$H$, delay, vote (\S\ref{sec:v2}) & $\beta_0/p_{\min}$ & $t_e+\Delta/A$ & yes, at any $\varepsilon$\\
\bottomrule
\end{tabular}
\end{table}

The two versions differ in three places and agree everywhere else including their reliance on \Cref{ass:source}. As shown in rows 2 and 4 of Table~\ref{tab:primitives}, both attain \Cref{goal:unbias} with $\beta_0/p_{\min}$ and \Cref{goal:unpred} at $t_e+\Delta/A$, the consensus phase leaving those two where it found them (see \Cref{app:primitives}). We now explain what the differences are. The basic version needs \Cref{ass:pub}, a hypothesis about a designated
party that no observer can test and whose failure is silent, where the consensus
version needs \Cref{ass:agree}, testable through \Cref{goal:verif} and failing by
\Cref{prop:degrade} rather than totally. The delay must outlast $\tau$ rather
than $\tau+\rho$. And \Cref{goal:verif} is reached only at $\varepsilon=0$ (i.e., if feeds coincide and two honest users download the same string) rather than at any $\varepsilon$; by \Cref{prop:noaudit} that last is the whole of the difference, so if honest readings never disagreed the round of messages would be redundant.

The majority is necessary in the consensus version: \Cref{prop:nomaj} runs the argument
of \Cref{prop:noaudit} with a reporter set and shows $\theta=\lfloor n/2\rfloor+1$
to be the least threshold at which the consensus phase works. Without it, \Cref{goal:verif} is lost, but unbiasability degrades rather than disappears: however many reporters deviate, \Cref{prop:degrade} bounds the probability of every event $E$ by $(q_E+1)p_E$, up to $\epsilon_{\mathrm{seq}}$ and the factor $1/p_{\min}$. The coalition can at best choose the most favourable of $q_E+1$ candidate values. This keeps out of its reach every event much rarer than $1/(q_E+1)$, such as a value fixed beforehand, though not the events of probability $1/(q_E+1)$ or more, so the bound is weaker than \Cref{goal:unbias}. What the majority buys is that the string the epoch uses is the market's own reading, and with it the additive bound of \Cref{thm:v2}. Even with every reporter deviating, the coalition cannot set the value: it can choose among its candidates or close the gate, and a closed gate is a signed public refusal. This is where the design parts from the beacons that draw randomness from their participants, where corrupting the threshold yields the output itself: \Cref{app:cointoss} compares them.

\subsection{Specification of the consensus beacon}
\label{app:spec}
This subsection carries the notation, the protocol in full and the verification
procedure, together with the attack vectors the design answers. \Cref{sec:v2} states what is added to
the basic beacon of \Cref{sec:v1}; here it is written out.

\subsubsection{Setting and Notation}
\label{sec:setting}
A set $\mathcal{P}=\{P_1,\dots,P_n\}$ of \emph{reporters} produces one value
$B_e\in\{0,1\}^{\lambda}$ per epoch $e$. Each reporter reads the market through a
provider of its own choosing and encodes the window by \eqref{eq:binarization},
obtaining a string $b_i$ in the space $\mathcal{B}=\{0,1\}^{*}$. The venue's own
record of the window determines one such string through \eqref{eq:binarization},
which we write $b^*$ and call the \emph{reading of the window}: it is what the
reporters are trying to report and what a verifier will later compare against. It
may happen that honest reporters obtain slightly different strings, and this is
why the protocol needs the consensus phase. A verifier, on the other hand, may
read the window later and through a provider of its own, and may then obtain a
string close to but not equal to the committed ones; it can run the verification
steps of \Cref{sec:verification} using a metric of its own choosing.

\paragraph{Adversary.}
In this version, an \emph{admissible} adversary has the same properties as the adversary described in \Cref{sec:primitives}, but it may control a subset of reporters, namely $\mathcal{A}\subset\mathcal{P}$ with $|\mathcal{A}|\leq f $. It sees every message on the broadcast as it is received, and may also trade in the observed tokens, hence both influences the source and observes it progressively as the window unfolds.

\Cref{tab:notation} collects the notation.

\begin{table}[htbp]
\centering\footnotesize
\caption{Notation.}
\label{tab:notation}
\begin{tabular}{@{}ll@{\hspace{0.4em}}ll@{}}
\toprule
\multicolumn{2}{@{}l}{\emph{Parties, data, verification}} & \multicolumn{2}{l}{\emph{Timing, resources, objects}}\\
\midrule
$\mathcal{P},n$        & reporters, $|\mathcal{P}|=n$                & $\tau,\rho$   & window, consensus phase\\
$\mathcal{A},f$        & deviating reporters, $|\mathcal{A}|\le f$   & $T^{*}_e$ & commitment deadline\\
$\theta$               & $\lfloor n/2\rfloor+1$                      & $T^{\min}_e$  & earliest $B_e$\\
$T_1,\dots,T_J$        & observed tokens                             & $\rho_{\mathrm{fin}}$ & lag before verifying\\
$\mathcal{B}$          & strings, $\{0,1\}^{*}$                      & $t_e$         & start of epoch $e$\\
$\mathsf{Audit}$       & the verification of \Cref{sec:verification} & $\Delta,A,K$  & delay, speedup, parallelism\\
$\Hinf$                & conditional min-entropy                     & $W$           & preparation interval, \eqref{eq:qE}\\
$b_i$                  & string committed by $P_i$                   & $q_H,q_E$     & hash queries, evaluations\\
$b^*$              & reading of the window                       & $\kappa,\lambda$ & security, output length\\
$b_V$                  & a verifier's own string                     & $\epsilon_{\mathrm{seq}},\epsilon_{\mathrm{uniq}}$ & VDF errors\\
$\hat b$               & reference string, \eqref{eq:vote}           & $s_e,y_e,\pi_e,B_e$ & seed, output, proof, value\\
$S,\nu(b)$             & admitted set, its counts                    & $g_e$         & gate, \eqref{eq:gate}\\
$\epsilon_{\mathrm{ext}}$ & extraction distance                      & $N,\hat h,\hat\sigma_h$ & length, rate, error\\
$\view_{\mathcal{A}}(t)$ & adversary's view at $t$                   & $\Ecal,\Ecal^{-}$ & entropy, \eqref{eq:Ecal}\\
$d,\varepsilon$        & metric and radius, \Cref{ass:eps}           & $\eta,\eta^{*}$ & threshold, entropy floor\\
$t_{\mathrm{aud}},\mathcal{V}$ & tolerance, verified set             & $\mathcal{Y},\Lambda$ & precomputed seeds, bad event\\
$m_{\min},L_{\max}$    & minimum posts, size cap                     & $E,p_E$       & target set, $|E|/2^{\lambda}$\\
$pp,(\Stp,\Eval,\Vf)$  & delay function, \Cref{sec:primitives}       & $\beta_0,p_{\min}$ & \eqref{eq:beta0}, $\Pr[g_e=1]$ floor\\
$H$                    & hash into $\{0,1\}^{\lambda}$               & $\mathsf{dom}_0,\mathsf{dom}_1,\mathsf{dom}_2$ & domain separators\\
$\langle e\rangle$     & encoding of $e$; $\|$ concatenation         & $\bot$        & the vote declines\\
\bottomrule
\end{tabular}
\end{table}

\subsubsection{The Protocol}
\label{subsec:protocol}
\paragraph{Public parameters.}
The tokens $T_1,\dots,T_J$, the window of length $\tau$ and the encoding
\eqref{eq:binarization}; the registry $\mathcal{P}$ with its verification keys;
$n$, $f$ and $\theta=\lfloor n/2\rfloor+1$; the consensus phase of length $\rho$;
VDF parameters $pp\leftarrow\Stp(1^{\kappa},\Delta)$ with $\Delta$ satisfying
\Cref{ass:delay}; the gate threshold $\eta$ and the minimum number of commitments
$m_{\min}\le\theta$; the size cap $L_{\max}$; the hash $H$ with a fixed-length
encoding $\langle\cdot\rangle$ of the epoch index and pairwise distinct domain
separators $\mathsf{dom}_0,\mathsf{dom}_1,\mathsf{dom}_2$; and a genesis seed $s_0$.
Epoch $e$ begins at $t_e$ and runs as in \Cref{fig:timeline}. We write
$T^{*}_e=t_e+\tau+\rho$ for the commitment deadline.

\begin{description}
\item[Phase 1: Observation, $[t_e,t_e+\tau)$.]
Each $P_i$ follows the trades of $T_1,\dots,T_J$ through its provider and builds
$b_i\in\mathcal{B}$ by \eqref{eq:binarization}. Nothing is published.

\item[Phase 2: Consensus, $[t_e+\tau,T^{*}_e)$.]
This is the phase in which the parties come to agree on a single input for the
delay function. It consists of one round of messages followed by a deterministic
public rule.

\emph{Commitment.} At $t_e+\tau$ each $P_i$ broadcasts its non-hiding commitment
(\Cref{def:commit}) to the string it observed, which is its proposal for the
epoch,
\begin{equation}
  c_i \;=\; \NHC\bigl(e,\ i,\ b_i\bigr).
  \label{eq:commit}
\end{equation}
By delivery of the broadcast every party holds it by $T^{*}_e$. The epoch index binds the proposal to
one epoch, so that it cannot be replayed into another. A commitment is admitted if
$|b_i|\le L_{\max}$ and if $P_i$ broadcast no other commitment for $e$; a reporter
that broadcast two distinct strings for the same epoch is excluded from the epoch,
and, since by agreement every party holds both of its messages, that exclusion is
publicly verifiable and attributable without adjudication. At the deadline
$T^{*}_e$ the messages held fix the admitted set $S$.

\emph{Derivation.} Every party computes the following from $S$ alone. Each step is
a deterministic function of its input, so every reader obtains the same result.
\begin{itemize}
    \item[D1] \emph{Vote.} Put the admitted strings to the agreement rule of
    \Cref{def:vote}: with $\nu(b)=\bigl|\{i\in S: b_i=b\}\bigr|$ the answer is
    the $\hat b$ of \eqref{eq:vote}, well defined because $\theta>n/2$, so that
    at most one string can meet the threshold.
\item[D2] \emph{Gate.} The epoch publishes iff a reference string exists, enough
reporters spoke, and that string clears the entropy target with the margin of
\Cref{sec:circuitbreaker}:
\begin{equation}
  g_e=1 \iff \hat b\neq\bot \ \text{ and }\ |S|\ge m_{\min} \ \text{ and }\
  \Ecal^{-}(\hat b)\ \ge\ \eta .
  \label{eq:gate}
\end{equation}
\item[D3] \emph{Seed.}
\begin{equation}
  s_e=
  \begin{cases}
    H\bigl(\mathsf{dom}_1\,\|\,\langle e\rangle\,\|\,s_{e-1}\,\|\,\hat b\bigr) & \text{if } g_e=1,\\[2pt]
    H\bigl(\mathsf{dom}_0\,\|\,\langle e\rangle\,\|\,s_{e-1}\bigr) & \text{if } g_e=0.
  \end{cases}
  \label{eq:seed}
\end{equation}
Every field but the last has a length fixed in the public parameters, so the
argument of $H$ determines the quantities it was built from, and the two
separators keep a published epoch and a skipped one apart. If $g_e=0$, no value is
published for the epoch and the seed chain advances anyway, so that the next epoch
is well defined.
\end{itemize}
\item[Phase 3: Delay, of length $\Delta$ from $T^{*}_e$.]
If $g_e=1$, at least one honest party computes
$(y_e,\pi_e)\leftarrow\Eval(pp,s_e)$ starting at $T^{*}_e$ and broadcasts the
pair. The value of the epoch is
\begin{equation}
  B_e \;=\; H\bigl(\mathsf{dom}_2\,\|\,\langle e\rangle\,\|\,s_e\,\|\,y_e\bigr)
  \;\in\;\{0,1\}^{\lambda}.
  \label{eq:value}
\end{equation}
Any party checks the pair by $\Vf(pp,s_e,y_e,\pi_e)$ and then recomputes
\eqref{eq:value} for itself; by delivery every party holds the pair by
$T^{*}_e+\Delta+\rho$. Since $s_e$ is fixed at $T^{*}_e$ and every input
to it is public, \emph{any} party can perform the evaluation, so the honest
evaluator of this phase need not be a reporter.
\end{description}

The final hash is what separates the output space of the delay function from the
beacon's own: $y_e$ ranges over whatever algebraic structure the VDF is built on
and is in no sense uniform on $\{0,1\}^{\lambda}$, so \Cref{goal:unbias} would be
unreachable for $B_e=y_e$. By \Cref{rem:lhl} this step consumes no source entropy.

Two facts about this phase are used throughout and proved in \Cref{app:proof}:
the admitted set, and with it $\hat b$, $g_e$ and $s_e$, is final at
$T^{*}_e$ and identical for every reader (\Cref{prop:binding}); and under
\Cref{ass:agree} the reference string is the reading of the window,
$\hat b=b^*$ (\Cref{prop:ref}).

\subsubsection{Verification}
\label{sec:verification}
A verifier acts after publication and at leisure, through a provider of its own
choosing; by \Cref{ass:eps} a reading made after $t_e+\tau+\rho_{\mathrm{fin}}$
yields a string $b_V$ within $\varepsilon$ of $b^*$. This is the first and only
point at which the lag $\rho_{\mathrm{fin}}$ and the radius $\varepsilon$ are
used. The verifier runs two checks, and only the second uses market data and a
metric of its own choosing.
\begin{description}
\item[V1 (arithmetic)]
Recompute $S$ from the messages held; recompute $\hat b$, $g_e$ and $s_e$ from
\eqref{eq:vote}, \eqref{eq:gate} and \eqref{eq:seed}; check
$\Vf(pp,s_e,y_e,\pi_e)=1$ and recompute $B_e$ by \eqref{eq:value}. This uses no
market data at all.
\item[V2 (anchoring)]
Fix a tolerance $t_{\mathrm{aud}}\ge2\varepsilon$ and compute
$\mathcal{V}=\{\,i\in S: d(b_V,b_i)\le t_{\mathrm{aud}}\,\}$. Accept the epoch as
anchored iff $|\mathcal{V}|\ge\theta$, and flag every $i\in S\setminus\mathcal{V}$.
\end{description}

Note that $B_e$ does not depend on $t_{\mathrm{aud}}$, which occurs nowhere in
\eqref{eq:vote}--\eqref{eq:value}. Each verifier chooses its metric and its
tolerance as it prefers, and neither choice affects other verifiers or the value
of the epoch.

\subsubsection{Attack Vectors}
\label{sec:attacks}

\cite{CMB23} organizes distributed beacons by the way entropy reaches the output
rather than by attack, and we are not aware of a taxonomy of attack vectors in the
literature that we could adopt. \Cref{tab:attacks} is therefore our own list of
the vectors a beacon read from a live market has to answer, together with the part
of the analysis that answers each. Where a vector already has a name in the
literature we use that name and give the reference; one of the entries is
particular to this setting and we mark it as ours.

\begin{table}[htbp]
\centering\small
\caption{Attack vectors and countermeasures. The right column names the
assumption, theorem or proposition that answers the vector. A2 has no prior name
that we are aware of.}
\label{tab:attacks}
\begin{tabular}{@{}lll@{}}
\toprule
 & Attack vector, and where it acts & Countermeasure\\
\midrule
A1 & market manipulation~\cite{CH10}, in $[t_e,t_e+\tau)$   & \Cref{ass:source}, \Cref{thm:v2}\\
A2 & strategic discrepancy (ours), its own commitment       & \Cref{prop:ref}, \Cref{prop:degrade}\\
A3 & feed corruption, the data provider                     & \Cref{ass:agree}; flagged by V2\\
A4 & withholding~\cite{BGB17}, its own commitment           & \Cref{prop:liveness}, \Cref{cor:abort}\\
A5 & equivocation, two commitments in one epoch             & \Cref{prop:binding}\\
A6 & precomputation~\cite{CCB24}, before $t_e$              & \Cref{ass:delay}, \Cref{lem:noeval}\\
A7 & replay, across epochs                                  & \eqref{eq:commit}, \eqref{eq:seed}\\
A8 & last-revealer~\cite{CCB24}, not applicable             & no reveal phase exists\\
\bottomrule
\end{tabular}
\end{table}

A1 is the vector financial beacons have to deal with: adversaries that trade in order to move the price and drive the outcome of the beacon. A4 is the analogue of the block
withholding of~\cite{BGB17}, where a miner discards a block whose beacon output
it dislikes, and of the coordinator censorship of~\cite{CCB24}; here it costs the
adversary nothing, and \Cref{cor:abort} is the statement that it also buys it
nothing. A6 is the precomputation phase of the adversary
of~\cite[\S3]{CCB24}, and it is the vector the delay function is aimed at.

A8 deserves a word, because its absence is a design decision rather than an
oversight. In a commit-and-reveal beacon a participant observes the other
openings, computes the outcome for each of its own choices, and withholds the
unfavourable one, which buys it a bit of bias~\cite{CCB24}. Here nothing is
hidden and therefore nothing is revealed: a commitment carries its string in the
clear (\Cref{def:commit}), so there is no second round in which a party learns
something and then chooses. What remains of that freedom is A4.

\section{Security analysis}
\subsection{Security analysis of the basic beacon}
\label{app:v1}

Throughout, $\overline{b}$ is the string the publisher broadcasts at $t_e+\tau$,
$\view_{\mathcal{A}}(t)$ is the adversary's view at time $t$, and we use the
properties of \Cref{sec:primitives} by the names given there. We write
\[
  x_e \;=\; \mathsf{dom}_1\,\|\,\langle e\rangle\,\|\,s_{e-1}\,\|\,\overline{b},
  \qquad
  x'_e \;=\; \mathsf{dom}_2\,\|\,\langle e\rangle\,\|\,s_e\,\|\,y_e
\]
for the two points at which the oracle is queried, so that $s_e=H(x_e)$ and
$B_e=H(x'_e)$. We prove \Cref{thm:v1} in four steps and \Cref{prop:freedom} in a
fifth. The first step is a consequence of the random oracle that both versions
use.

\subsubsection*{Step 1: what the oracle gives}

\begin{lemma}[Extraction]
\label{lem:ext1}
Let $u$ be a fixed prefix, $Z$ a random variable and $X$ a string with
$\Hinf(X\mid Z)\ge\eta^{*}$. An adversary whose view is $Z$ and which makes at
most $q_H$ oracle queries, queries $H$ at the point $u\|X$ with probability at
most $q_H2^{-\eta^{*}}$; on the complement, $H(u\|X)$ is uniform on
$\{0,1\}^{\lambda}$ and independent of $Z$.
\end{lemma}
\begin{proof}
Conditionally on $Z=z$ the query list is determined, so it contains $u\|X$ with
probability at most $q_H\max_x\Pr[X=x\mid Z=z]$; averaging over $z$ gives
$q_H2^{-\Hinf(X\mid Z)}\le q_H2^{-\eta^{*}}$. On the complement the point is
unqueried, and the value of $H$ at an unqueried point is uniform and independent
of the adversary's view. \qed
\end{proof}

By \Cref{ass:pub} the published string is a faithful reading of the window, so
\Cref{ass:source} gives
$\Hinf(\overline{b}\mid\view_{\mathcal{A}}(t_e))\ge\eta^{*}$ and the lemma
applies with $Z=\view_{\mathcal{A}}(t_e)$ and
$u=\mathsf{dom}_1\|\langle e\rangle\|s_{e-1}$. This is the only place \Cref{ass:pub} is
used, and \Cref{prop:freedom} is what remains without it.

One remark on the instant at which the lemma is applied, because every deadline
below inherits it. The conclusion bounds the probability of a query that is a
function of $\view_{\mathcal{A}}(t_e)$, that is of a query made before the window
opens, and it says nothing about a query made while the window runs. It cannot:
the conditional min-entropy of $\overline{b}$ given $\view_{\mathcal{A}}(t)$
decreases as $t$ grows, and by $t_e+\tau$ an adversary that has watched the feed
holds $\overline{b}$ itself and may query $x_e$ at once. \Cref{ass:source} is
stated at $t_e$, so $t_e$ is the instant from which the delay has to be counted,
and the horizon of \Cref{thm:v1} is $t_e+\Delta/A$ rather than
$t_e+\tau+\Delta/A$.

\subsubsection*{Step 2: nothing is evaluated in time}

\begin{lemma}
\label{lem:noeval1}
On the event that the adversary has not queried $H$ at $x_e$ strictly before
$t_e$, it does not output a list of at most $q_E$ candidates containing $y_e$
before $t_e+\Delta/A$, except with probability $\epsilon_{\mathrm{seq}}$.
\end{lemma}
\begin{proof}
The seed $s_e=H(x_e)$ is learned only by querying $H$ at $x_e$, so on the stated
event the adversary learns $s_e$ no earlier than $t_e$. Sequentiality in the
shortlist form of \Cref{sec:primitives}, for many challenges at once, then gives it $\Delta/A$ steps counted
from that instant, whichever of the seeds it learns during the window becomes $s_e$. \qed
\end{proof}

\Cref{ass:delay1} is what makes the conclusion useful rather than merely true.
With $\Delta>A\tau$ we have $\Delta/A>\tau$, so the interval $[t_e,t_e+\tau)$
during which the adversary may still trade is shorter than the time it needs, and
no candidate value can be evaluated while the window is open. Everything the
adversary chooses, its trades and whether the publisher it controls publishes at
all, is therefore chosen without the value in hand.

\subsubsection*{Step 3: unbiasability}

\begin{proposition}
\label{prop:unbias1}
Under the hypotheses of \Cref{thm:v1}, for every $E\subseteq\{0,1\}^{\lambda}$,
\[
  \bigl|\Pr[\,B_e\in E\ \wedge\ g_e=1\,]-p_E\Pr[\,g_e=1\,]\bigr|\;\le\;\beta_0 ,
  \]
  hence
  \[
  \bigl|\Pr[\,B_e\in E\mid g_e=1\,]-p_E\bigr|\;\le\;\beta_0/p_{\min}.
\]
\end{proposition}
\begin{proof}
Let $\mathcal{Y}$ be the set of seeds on which the adversary has completed an
evaluation of $\Eval$ strictly before $t_e$; it has at most $q_E$ elements and,
those evaluations being finished by then, it is a function of
$\view_{\mathcal{A}}(t_e)$. Three events can spoil the argument:
\begin{align*}
  \Lambda_1 &:\ \text{the adversary queries } H \text{ at } x_e
                \text{ strictly before } t_e ;\\
  \Lambda_2 &:\ s_e\in\mathcal{Y} ;\\
  \Lambda_3 &:\ \text{the adversary outputs a list containing } y_e
                \text{ strictly before } t_e+\Delta/A .
\end{align*}

By Step 1 and \Cref{lem:ext1}, $\Pr[\Lambda_1]\le q_H2^{-\eta^{*}}$. On
$\Lambda_1^{c}$ the point $x_e$ is unqueried at $t_e$\, but the adversary may still steer the window after $t_e$ towards a continuation whose seed it has computed by querying $H$. On $\Lambda_1^{c}$, then, $s_e$ is one of at most $q_H+1$ values of $H$ at points unqueried at $t_e$, namely the adversary's later queries and $x_e$ itself, each uniform on $\{0,1\}^{\lambda}$ and independent of $\view_{\mathcal{A}}(t_e)$, of which $\mathcal{Y}$ is a function; a union bound gives
$\Pr[\Lambda_2\cap\Lambda_1^{c}]\le(q_H+1)q_E2^{-\lambda}$. On $\Lambda_1^{c}$,
\Cref{lem:noeval1} gives
$\Pr[\Lambda_3\cap\Lambda_1^{c}]\le\epsilon_{\mathrm{seq}}$. Writing
$\Lambda=\Lambda_1\cup\Lambda_2\cup\Lambda_3$ and using
$\Lambda=\Lambda_1\cup(\Lambda_2\cap\Lambda_1^{c})\cup(\Lambda_3\cap\Lambda_1^{c})$,
the three bounds add and $\Pr[\Lambda]\le\beta_0$.

Outside $\Lambda$ the adversary neither precomputed $y_e$ nor produced it in
time, so it does not hold $y_e$ at $t_e+\tau$; writing down the query point
$x'_e$ requires exhibiting $y_e$, so that point is unqueried at $t_e+\tau$ and
$B_e=H(x'_e)$ is uniform on $\{0,1\}^{\lambda}$ and independent of
$\view_{\mathcal{A}}(t_e+\tau)$. The gate $g_e$ is a function of $\overline{b}$,
hence of that view, and so is $\Lambda$. Therefore
$\Pr[B_e\in E,\ g_e=1,\ \Lambda^{c}]=p_E\Pr[g_e=1,\ \Lambda^{c}]$, and
\begin{align*}
  \bigl|\Pr[B_e\in E,\ g_e=1]-p_E\Pr[g_e=1]\bigr|
  &\;=\;\bigl|\Pr[B_e\in E,\ g_e=1,\ \Lambda]-p_E\Pr[g_e=1,\ \Lambda]\bigr|\\
  &\;\le\;\Pr[\Lambda]\;\le\;\beta_0 .
\end{align*}
Dividing by $\Pr[g_e=1]\ge p_{\min}$ gives the second form. \qed
\end{proof}

The division is where the gate is paid for, and it is needed for the same reason
in both versions. The publisher decides whether to publish, and that decision is
a function of the string; by \Cref{lem:noeval1} it cannot also be a function of
the value, since nobody holds the value at the moment the decision is taken. What
the division covers is the residue: an adversary able to close the gate exactly
on the epochs where one of the three events had occurred would bias the
conditional distribution even though the joint bound held. Closing the gate buys little. A party that trades can move the gate, since the gate is a function of a string it helps to write, but it cannot make the gate depend on the value, and the division by $p_{\min}$ is all that closing the gate can buy on the value: $\beta_0$ being negligible (\Cref{app:v2}), even a party that forced a refusal in nine epochs out of ten would only multiply the bias bound by ten. Closing the gate is therefore an attack on liveness, and a visible one: a refusal is signed and public, honest refusals follow the clock (\Cref{sec:circuitbreaker}), and in the configuration of \Cref{subsec:bound} the pooled basket had none in 2025, its least favourable five-minute window clearing the target by $189$ bits (\Cref{tab:mixedbound}).

\subsubsection*{Step 4: unpredictability}

To output $B_e$ the adversary either does not query $H$ at $x'_e$, in which case
each of its at most $q_H$ outputs matches a uniform unqueried value with
probability $2^{-\lambda}$; or it queries $x'_e$, which exhibits $y_e$ and
therefore places the execution in $\Lambda$. Adding the two gives
$\beta_0+q_H2^{-\lambda}$, which is \Cref{goal:unpred} with $T=t_e+\Delta/A$ and
completes the proof of \Cref{thm:v1}. The honest publisher holds $B_e$ at
$t_e+\tau+\Delta$, so the adversary may be ahead of the consumers by up to
$\tau+\Delta(1-1/A)$; this is inherent to any beacon built on a delay
function~\cite{BGB17,LW15,CCB24}, and it is the reason $A$ appears in
\Cref{ass:delay1} rather than being assumed away.

\subsubsection*{Step 5: what remains without Assumption 1a}

\Cref{prop:v1audit} and \Cref{prop:noaudit} are stated and proved in \Cref{subsec:v1thm}; we use them here without repeating the arguments.

So \Cref{prop:v1audit} decides whether the record was derived from the verifier's
own reading, and \Cref{prop:noaudit} shows that no audit reading the record and
one string decides faithfulness exactly. The basic version has no tolerance to
set: its audit compares hashes, and a verifier whose feed dropped one record is
told no by an honest publisher. One may ask what the nearest tolerant audit would
buy, since it is the shape the consensus version generalises: let the publisher
broadcast $\overline{b}$ when the window closes, at $t_e+\tau$, as in \Cref{sec:v1}, and let the verifier accept when
$\overline{b}$ reproduces the record and $d(b_V,\overline{b})\le
t_{\mathrm{aud}}$. That audit is complete, an honest publisher having
$\overline{b}=b^{*}$ and $d(b_V,b^{*})\le\varepsilon\le t_{\mathrm{aud}}$ by
\Cref{ass:eps}, and when it accepts,
$d(\overline{b},b^{*})\le d(\overline{b},b_V)+d(b_V,b^{*})\le
t_{\mathrm{aud}}+\varepsilon$. What it does not do is pin the string to the
market's reading, so it does not attain \Cref{goal:verif} either, and
\Cref{prop:freedom} measures what is left. Only \Cref{ass:agree} closes the gap,
which is why one honest reader is not enough however the audit is set up.

\begin{proposition}
\label{prop:freedom}
Under the audit above, and for any tolerance $t_{\mathrm{aud}}\ge2\varepsilon$, a
publisher that violates \Cref{ass:pub} may derive the record from any string in
the ball of radius $t_{\mathrm{aud}}-\varepsilon$ about $b^{*}$ and be accepted
by every verifier. If it completes at most $q_E$ evaluations of the delay function by the time it broadcasts its string, then
\[
  \Pr[\,B_e\in E\mid g_e=1\,]\;\le\;
  \frac{(q_E+1)\,p_E\;+\;\epsilon_{\mathrm{seq}}}{p_{\min}} ,
\]
with $q_E$ bounded by \eqref{eq:qE}.
\end{proposition}
\begin{proof}
Let $b'$ satisfy $d(b',b^{*})\le t_{\mathrm{aud}}-\varepsilon$ and let the
publisher derive the record from $b'$ and broadcast $b'$. Every verifier reads
some $b_V$ with $d(b_V,b^{*})\le\varepsilon$ by \Cref{ass:eps}, so
$d(b_V,b')\le t_{\mathrm{aud}}$ and the audit accepts; the first check is
satisfied because the record was in fact derived from the broadcast string. The
accepted strings therefore contain the ball of radius
$t_{\mathrm{aud}}-\varepsilon$ about $b^{*}$, which is non-empty beyond
$b^{*}$ itself whenever $t_{\mathrm{aud}}\ge2\varepsilon$ and $\varepsilon>0$.

For the bias, let $\mathcal{Y}$ be the set of seeds on which the publisher has completed an evaluation of $\Eval$ by $t_e+\tau$, when it broadcasts its string. Except with probability $\epsilon_{\mathrm{seq}}$, counted below, each of them was learned before $t_e$, since an evaluation takes $\Delta/A>\tau$ by \Cref{ass:delay1}; so $|\mathcal{Y}|\le q_E$ with $q_E$ bounded by \eqref{eq:qE}, and $\mathcal{Y}$ was
chosen without knowledge of the window. Let $G$ be the event that some
$s\in\mathcal{Y}$ has $H(\mathsf{dom}_2\|\langle e\rangle\|s\|\Eval(pp,s))\in E$; each
such point is a fresh oracle point, so $\Pr[G]\le q_Ep_E$. On $G^{c}$ the
publisher gains nothing from $\mathcal{Y}$: if it imposes a candidate whose seed
lies in $\mathcal{Y}$ the value is not in $E$, and otherwise $s_e\notin\mathcal{Y}$
and by the definition of $\mathcal{Y}$ and sequentiality it does not hold $y_e$ at $t_e+\tau$, so $B_e$ is
uniform and lands in $E$ with probability $p_E$. Adding
$\epsilon_{\mathrm{seq}}$ for the event that sequentiality fails and
dividing by $\Pr[g_e=1]\ge p_{\min}$ gives
\[
  \Pr[\,B_e\in E\mid g_e=1\,]\;\le\;
  \frac{(q_E+1)\,p_E\;+\;\epsilon_{\mathrm{seq}}}{p_{\min}} . \eqno{\qed}
\]
\end{proof}

Two things are worth noting about this bound. The first is that it is the same
bound as \Cref{prop:degrade}, proved by the same argument: a publisher free
inside a ball and a set of reporters that controls the vote have exactly the same
freedom, and the delay prices it identically. The second is that the radius of
the ball does not appear in it. What bounds the attack is the number of
candidates the attacker can evaluate, not the number the audit would accept, so
tightening the verifier's tolerance does not improve the bias bound; only
lengthening the delay does. This is why \Cref{sec:v2} removes the freedom rather
than shrinking it. The bound is also multiplicative: it keeps out of the publisher's reach every event much rarer than $1/(q_E+1)$, such as a value fixed in advance, though not every event, so it is not \Cref{goal:unbias} with a small $\beta$. And it holds because the publisher broadcasts its string when the window closes (\Cref{sec:v1}): a publisher allowed to announce the string only together with the record could evaluate several strings of the ball in parallel after the window closes, and keep the one it prefers.

\subsection{Security analysis of the consensus beacon}
\label{app:v2}
\label{app:proof}

Throughout, $T^{*}_e=t_e+\tau+\rho$ is the commitment deadline and
$\view_{\mathcal{A}}(t)$ the adversary's view at time $t$. We use the properties
of \Cref{sec:primitives} by the names given there: \emph{delivery},
\emph{agreement}, \emph{attribution} and \emph{persistence} for the broadcast,
\emph{agreement} and \emph{validity} for the vote, \emph{recoverable} and
\emph{binding} for the commitment, and \emph{correct}, \emph{sound} and
\emph{sequential} for the delay function.

\subsubsection*{Step 1: the epoch is fixed at the deadline}

\begin{proposition}
\label{prop:binding}
The string carried by $c_i$ cannot be altered after $c_i$ is received; whether
$c_i$ is admitted is decided at $T^{*}_e$ and is then final; for each pair
$(e,i)$ at most one string is admitted; and the admitted set $S$, hence $\hat b$,
$g_e$ and $s_e$, is identical for every reader.
\end{proposition}
\begin{proof}
The string is recoverable from $c_i$, and it is the only one $\Ext$ returns, so
$c_i$ opens to one triple; by persistence the message survives, so that triple
cannot be altered. Admission depends only on the messages held at the deadline,
which by agreement of the broadcast are the same for every party, so the predicate
``$P_i$ broadcast no other commitment for $e$'' is decided identically by every
reader and is final at $T^{*}_e$; and by attribution a message ascribed to an
honest $P_i$ was sent by $P_i$. The remaining quantities are deterministic
functions of $S$: $\hat b$ by agreement of the vote, then $g_e$ and $s_e$ by
\eqref{eq:gate} and \eqref{eq:seed}. \qed
\end{proof}

\subsubsection*{Step 2: the reference string is the market's}

\begin{proposition}
\label{prop:ref}
Under \Cref{ass:agree}, $\hat b=b^*$, whatever the other reporters commit to
and whenever they choose it.
\end{proposition}
\begin{proof}
By \Cref{ass:agree}, $\nu(b^*)\ge\theta$, so by validity of the vote the
answer is not $\bot$ and satisfies $\nu(\hat b)\ge\theta$. Two strings with
$\theta>n/2$ supporters each would need more than $n$ reporters between them,
while $|S|\le n$; hence $\hat b=b^*$. \qed
\end{proof}

These two give the arithmetic half of the verification. Indeed $S$ is final and
common at $T^{*}_e$ by \Cref{prop:binding}, and $\hat b$, $g_e$, $s_e$ are
deterministic functions of $S$; if a pair $(y,\pi)$ satisfies
$\Vf(pp,s_e,y,\pi)=1$ then $y=y_e$ by soundness, except with probability
$\epsilon_{\mathrm{uniq}}$, so \eqref{eq:value} returns one value. Any two parties
that read the broadcast at $T^{*}_e$ therefore accept the same $B_e$, and no step
of the check V1 reads market data.

\subsubsection*{Step 3: nothing is evaluated in time}

\begin{lemma}
\label{lem:noeval}
Under \Cref{ass:delay}, let $s$ be a value the adversary learns no earlier than
$t_e$. Then it outputs a list containing $\Eval(pp,s)$ before $T^{*}_e$ with
probability at most $\epsilon_{\mathrm{seq}}$.
\end{lemma}
\begin{proof}
By sequentiality, for many challenges at once (\Cref{sec:primitives}), the adversary needs time $\Delta/A$ from the instant it learns
$s$, whichever of the values it learns after $t_e$ turns out to matter, and $\Delta/A>\tau+\rho$ by hypothesis. The interval $[t_e,T^{*}_e)$ has
length $\tau+\rho$, so it is too short. \qed
\end{proof}

This is the step the delay function exists for. Everything the adversary chooses,
its trades during the window, the strings it commits to and whether it commits at
all, is fixed by $T^{*}_e$; \Cref{lem:noeval} says that at that instant the
only candidate values it holds are those it prepared before $t_e$, which it chose
without having seen the window.

\subsubsection*{Step 4: unbiasability}

\begin{proposition}
\label{prop:unbias}
Under the hypotheses of \Cref{thm:v2}, for every
$E\subseteq\{0,1\}^{\lambda}$,
\begin{equation}
  \label{eq:unbias-joint}
  \bigl|\Pr[\,B_e\in E \ \wedge\ g_e=1\,]\;-\;p_E\,\Pr[\,g_e=1\,]\bigr|
  \;\le\;\beta_0 .
\end{equation}
\end{proposition}
\begin{proof}
By \Cref{prop:ref} the reference string is $b^*$, so on $g_e=1$ the seed is
$s_e=H(x_e)$ with $x_e=\mathsf{dom}_1\|\langle e\rangle\|s_{e-1}\|b^*$, and by
\Cref{ass:source} the string $b^*$ carries $\eta^{*}$ bits of min-entropy
given $\view_{\mathcal{A}}(t_e)$. Let $\mathcal{Y}$ be the set of seeds on which
the adversary completed an evaluation before $t_e$; it has at most $q_E$ elements
and is a function of $\view_{\mathcal{A}}(t_e)$. Three things can go wrong.

The adversary may query $H$ at $x_e$ before the window opens. Such a query is a
function of $\view_{\mathcal{A}}(t_e)$, so by the extraction bound of
\Cref{sec:primitives} it happens with probability at most $q_H2^{-\eta^{*}}$.
Failing that, the seed may nevertheless land in $\mathcal{Y}$, possibly because the adversary steered the window towards it: $s_e$ is then one of at most $q_H+1$ values of $H$ at points unqueried at $t_e$, each uniform and independent of $\mathcal{Y}$, so this happens with probability at most $(q_H+1)q_E2^{-\lambda}$. Failing both, the adversary
may still produce $y_e$ during the epoch, which by \Cref{lem:noeval} it does with
probability at most $\epsilon_{\mathrm{seq}}$, since it learns $s_e$ no earlier
than $t_e$. Call $\Lambda$ the union of the three, so that $\Pr[\Lambda]\le\beta_0$.

Outside $\Lambda$ the adversary does not hold $y_e$ at the deadline, having neither
precomputed nor computed it, and a query at
$x'_e=\mathsf{dom}_2\|\langle e\rangle\|s_e\|y_e$ would exhibit $y_e$ and is therefore part
of the third event. So $x'_e$ is unqueried at $T^{*}_e$, and $B_e=H(x'_e)$ is
uniform on $\{0,1\}^{\lambda}$ and independent of
$\view_{\mathcal{A}}(T^{*}_e)$.

Every choice the adversary makes is a function of $\view_{\mathcal{A}}(T^{*}_e)$;
so are $g_e$, by \eqref{eq:gate} and \Cref{prop:binding}, and $\Lambda$. Hence
$\Pr[B_e\in E,\ g_e=1,\ \Lambda^{c}]=p_E\Pr[g_e=1,\ \Lambda^{c}]$, and the two
remaining terms differ by at most $\Pr[\Lambda]$:
\begin{align*}
  \bigl|\Pr[B_e\in E,\ g_e=1]-p_E\Pr[g_e=1]\bigr|
  &\;=\;\bigl|\Pr[B_e\in E,\ g_e=1,\ \Lambda]-p_E\Pr[g_e=1,\ \Lambda]\bigr|\\
  &\;\le\;\Pr[\Lambda]\;\le\;\beta_0 . \tag*{\qed}
\end{align*}
\end{proof}

Dividing \eqref{eq:unbias-joint} by $\Pr[g_e=1]\ge p_{\min}$ gives the bias bound
of \Cref{thm:v2}. The division is the only place $p_{\min}$ is used, and it is
needed for the same reason as in \Cref{prop:unbias1}: an adversary that could
close the gate selectively, on the epochs where one of the three events had
occurred, would bias the conditional distribution even though the joint bound
held. No reporter can do this as long as enough honest ones agree, by \Cref{prop:liveness}; a party that trades can, with the limited effect discussed after \Cref{prop:unbias1}.

\begin{corollary}
\label{cor:abort}
An adversary that decides whether to withhold its commitments, or which strings
to commit to, after observing every other message of the epoch, still satisfies
\Cref{prop:unbias}.
\end{corollary}
\begin{proof}
Such a decision is a function of $\view_{\mathcal{A}}(T^{*}_e)$, and outside
$\Lambda$ the value $B_e$ is independent of that view. A rushing adversary may therefore
learn $s_e$ before it acts, and this changes nothing: what it would need in order
to act usefully is $B_e$, and by \Cref{lem:noeval} the map $s_e\mapsto B_e$ is
unavailable to it until long after the deadline. \qed
\end{proof}

\subsubsection*{Step 5: unpredictability}

\begin{proposition}
\label{prop:unpred}
Under the hypotheses of \Cref{thm:v2}, no admissible adversary outputs $B_e$
before $T^{\min}_e=t_e+\Delta/A$ except with probability
$\beta_0+q_H2^{-\lambda}$; and every party holds $B_e$ by
$T^{*}_e+\Delta+\rho$.
\end{proposition}
\begin{proof}
To output $B_e$ the adversary either does not query $H$ at
$x'_e=\mathsf{dom}_2\|\langle e\rangle\|s_e\|y_e$, in which case each of its at most $q_H$
outputs matches a uniform unqueried value with probability $2^{-\lambda}$; or it
queries $x'_e$, which exhibits $y_e$. In the latter case one of the three events
of \Cref{prop:unbias} has occurred, read against the later deadline
$T^{\min}_e$: either $x_e$ was queried before $t_e$, of probability at most
$q_H2^{-\eta^{*}}$; or $s_e$ was among the at most $q_E$ seeds evaluated
before $t_e$, of probability at most $(q_H+1)q_E2^{-\lambda}$, the adversary possibly having steered the window towards it; or the adversary learned
$s_e$ no earlier than $t_e$ and exhibited $y_e=\Eval(pp,s_e)$ before
$t_e+\Delta/A$, which sequentiality bounds by $\epsilon_{\mathrm{seq}}$. The last
claim is correctness together with the fact that $s_e$ is public at $T^{*}_e$,
that Phase~3 has an honest evaluator, and that delivery carries its broadcast
within $\rho$. \qed
\end{proof}

Finally $T^{\min}_e-T^{*}_e=\Delta/A-(\tau+\rho)$, positive by
\Cref{ass:delay}, which is the unpredictability claim of \Cref{thm:v2}. Two remarks on the horizon. It is
measured from $t_e$ rather than from the deadline because \Cref{ass:source} is
stated at $t_e$ and says nothing about what the adversary learns as the window
unfolds; a window whose last trades are sparse has its string effectively
determined well before it closes, and we do not want the bound to depend on that.
And the adversary finishes at $T^{\min}_e$ while honest parties finish at
$T^{*}_e+\Delta$, so there is an interval of length
$(\tau+\rho)+\Delta(1-1/A)$ during which it knows $B_e$ and the consumers do not.
This is inherent to any beacon built on a delay
function~\cite{BGB17,LW15,CCB24}, but a protocol that consumes $B_e$ needs to
know it, and it is the reason $A$ appears in \Cref{ass:delay} rather than being
assumed away.

\subsubsection*{Step 6: verifiability and liveness}

\begin{proposition}
\label{prop:audit}
Under \Cref{ass:eps}, let the verifier choose any tolerance
$t_{\mathrm{aud}}\ge2\varepsilon$. Then, under \Cref{ass:agree}, every reporter
that committed to $b^*$ belongs to $\mathcal{V}$, so $|\mathcal{V}|\ge\theta$
and V2 accepts; every flagged $i$ satisfies
$d(b_i,b^*)>t_{\mathrm{aud}}-\varepsilon\ge\varepsilon$; and if V2 accepts
then $d(\hat b,b^*)\le t_{\mathrm{aud}}+\varepsilon$, whether or not
\Cref{ass:agree} holds.
\end{proposition}
\begin{proof}
If $b_i=b^*$ then $d(b_V,b_i)=d(b_V,b^*)\le\varepsilon\le
t_{\mathrm{aud}}$ by \Cref{ass:eps}, so $P_i\in\mathcal{V}$; there are at least
$\theta$ such reporters by \Cref{ass:agree}. If $i$ is flagged then
$d(b_V,b_i)>t_{\mathrm{aud}}$, while $d(b_V,b^*)\le\varepsilon$, so the
triangle inequality gives $d(b_i,b^*)\ge d(b_V,b_i)-d(b_V,b^*)>
t_{\mathrm{aud}}-\varepsilon$, which is at least $\varepsilon$ because
$t_{\mathrm{aud}}\ge2\varepsilon$. For the converse, acceptance gives
$|\mathcal{V}|\ge\theta$ and validity of the vote gives
$\bigl|\{i\in S: b_i=\hat b\}\bigr|\ge\theta$; two subsets of $S$ of size
$\theta>n/2$ with $|S|\le n$ must meet, so some $i$ has $b_i=\hat b$ and
$d(b_V,\hat b)\le t_{\mathrm{aud}}$, whence
$d(\hat b,b^*)\le d(\hat b,b_V)+d(b_V,b^*)\le
t_{\mathrm{aud}}+\varepsilon$. \qed
\end{proof}

Under \Cref{ass:agree} this gives \Cref{goal:verif} exactly, and not at a
tolerance. The first half is completeness: a faithful epoch is accepted whatever
the verifier's feed did to its own reading, which is the property the basic
version loses as soon as $\varepsilon>0$. Soundness is \Cref{prop:ref} together
with V1: the reference string is $b^{*}$ itself, so a record derived from
anything else fails the arithmetic check. The tolerance
$t_{\mathrm{aud}}$ is the verifier's own parameter and appears in neither
\eqref{eq:vote} nor \eqref{eq:value}; what it buys is room for the
disagreement between honest feeds, which is why the consensus version has one and
the basic version, comparing hashes, cannot have one at all.

The third clause is separate and is the half a verifier actually wants, because
it is the only one that does not assume the reporters are mostly honest: an
accepted epoch has a reference string within $t_{\mathrm{aud}}+\varepsilon$ of
the market's own reading whatever the reporters did. That is a statement about
the record rather than about the participants, and it is what makes A2 and A3 of
\Cref{tab:attacks} detectable rather than merely assumed away. When
\Cref{ass:agree} holds, \Cref{prop:ref} collapses that ball to the single point
$b^{*}$; when it fails the ball reopens, the audit has no basis on which to
reject any string inside it, and \Cref{prop:degrade} then prices what a coalition
can do with the freedom.

\begin{proposition}
\label{prop:liveness}
Under \Cref{ass:agree}, and with $m_{\min}\le\theta$, $g_e=1$ whenever
$\Ecal^{-}(b^*)\ge\eta$. In particular the deviating reporters cannot force a
skip, whether by withholding their commitments, by equivocating, or by committing
to strings of their own.
\end{proposition}
\begin{proof}
By \Cref{ass:agree} at least $\theta$ admitted commitments carry $b^*$, so
$|S|\ge\theta\ge m_{\min}$ and, by \Cref{prop:ref}, $\hat b=b^*\neq\bot$. The
third conjunct of \eqref{eq:gate} is then the hypothesis. Withheld commitments
are absent from $S$ and were never counted towards $b^*$; equivocation
removes the equivocating reporter by \Cref{prop:binding}; and a string committed
by fewer than $\theta$ reporters cannot win \eqref{eq:vote}. \qed
\end{proof}

This is the liveness claim of \Cref{thm:v2}, and it also shows that the second
conjunct of \eqref{eq:gate} never binds: by validity of the vote, $\hat b\neq\bot$ already forces
$|S|\ge\theta\ge m_{\min}$. We keep $m_{\min}$ in the gate because a deployment
may instantiate the vote with a threshold other than a strict majority, and there
the floor does work; but under the instantiation of \Cref{def:vote} it is inert.
This is why the public parameters fix $m_{\min}\le\theta$: a floor above the
threshold would be a denial-of-service lever for exactly the reporters
\Cref{ass:agree} is meant to neutralize.

This completes the proof of \Cref{thm:v2}. Its four claims are
\Cref{prop:unbias} for \Cref{goal:unbias}, \Cref{prop:unpred} for
\Cref{goal:unpred}, \Cref{prop:audit} together with \Cref{prop:ref} for
\Cref{goal:verif}, whatever tolerance $t_{\mathrm{aud}}\ge2\varepsilon$ the
verifier chooses, and \Cref{prop:liveness} for liveness; Steps
1 and 2 are what they rest on and are the only two steps \Cref{app:v1} does not
contain.

\paragraph{The bounds as numbers.}
The symbols hide how much room the parameters leave, so we evaluate them once.
Take the configuration of \Cref{subsec:bound}, $\eta^{*}=512$ and $\lambda=256$,
and grant the adversary $q_H\le2^{128}$ oracle queries and $q_E\le2^{40}$
completed evaluations, the latter being generous: by \eqref{eq:qE} it would take
a million processors running a year at $A=10$ against a five-minute window to
approach it. Then
\[
  \beta_0 \;=\; q_H2^{-\eta^{*}}+(q_H+1)\,q_E2^{-\lambda}+\epsilon_{\mathrm{seq}}
  \;\approx\; \epsilon_{\mathrm{seq}}+2^{-88},
\]
the three terms being $2^{-384}$, about $2^{-88}$ and $\epsilon_{\mathrm{seq}}$, and the prediction bound is
$\beta_0+q_H2^{-\lambda}\approx\epsilon_{\mathrm{seq}}+2^{-88}$ as well. Three things follow. The entropy target controls only the first term, which at $\eta^{*}=2\lambda$ is $2^{-384}$, so far below the others that doubling the target again would change nothing; this is the margin of \Cref{rem:lhl} seen from the other side. The second term pays for an adversary that steers the window towards a seed it evaluated in advance; it is the price of the hardware of \eqref{eq:qE}, and at $2^{-88}$ it is negligible. And every bound reduces to $\epsilon_{\mathrm{seq}}$ up to that term, $\epsilon_{\mathrm{seq}}$ being the one term that is assumed of the delay function rather than derived, which is why \Cref{sec:primitives} states it as a property of the primitive. The division by $p_{\min}$ costs little. In the configuration of \Cref{subsec:bound} the gate fired on no interval of 2025, and on a single high-capitalization ticker at five minutes it fired on one interval in five, a factor of about $1.25$; since $\beta_0$ is negligible, the bound remains negligible for any $p_{\min}$ that is not itself negligible, including against a party that forces refusals by trading (see the discussion after \Cref{prop:unbias1}).

\subsubsection*{What remains without Assumption 1b}

\Cref{thm:v2} assumes \Cref{ass:agree}. What follows is what survives when it
fails, which is the claim \Cref{subsec:v2thm} makes in prose.

\begin{proposition}
\label{prop:degrade}
Suppose \Cref{ass:agree} fails and the deviating reporters determine $\hat b$.
Then under \Cref{ass:delay}, with $H$ a random oracle and $p_{\min}$ as in
\Cref{thm:v2},
\[
  \Pr[\,B_e\in E \mid g_e=1\,]
  \;\le\;
  \frac{(q_E+1)\,p_E\;+\;\epsilon_{\mathrm{seq}}}{p_{\min}} ,
\]
with $q_E$ bounded by \eqref{eq:qE}. The multiplicative advantage is at most
$q_E+1$ however many reporters deviate.
\end{proposition}
\begin{proof}
The coalition fixes its commitments by $T^{*}_e$, so the string it imposes is
a function of $\view_{\mathcal{A}}(T^{*}_e)$. By \Cref{lem:noeval} the pairs
$(s,\Eval(pp,s))$ it holds at that instant are, except with probability
$\epsilon_{\mathrm{seq}}$, those from evaluations completed before $t_e$; call
that set $\mathcal{Y}$, with $|\mathcal{Y}|\le q_E$ and $\mathcal{Y}$ chosen
without knowledge of the window. For $s\in\mathcal{Y}$ the point
$\mathsf{dom}_2\|\langle e\rangle\|s\|\Eval(pp,s)$ is a fresh oracle point, so the
corresponding value lands in $E$ independently with probability $p_E$, and the
probability that some $s\in\mathcal{Y}$ has its value in $E$ is at most $q_Ep_E$.
If no such $s$ exists, then either the string the coalition imposes has its seed in
$\mathcal{Y}$, and the value of that seed is not in $E$; or the seed lies outside
$\mathcal{Y}$, the coalition does not hold $y_e$ by \Cref{lem:noeval}, and $B_e$ is
uniform and lands in $E$ with probability $p_E$. Adding $\epsilon_{\mathrm{seq}}$
for the event that \Cref{lem:noeval} fails and dividing by
$\Pr[g_e=1]\ge p_{\min}$ gives the bound, which is that of \Cref{prop:freedom}
with the same proof. \qed
\end{proof}

\subsection{Tightness of the majority}
\label{app:nomaj}

\Cref{prop:noaudit} rules out a single reader. The same argument run with a
reporter set shows that a majority is the least threshold at which the consensus
phase does anything, and it shows it in the form that matters here: below a
majority the protocol returns to the situation of \Cref{prop:freedom}, in which
whoever determines the reference string is free inside a ball and the delay is
all that prices the freedom.

\begin{proposition}
\label{prop:nomaj}
Suppose the adversary controls $f$ of the $n$ reporters. If $f\ge\theta$ it
determines the reference string, and may set it to any string within
$t_{\mathrm{aud}}-\varepsilon$ of the reading of the window with every verifier's
audit accepting; if $n-f<\theta$ it forces $g_e=0$ on any epoch of its choosing.
\end{proposition}
\begin{proof}
For (1), let the $f\ge\theta$ deviating reporters all commit $b'$. By validity of
the vote $\hat b=b'$. By \Cref{ass:eps} every verifier reads a $b_V$ with
$d(b_V,b^{*})\le\varepsilon$, so
$d(b_V,b')\le d(b_V,b^{*})+d(b^{*},b')\le t_{\mathrm{aud}}$ and all $f\ge\theta$
of them are admitted, whence $|\mathcal{V}|\ge\theta$ and V2 accepts. The
candidates and the bound are those of \Cref{prop:freedom}, with the coalition in
place of the publisher.

For (2), let the $f>n-\theta$ deviating reporters commit pairwise distinct
strings, none of them $b^{*}$. At most $n-f<\theta$ reporters commit any one
string, so no string reaches the threshold, $\hat b=\bot$ by \Cref{def:vote} and
$g_e=0$ by \eqref{eq:gate}. \qed
\end{proof}

Part (1) is not a failure of the audit. The string the coalition imposes lies
within $t_{\mathrm{aud}}+\varepsilon$ of the reading of the window, so the audit
is right to accept it and the third clause of \Cref{prop:audit} still holds; what
no longer holds is \Cref{goal:verif} itself, because the accepted record need no
longer be the market's. What is
lost is \Cref{prop:ref}, and with it the pinning that made the ball a point; the
audit has no basis on which to reject, because every string in the ball is one
that an honest feed might have delivered. This is \Cref{prop:noaudit} again, in
the form the consensus version takes: an honest message is a claim carrying no
mark of its honesty, the encoding \eqref{eq:binarization} being public, so a
deviating reporter computes exactly what an honest reporter reading a slightly
different window would have sent. Escaping it requires evidence that is not a
claim.

Part (2) is the converse of \Cref{prop:liveness}, which says that above the
threshold no behaviour of the reporters closes the gate. Below it, closing the
gate is the one thing they can always do, and the protocol degrades towards a bounded bias and denial of service rather than towards a chosen output; the two failure modes are answered separately, the first by \Cref{prop:degrade} and the second by the fact that a refusal is signed and public.

\subsection{What each primitive buys}
\label{app:primitives}

Beyond the encoding \eqref{eq:binarization} the construction uses three
primitives: a hash function, a delay function and a vote. \Cref{tab:primitives}
records which goals each combination attains, and the division of labour is
clean. The delay decides \Cref{goal:unbias,goal:unpred} and does nothing for
\Cref{goal:verif}; the vote decides \Cref{goal:verif} and does not improve
$\beta_0$ at all. Each of the two problems this paper treats, manipulation of the
market and disagreement of the feeds, is answered by one primitive and by neither
of the others.

\paragraph{Hashing alone.}
Against a party that reads the window but does not trade in it, the first row
would give \Cref{goal:unbias} with $\beta=q_H2^{-\eta^{*}}$ by \Cref{lem:ext1}
and \Cref{goal:unpred} with $T=t_e+\tau$, and the second is already useless,
since every party holding a feed learns the value at that same instant. Against
the adversary of \Cref{app:spec}, which trades inside the window, both fail
outright. At $t_e+\tau^{-}$ it holds the state of the venue, enumerates the $m$
continuations its own remaining actions realise, computes the value of each with
$m$ oracle queries and plays the one it prefers, so that
$\Pr[B_e\in E]\approx1-(1-p_E)^m$; a single output bit is bought with $m=2$. This is the look-ahead attack that rules out deterministic extraction from sources whose adversarial parts may depend on the parts that precede them~\cite{AORSV20}.
Nothing in the protocol bounds $m$, only the cost of trading, which is not a
protocol parameter. This is what the delay function repairs, and all it repairs:
an adversary that cannot evaluate it inside the window must fix its candidate
seeds before the window opens, at most the $q_E$ of \eqref{eq:qE}, priced in hardware, and it can steer the window towards one of them only if some continuation it can realise hashes into that set, which the second term of \eqref{eq:beta0} bounds; that term and the third are what remains of the attack.

\paragraph{The vote without the delay.}
The third row loses \Cref{goal:unbias,goal:unpred} twice over. It loses them to
the trading adversary as the first row does, and it loses them again to the
ordering of the commitment round: the commitments of \Cref{def:commit} are not
hiding, so the reporter that speaks last reads the others, enumerates the
reference strings its own message can still produce and computes the value of
each. Under \Cref{ass:delay} that second attack is priced at $q_E$ like the
first, its candidate seeds having to be prepared before a message it had not yet
seen. The commitment is safe in being non-hiding because of the delay, and not
in spite of it.

\paragraph{Assumptions.}
\Cref{tab:assgoals} is the same accounting for the assumptions. The map from
assumptions to goals is identical in the two versions; what changes is which
assumption plays the first role and which radius the fourth is allowed to take.

\begin{table}[htbp]
\centering\footnotesize
\caption{Which assumptions each goal needs, and what each version attains. Row~1
is \Cref{ass:pub} in the basic version and \Cref{ass:agree} in the consensus
version, and row~2 is \Cref{ass:delay1} and \Cref{ass:delay}; rows~3 and~4 are
the same assumption in both.}
\label{tab:assgoals}
\begin{tabular}{@{}llll@{}}
\toprule
 & unbiasability & unpredictability & verifiability\\
\midrule
1\ \ who reports      & yes                & yes                 & yes\\
2\ \ delay dominance  & yes                & yes                 & ---\\
3\ \ source entropy   & yes                & yes                 & ---\\
4\ \ reading accuracy & ---                & ---                 & yes\\
\midrule
attained, basic       & $\beta_0/p_{\min}$ & $t_e+\Delta/A$ & only if $\varepsilon=0$\\
attained, consensus   & $\beta_0/p_{\min}$ & $t_e+\Delta/A$ & yes, at any $\varepsilon$\\
\bottomrule
\end{tabular}
\end{table}

Three things are visible in the rows. Assumption~1 is the only one all three
goals need: it is what puts the market's reading inside the hash, so that
\Cref{ass:source} applies to the string actually queried, and it is what lets the
audit tell a faithful record from an unfaithful one. It is also the assumption whose content, and not merely whose parameter, differs
between the versions, which is why the whole of the difference between them is
carried by \Cref{ass:pub} against \Cref{ass:agree}.
Assumptions~2 and~3 buy \Cref{goal:unbias,goal:unpred} and are irrelevant to
\Cref{goal:verif}, while \Cref{ass:eps} buys \Cref{goal:verif} and is irrelevant
to the other two: no assumption serves both of the paper's two problems, just as
no primitive does. And the factor $1/p_{\min}$ is the price of the gate in both versions rather than
a cost of the vote, so the consensus phase leaves unbiasability and
unpredictability exactly where it found them; what it buys is the last column.

The last column deserves one more line, because \Cref{goal:verif} is the one goal
whose failure is not graceful in the same way. Below \Cref{ass:agree} the two
versions become the same object: whoever determines the reference string may
choose any string an honest feed might plausibly have delivered, a ball of radius
$t_{\mathrm{aud}}-\varepsilon$ about $b^{*}$ for the coalition of
\Cref{prop:nomaj} and about $\overline{b}$ for the publisher of
\Cref{prop:freedom}, and the audit has no basis on which to reject anything
inside it. \Cref{ass:agree} is what shrinks the ball to the point $b^{*}$, by
\Cref{prop:ref}, and \Cref{prop:degrade} is what prices the freedom when it does
not. So the vote does not widen what the audit tolerates, which is the verifier's
own choice of $t_{\mathrm{aud}}$ and is the same either way: it decides whether
the string the audit accepts is the market's reading or merely close to it.

\section{Comparison with contributory beacons}
\label{app:cointoss}
\label{subsec:cointoss}
Most deployed beacons draw their randomness from the participants rather than
from a source, and \Cref{subsec:rwprotocol} placed that family and the bound of
Cleve~\cite{Cle86} under which it lives. \Cref{tab:beacons} says what the two
versions of this paper obtain in exchange for \Cref{ass:source}. Its first column
is there to show parity, counting honest parties not separating the designs: the
consensus version needs a majority, and so does a commit--reveal beacon that
recovers withheld values. The separation is in the other three.

\begin{table}[htbp]
\centering\footnotesize
\caption{Where the two versions sit among beacons. A contribution carries no mark
of its unpredictability, so the third column asks whether the transcript lets
anyone decide afterwards that a participant behaved; the fourth is what an
adversary that corrupts every participant obtains while passing every audit}.
\label{tab:beacons}
\begin{tabular}{@{}llccl@{}}
\toprule
 & honest parties & honesty is & auditable? & all corrupt\\
\midrule
commit--reveal, punished~\cite{Blu83,ethspec} & $1$        & contributing entropy & no  & the output\\
commit--reveal, recovered~\cite{SJK+17,CD17}  & majority   & contributing entropy & no  & the output\\
threshold signature~\cite{drand}              & $t$ of $n$ & contributing entropy & no  & the output\\
\midrule
ours, basic (\Cref{sec:v1})                   & publisher  & reporting faithfully & if $\varepsilon=0$ & $q_E+1$ candidates\\
ours, consensus (\Cref{sec:v2})               & majority   & reporting faithfully & yes & $q_E+1$ candidates\\
\bottomrule
\end{tabular}
\end{table}

A participant in a coin-tossing beacon is honest only if it follows the protocol
\emph{and} draws a contribution no one else can predict. The second half is a
claim about private state that no transcript can check: any value being a valid
contribution, a participant whose generator is weak or whose seed is shared with
its peers is indistinguishable from an honest one while contributing nothing. A
reporter makes no such claim. It applies the public map \eqref{eq:binarization}
to public data, holds no secret, and what it did is settled afterwards by
\Cref{goal:verif} against a reading anyone may take. Honesty is a property of the
participant there and of the message here, and only the second is evidence.

The two families also fail differently. When every participant of a coin-tossing
beacon is corrupt the output is theirs to choose, at once and at no cost; when
every reporter here is corrupt \Cref{prop:degrade} still applies: the coalition must have evaluated its candidates before the window opened, and the delay leaves it the best of $q_E+1$ of them, a leverage bounded by its hardware that keeps every event much rarer than $1/(q_E+1)$ out of its reach. What it can always do instead is close the gate, by the
second half of \Cref{prop:nomaj}, and a closed gate is a signed public refusal
rather than a value that looks like any other. A beacon that can be made to stop
fails more visibly than one that can be made to lie.

\section{Conclusions}

We make our data analysis code, consisting of Python and shell scripts, together with some Jupyter notebooks, available.~\footnote{GitHub: \url{https://github.com/silviaonofri/High-Frequency-Financial-Data-for-Public-Randomness.git}} High-frequency trading data can feed a randomness beacon far more often than daily closing prices. On the 2025 Binance data, eleven tickers pooled into one string yield thousands of bits of entropy every five minutes, and the conservative bound never fell below the $512$-bit target in any five-minute window of the year (RQ1, RQ2). Since market activity is outside the beacon's control, the design must be able to refuse, and our circuit breaker does so publicly, at hours that are predictable from the clock (RQ3). The delay-based construction of Landis and Bonneau keeps its guarantees in this setting (RQ4), and we extend it to data feeds that disagree, a problem already observed with daily prices~\cite{Adi09}. Future work could test whether these entropy rates generalize across exchanges, asset classes and market conditions.

\bibliographystyle{splncs04}
\bibliography{pulp.bib, bib.bib}

\clearpage
\begin{appendix}
\section{The circuit breaker}
\label{app:cb}

\begin{figure}[p]
    \centering
    \begin{subfigure}{0.35\textwidth}
        \centering
        \includegraphics[width=\linewidth]{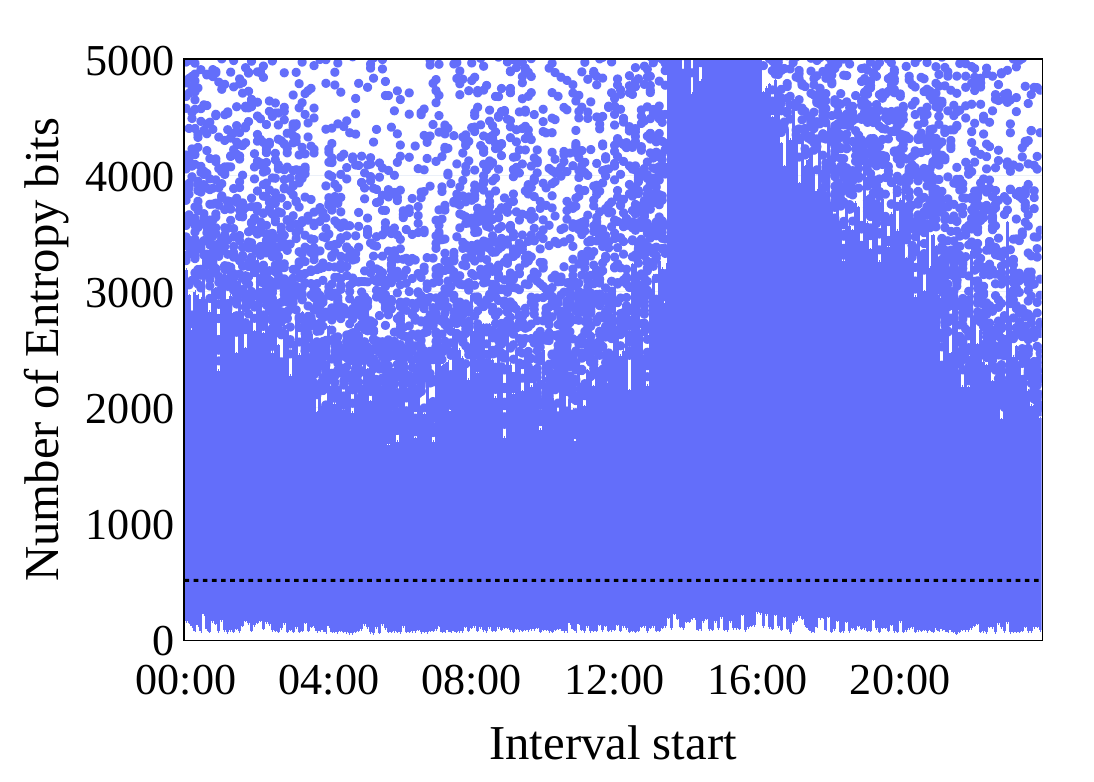}
        \caption{5 minutes}
        \label{fig:BTC_5_combined}
    \end{subfigure}
    \hfill
    \begin{subfigure}{0.35\textwidth}
        \centering
        \includegraphics[width=\linewidth]{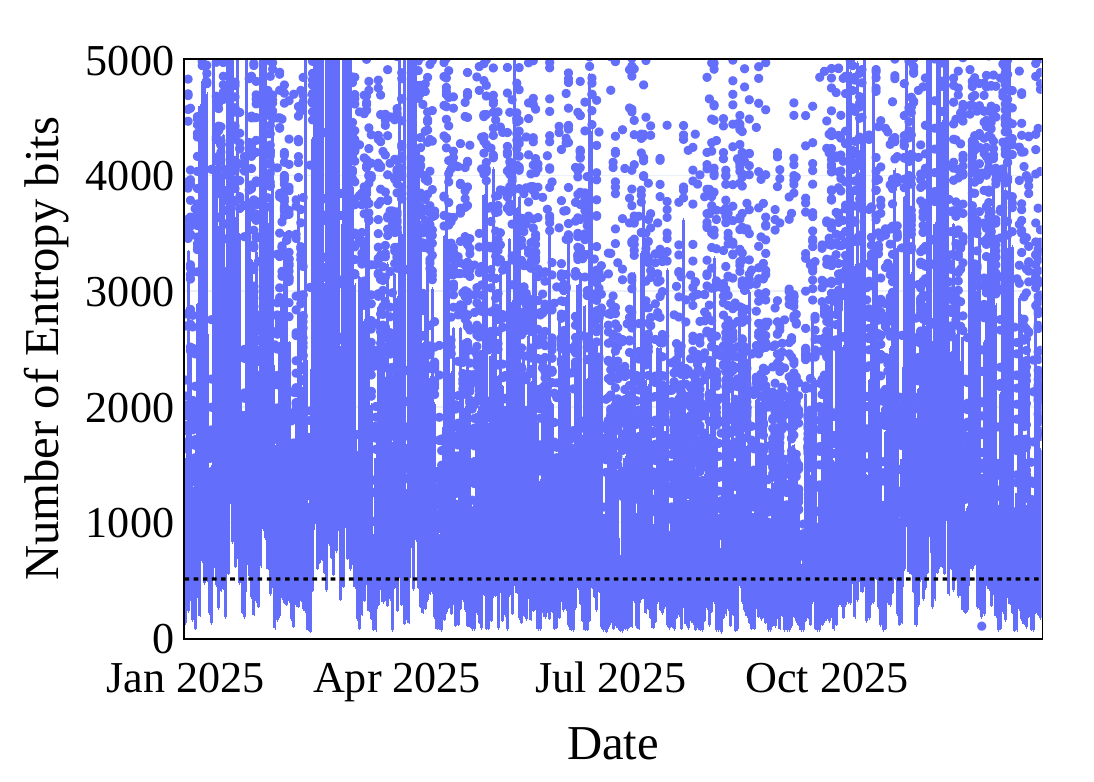}
        \caption{5 minutes, per-day}
        \label{fig:BTC_bd_5_combined}
    \end{subfigure}
    \vspace{0.2cm}
    \begin{subfigure}{0.35\textwidth}
        \centering
        \includegraphics[width=\linewidth]{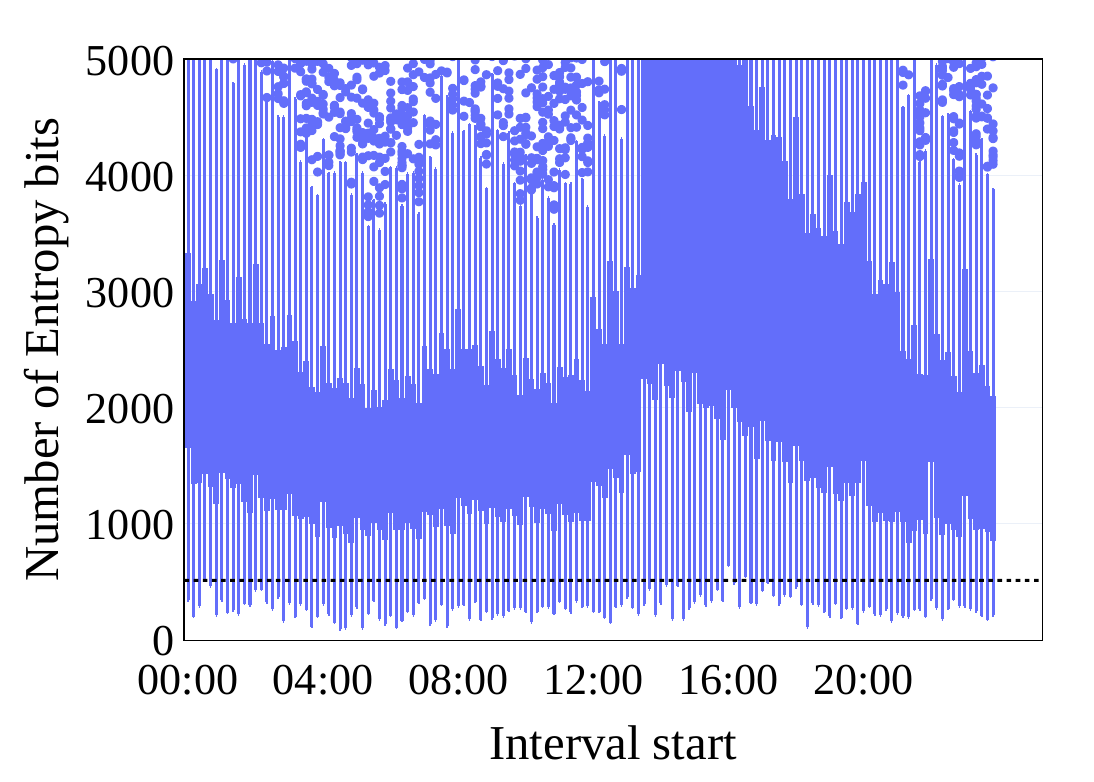}
        \caption{10 minutes}
        \label{fig:BTC_10_combined}
    \end{subfigure}
    \hfill
    \begin{subfigure}{0.35\textwidth}
        \centering
        \includegraphics[width=\linewidth]{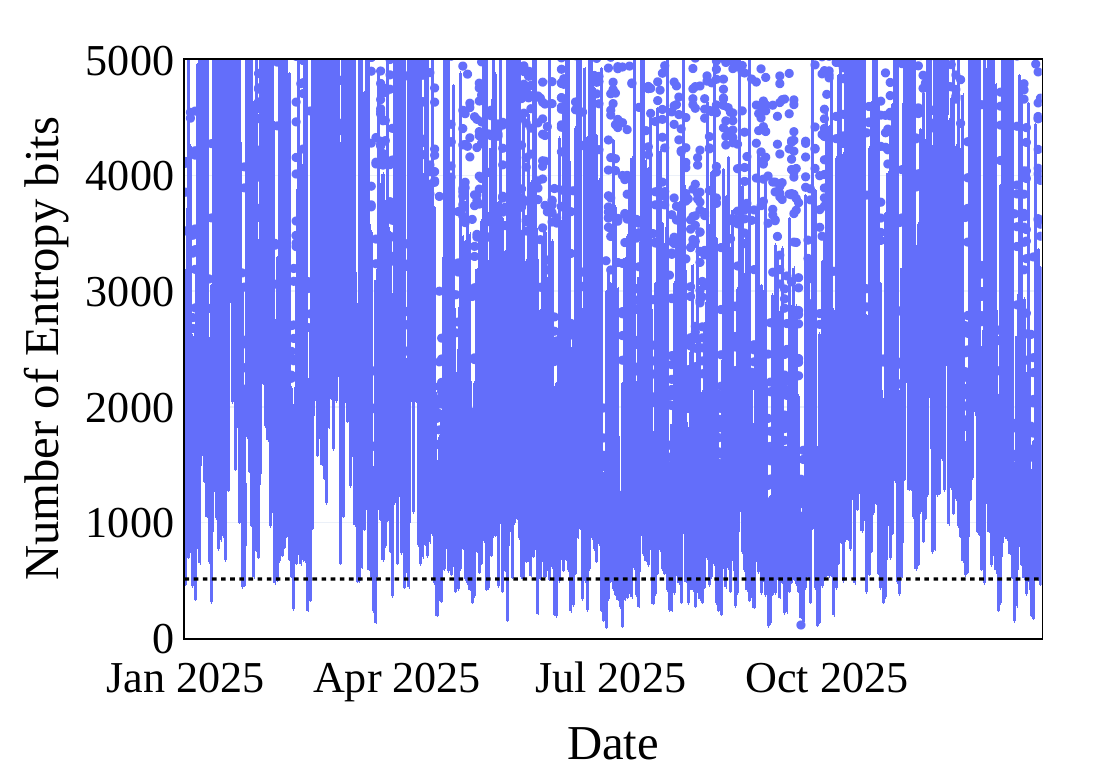}
        \caption{10 minutes, per-day}
        \label{fig:BTC_bd_10_combined}
    \end{subfigure}
    \vspace{0.2cm}
    \begin{subfigure}{0.35\textwidth}
        \centering
        \includegraphics[width=\linewidth]{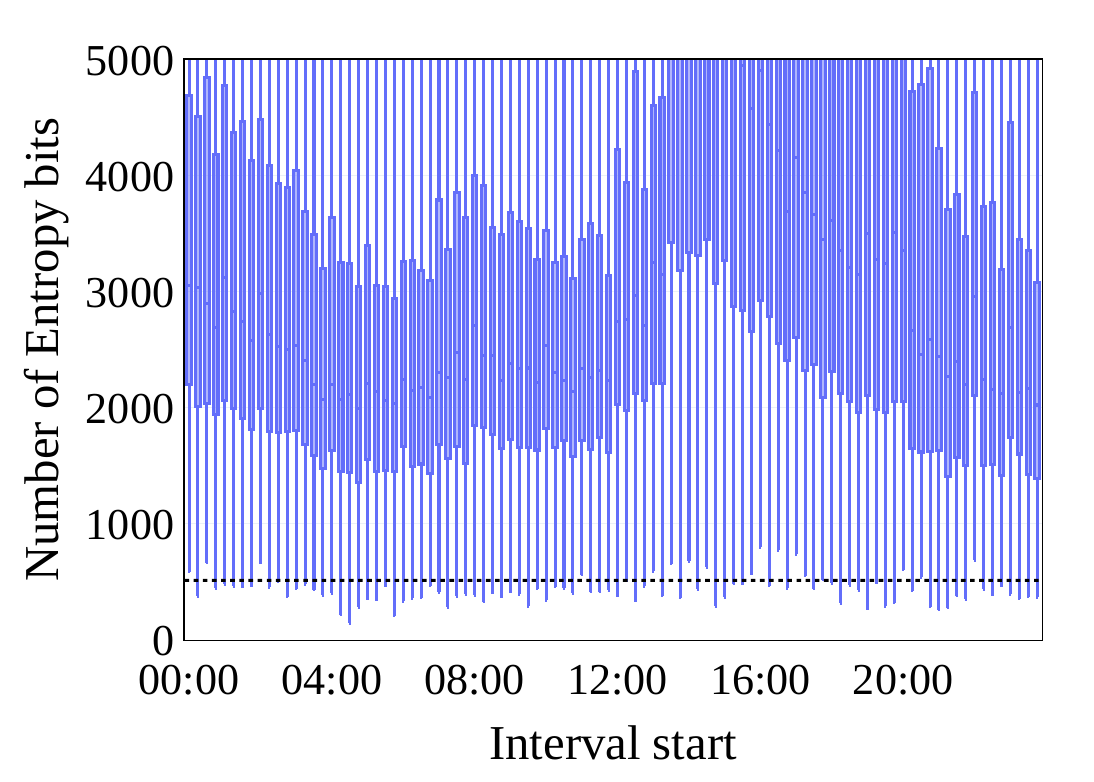}
        \caption{15 minutes}
        \label{fig:BTC_15_combined}
    \end{subfigure}
    \hfill
    \begin{subfigure}{0.35\textwidth}
        \centering
        \includegraphics[width=\linewidth]{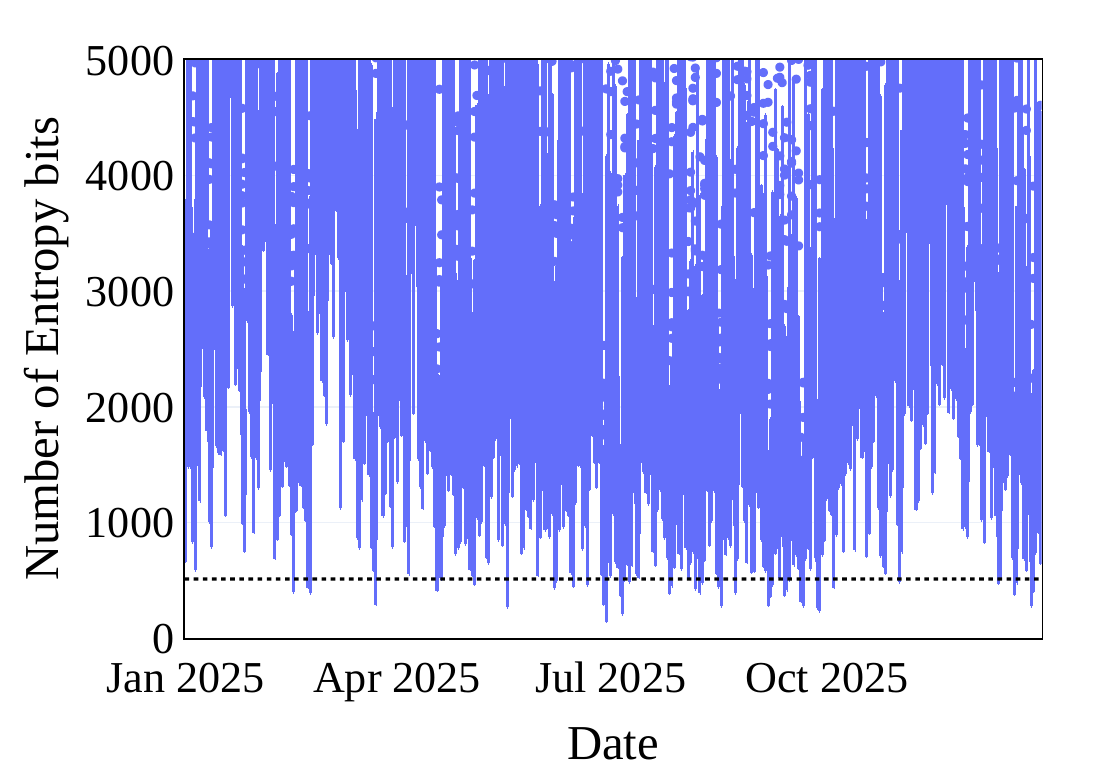}
        \caption{15 minutes, per-day}
        \label{fig:BTC_bd_15_combined}
    \end{subfigure}
    \vspace{0.2cm}
    \begin{subfigure}{0.35\textwidth}
        \centering
        \includegraphics[width=\linewidth]{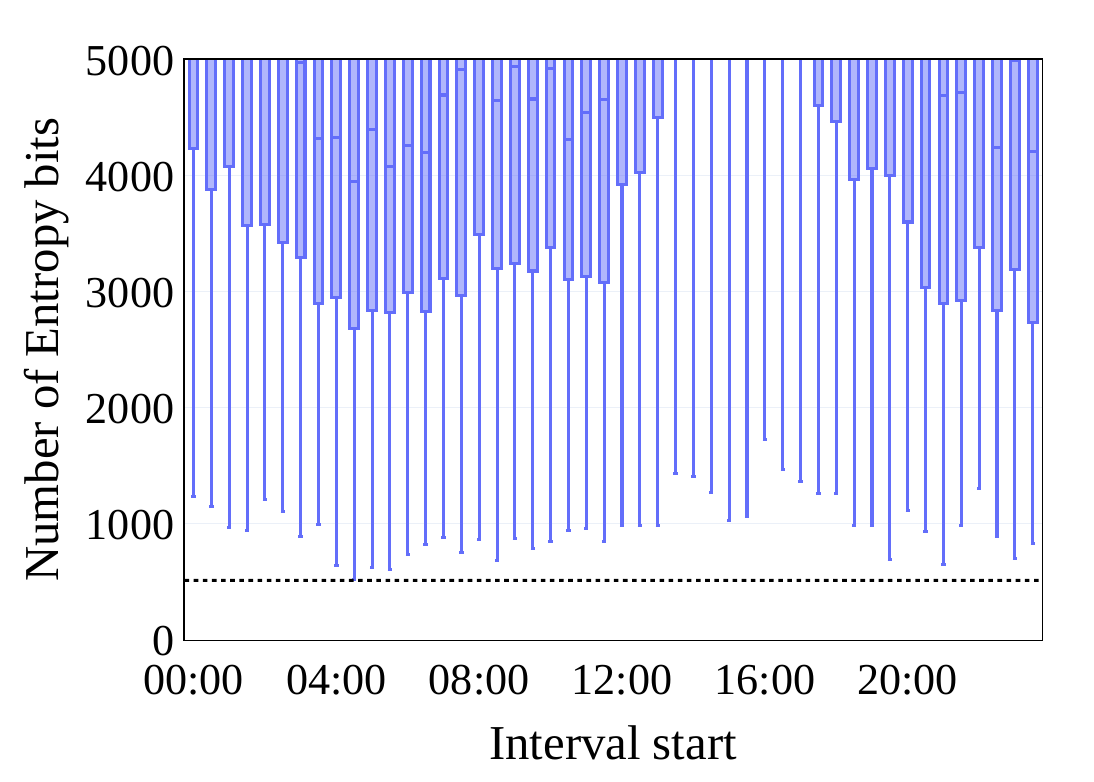}
        \caption{30 minutes}
        \label{fig:BTC_30_combined}
    \end{subfigure}
    \hfill
    \begin{subfigure}{0.35\textwidth}
        \centering
        \includegraphics[width=\linewidth]{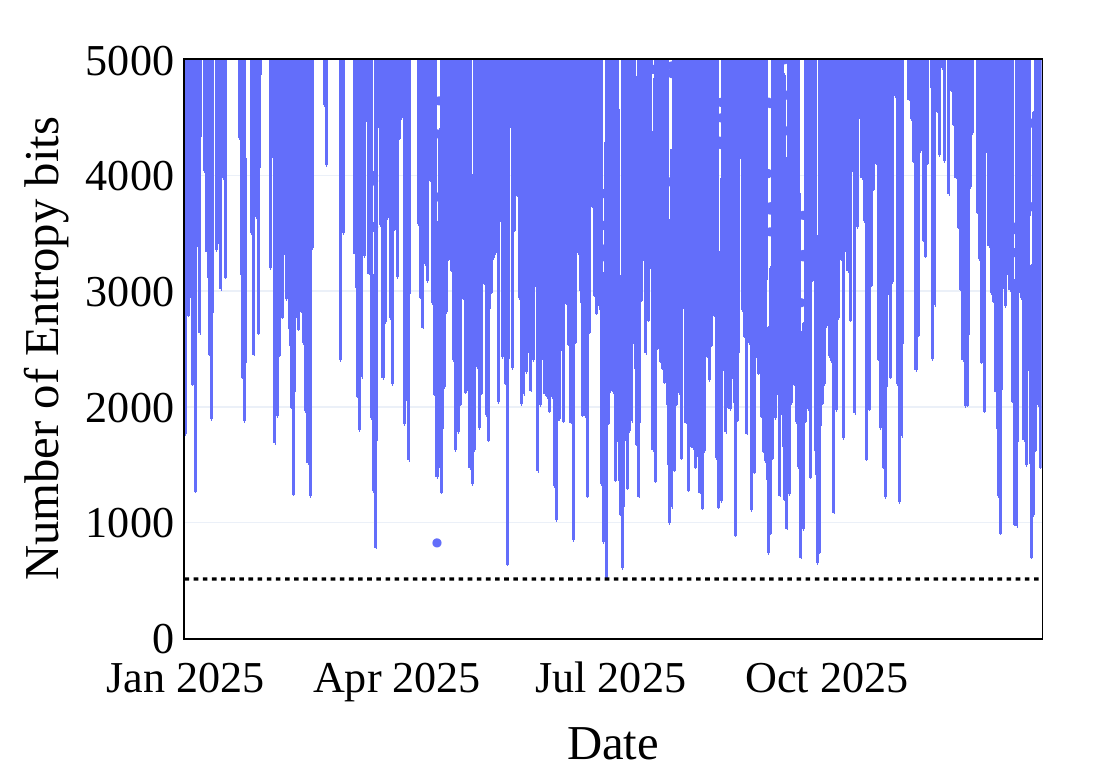}
        \caption{30 minutes, per-day}
        \label{fig:BTC_bd_30_combined}
    \end{subfigure}
    \caption{Number of entropy bits $n_E$ created by strings of BTCUSDT in intervals
    of $5$, $10$, $15$, $30$ minutes. Left column: distribution by interval start;
    right column: per-day distribution over 2025. The dotted line marks the $512$-bit
    target.}
    \label{fig:BTC_entropy_4x2_combined}
\end{figure}

\begin{figure}[p]
    \centering
    \begin{subfigure}{0.35\textwidth}
        \centering
        \includegraphics[width=\linewidth]{figures/nentrbits_std/nentrbits_lb3_BTCUSDT_5.pdf}
        \caption{5 minutes}
        \label{fig:BTC_lb3_5}
    \end{subfigure}
    \hfill
    \begin{subfigure}{0.35\textwidth}
        \centering
        \includegraphics[width=\linewidth]{figures/nentrbits_std/failrate_bydate_BTCUSDT_5.pdf}
        \caption{5 minutes, per-day failure rate}
        \label{fig:BTC_lb3_bd_5}
    \end{subfigure}
    \vspace{0.2cm}
    \begin{subfigure}{0.35\textwidth}
        \centering
        \includegraphics[width=\linewidth]{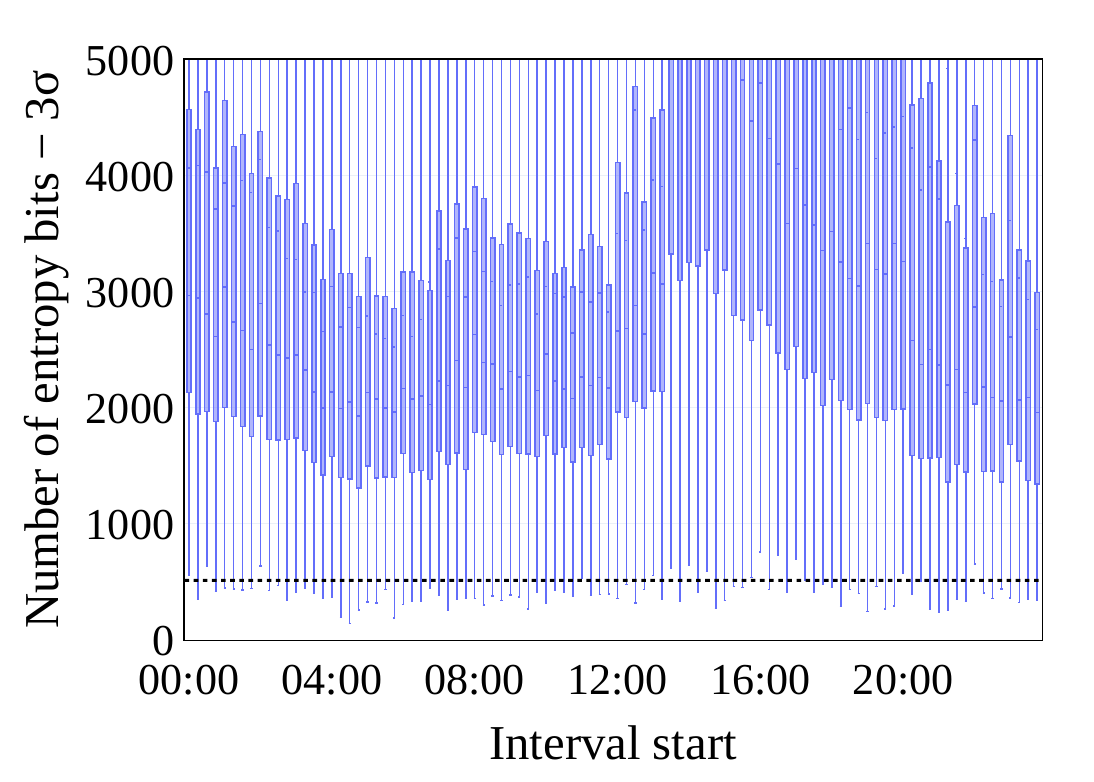}
        \caption{15 minutes}
        \label{fig:BTC_lb3_15}
    \end{subfigure}
    \hfill
    \begin{subfigure}{0.35\textwidth}
        \centering
        \includegraphics[width=\linewidth]{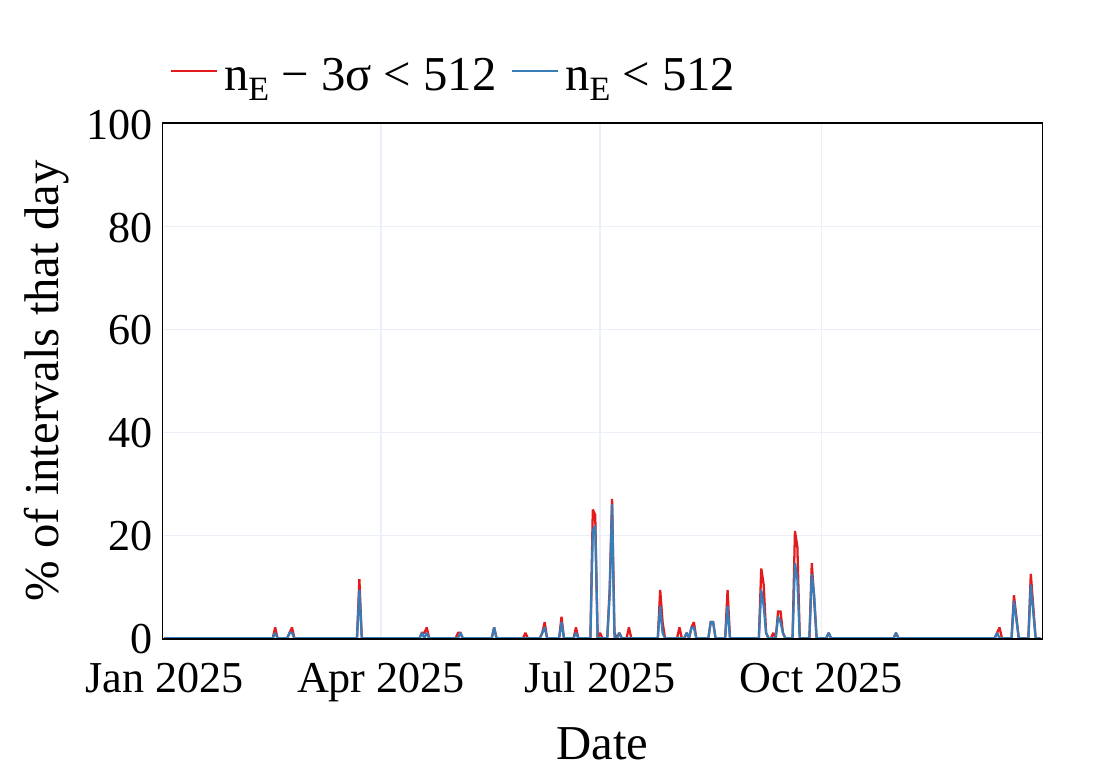}
        \caption{15 minutes, per-day failure rate}
        \label{fig:BTC_lb3_bd_15}
    \end{subfigure}
    \vspace{0.2cm}
    \begin{subfigure}{0.35\textwidth}
        \centering
        \includegraphics[width=\linewidth]{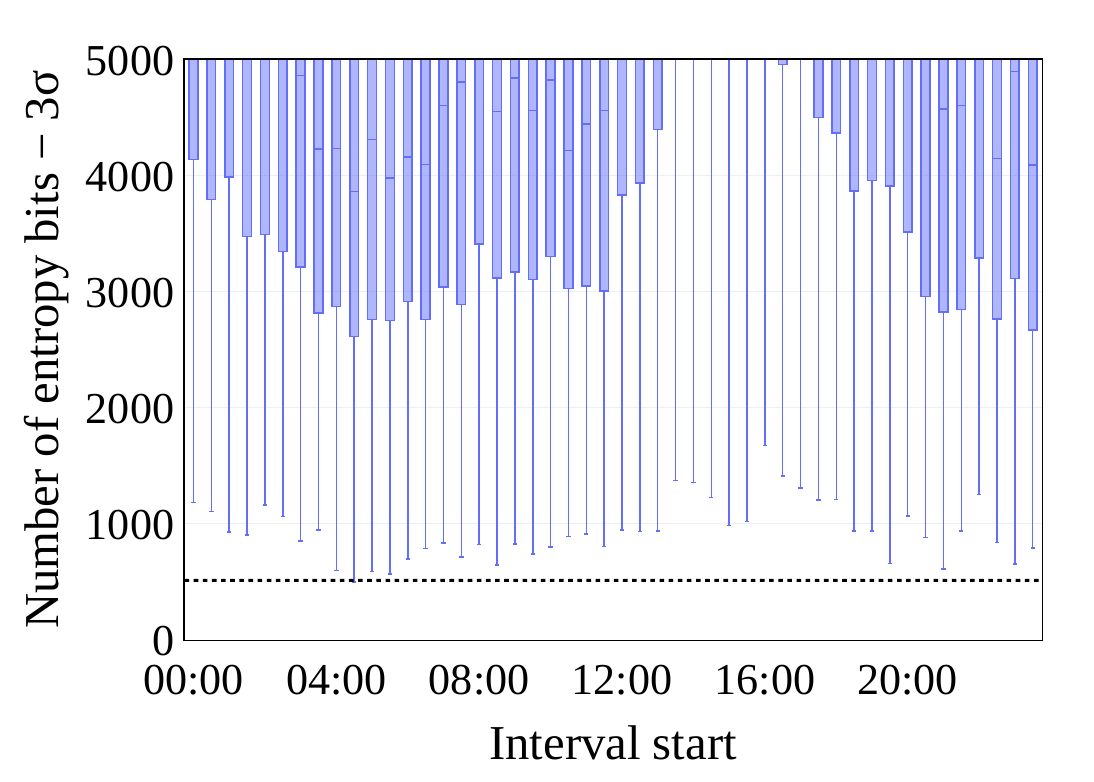}
        \caption{30 minutes}
        \label{fig:BTC_lb3_30}
    \end{subfigure}
    \hfill
    \begin{subfigure}{0.35\textwidth}
        \centering
        \includegraphics[width=\linewidth]{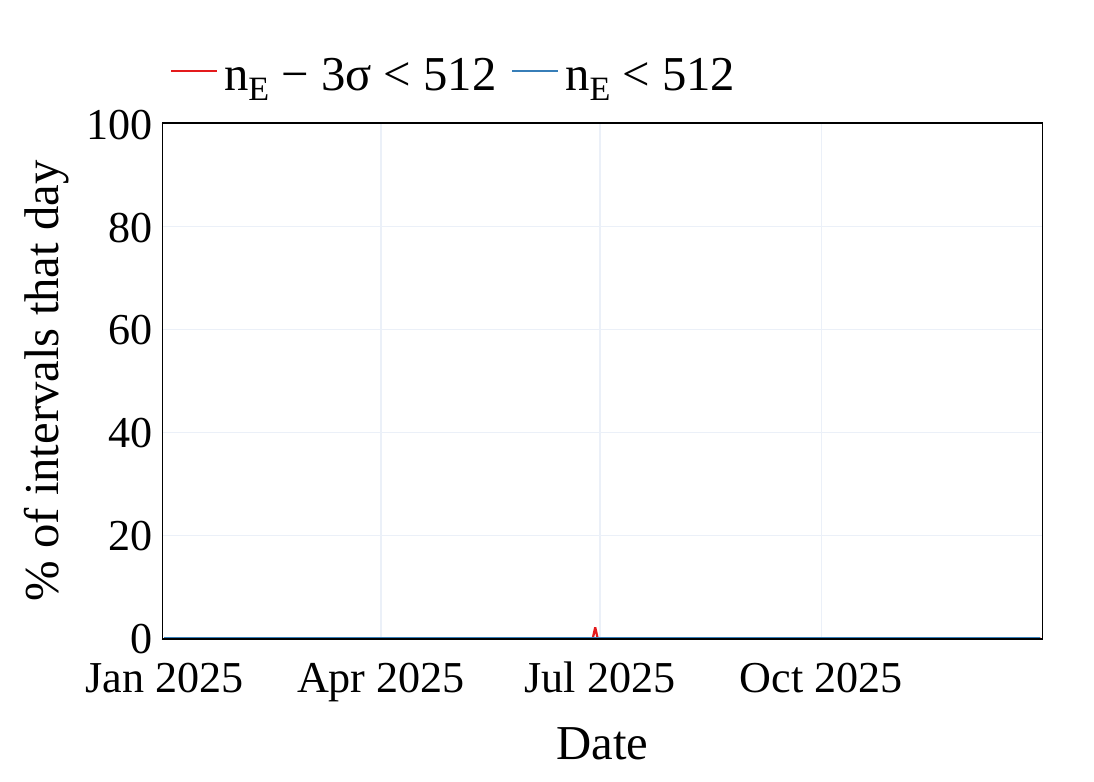}
        \caption{30 minutes, per-day failure rate}
        \label{fig:BTC_lb3_bd_30}
    \end{subfigure}
    \caption{Conservative counterpart of \Cref{fig:BTC_entropy_4x2_combined} for
    BTCUSDT. Left column: distribution of $n_E-3\sigma_{n_E}$ by interval start,
    against the $512$-bit target (dotted). Right column: share of that day's intervals
    falling below the target, under the plain criterion (blue) and the conservative
    one (red); the gap between the two curves is what the standard error costs.}
    \label{fig:BTC_lb3}
\end{figure}

\begin{figure}[p]
    \centering
    \begin{subfigure}{0.35\textwidth}
        \centering
        \includegraphics[width=\linewidth]{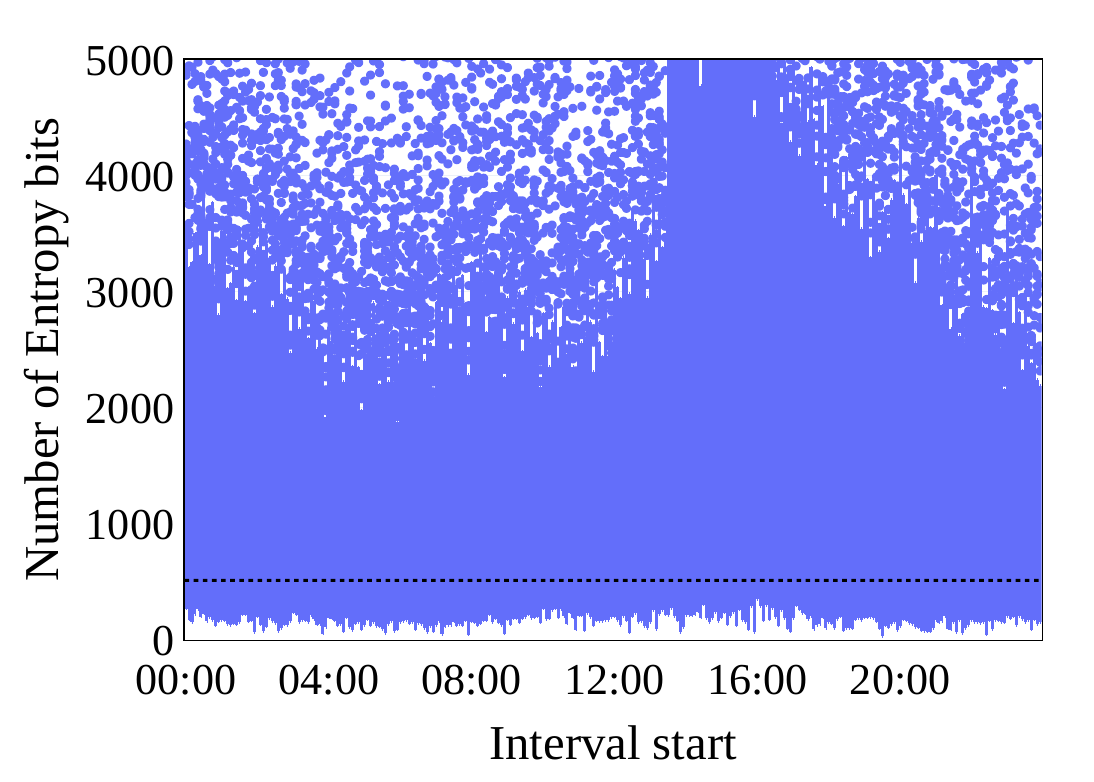}
        \caption{5 minutes}
        \label{fig:ETH_5_combined}
    \end{subfigure}
    \hfill
    \begin{subfigure}{0.35\textwidth}
        \centering
        \includegraphics[width=\linewidth]{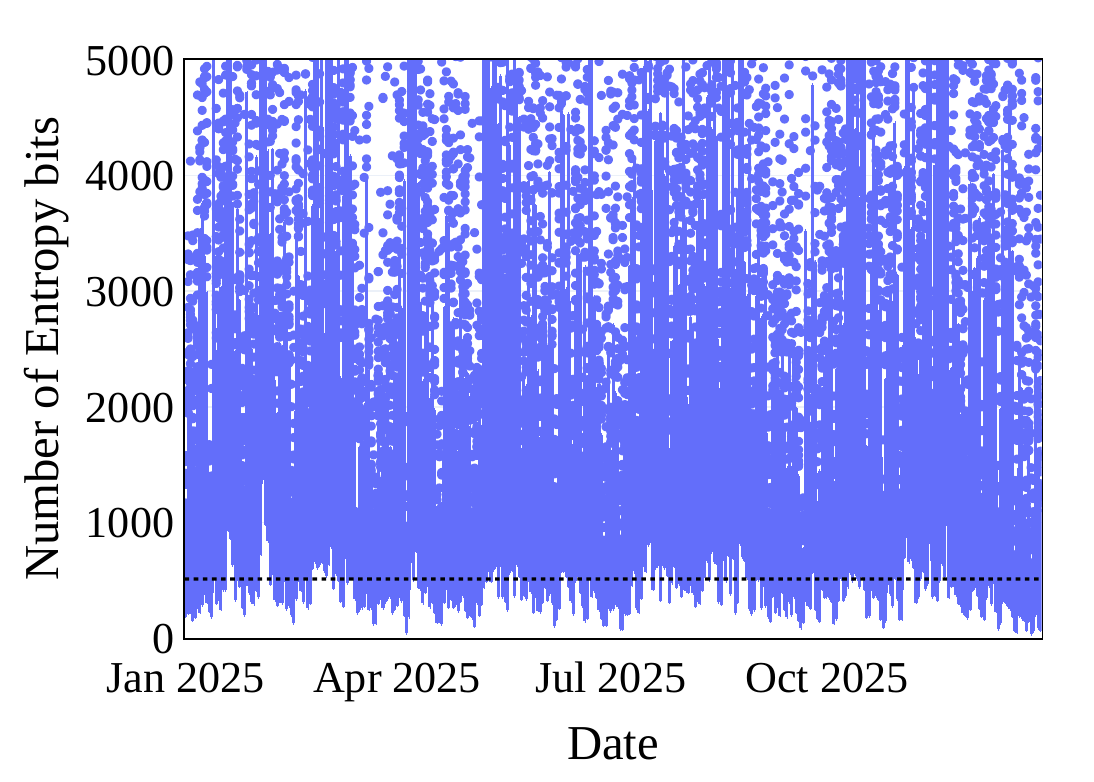}
        \caption{5 minutes, per-day}
        \label{fig:ETH_bd_5_combined}
    \end{subfigure}
    \vspace{0.2cm}
    \begin{subfigure}{0.35\textwidth}
        \centering
        \includegraphics[width=\linewidth]{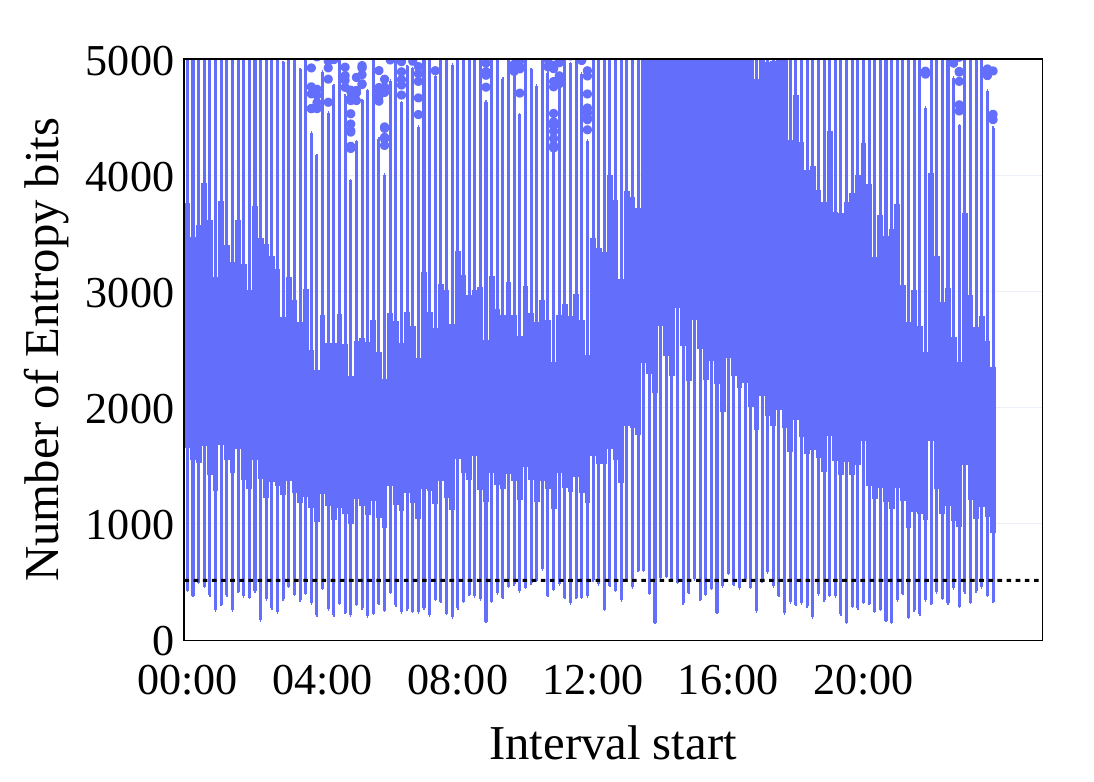}
        \caption{10 minutes}
        \label{fig:ETH_10_combined}
    \end{subfigure}
    \hfill
    \begin{subfigure}{0.35\textwidth}
        \centering
        \includegraphics[width=\linewidth]{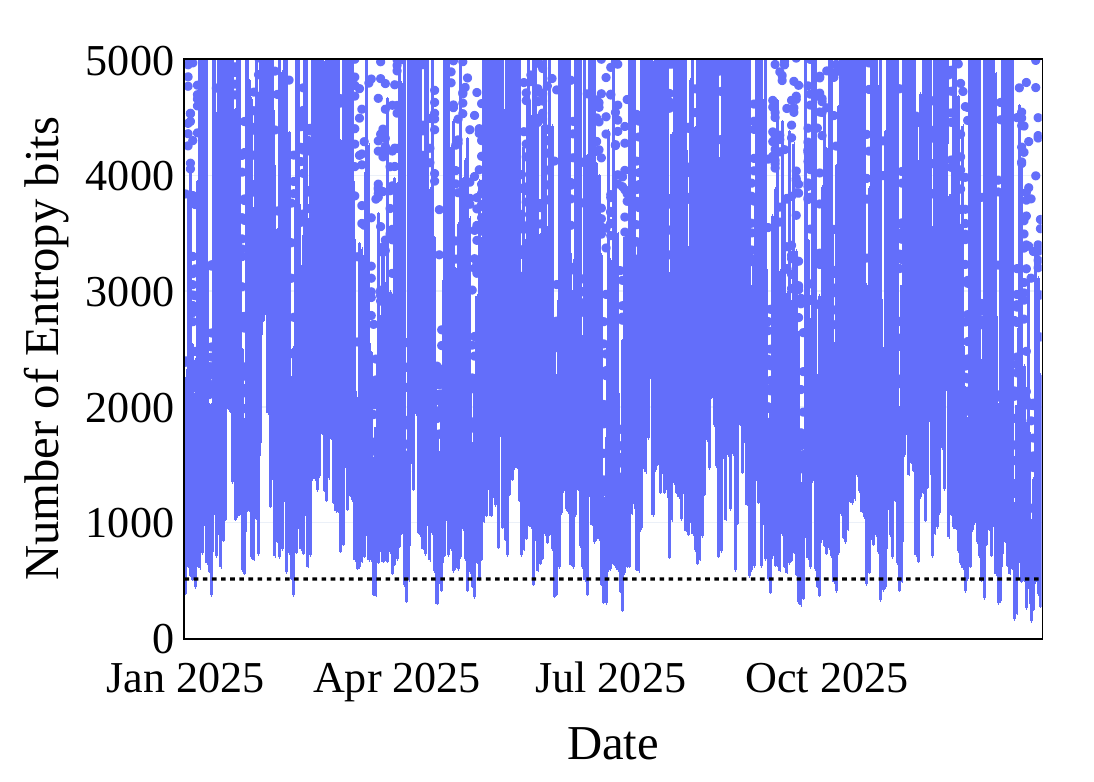}
        \caption{10 minutes, per-day}
        \label{fig:ETH_bd_10_combined}
    \end{subfigure}
    \vspace{0.2cm}
    \begin{subfigure}{0.35\textwidth}
        \centering
        \includegraphics[width=\linewidth]{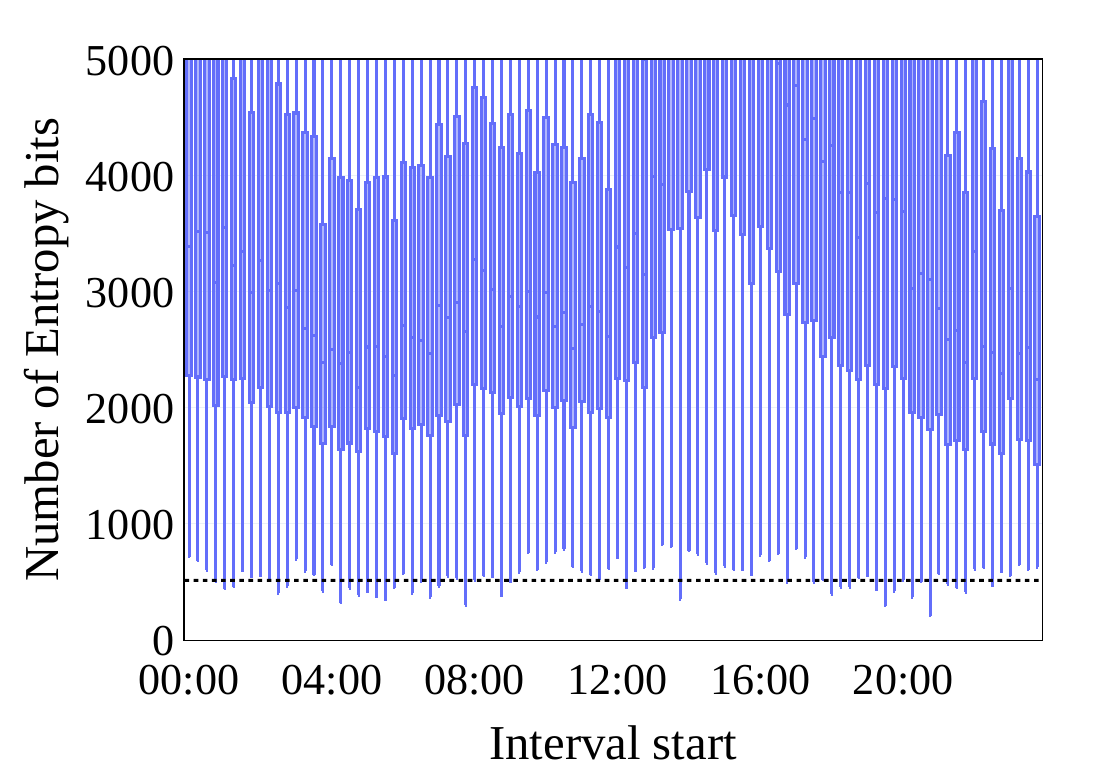}
        \caption{15 minutes}
        \label{fig:ETH_15_combined}
    \end{subfigure}
    \hfill
    \begin{subfigure}{0.35\textwidth}
        \centering
        \includegraphics[width=\linewidth]{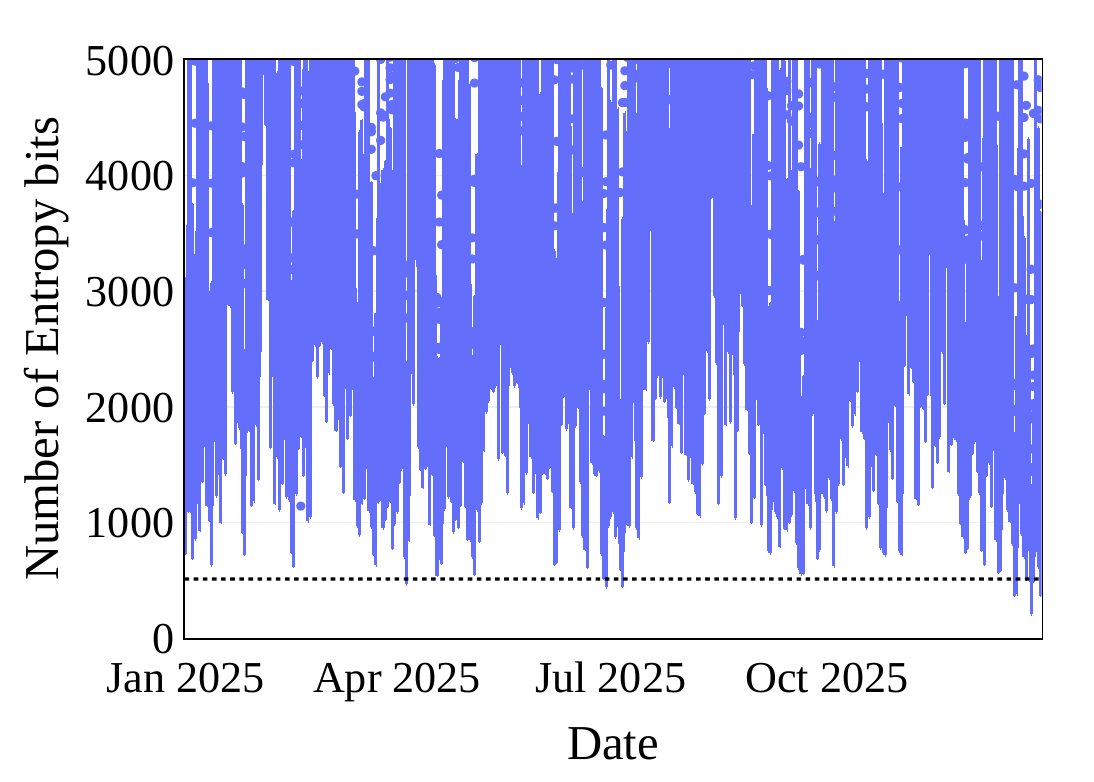}
        \caption{15 minutes, per-day}
        \label{fig:ETH_bd_15_combined}
    \end{subfigure}
    \vspace{0.2cm}
    \begin{subfigure}{0.35\textwidth}
        \centering
        \includegraphics[width=\linewidth]{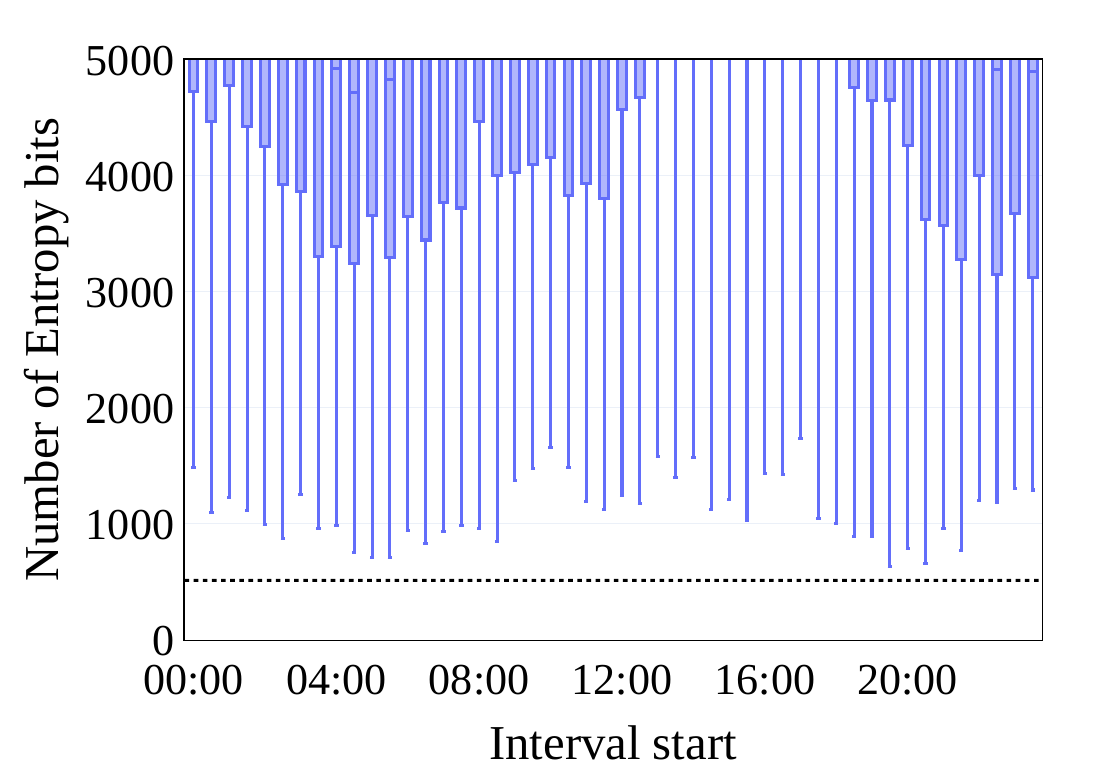}
        \caption{30 minutes}
        \label{fig:ETH_30_combined}
    \end{subfigure}
    \hfill
    \begin{subfigure}{0.35\textwidth}
        \centering
        \includegraphics[width=\linewidth]{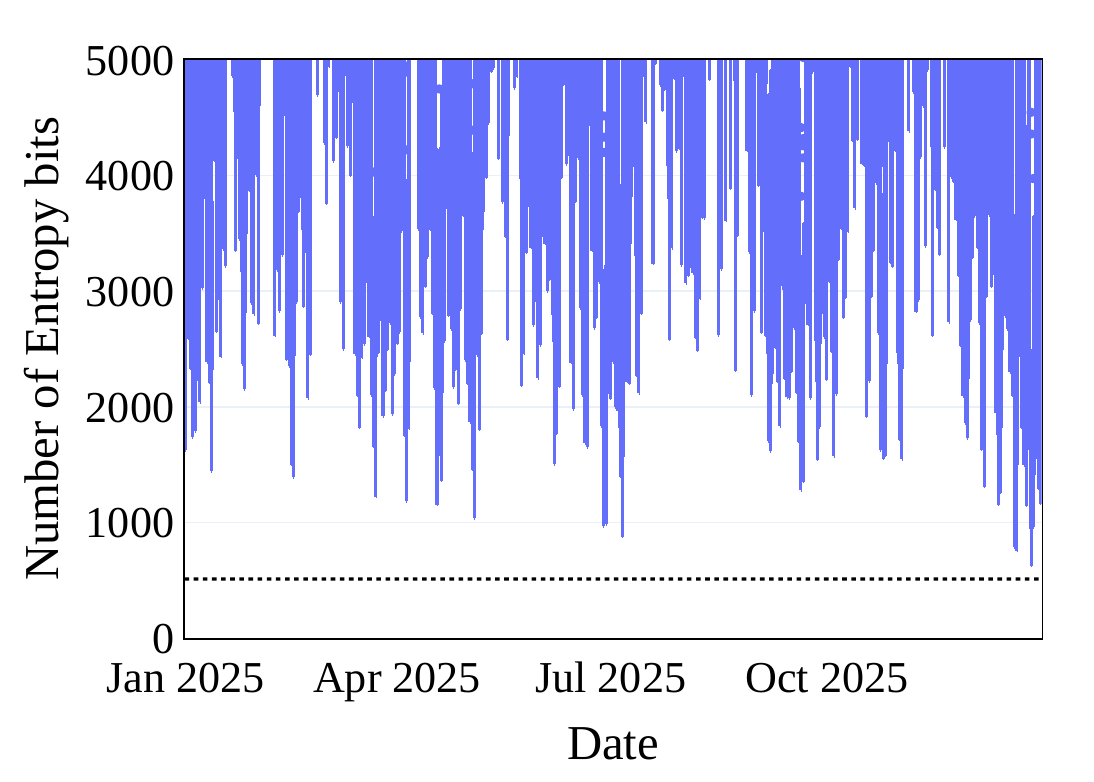}
        \caption{30 minutes, per-day}
        \label{fig:ETH_bd_30_combined}
    \end{subfigure}
    \caption{Number of entropy bits $n_E$ created by strings of ETHUSDT in intervals
    of $5$, $10$, $15$, $30$ minutes. Left column: distribution by interval start;
    right column: per-day distribution over 2025. The dotted line marks the $512$-bit
    target.}
    \label{fig:ETH_entropy_4x2_combined}
\end{figure}

\begin{figure}[p]
    \centering
    \begin{subfigure}{0.35\textwidth}
        \centering
        \includegraphics[width=\linewidth]{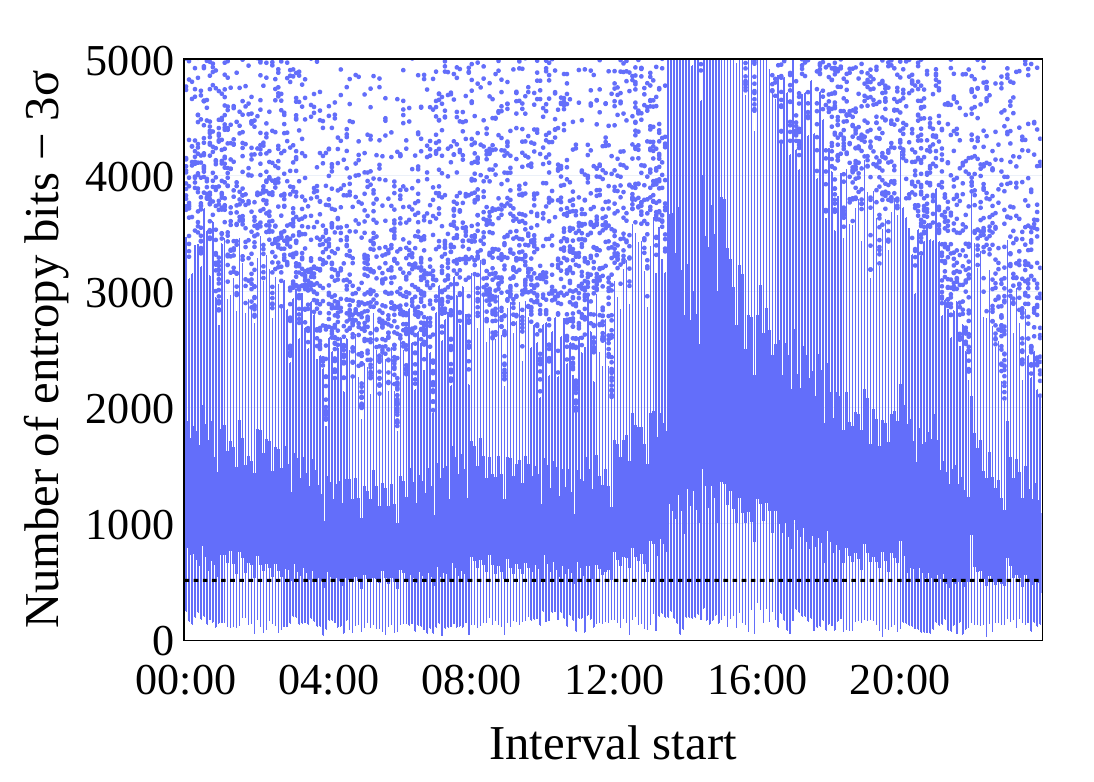}
        \caption{5 minutes}
        \label{fig:ETH_lb3_5}
    \end{subfigure}
    \hfill
    \begin{subfigure}{0.35\textwidth}
        \centering
        \includegraphics[width=\linewidth]{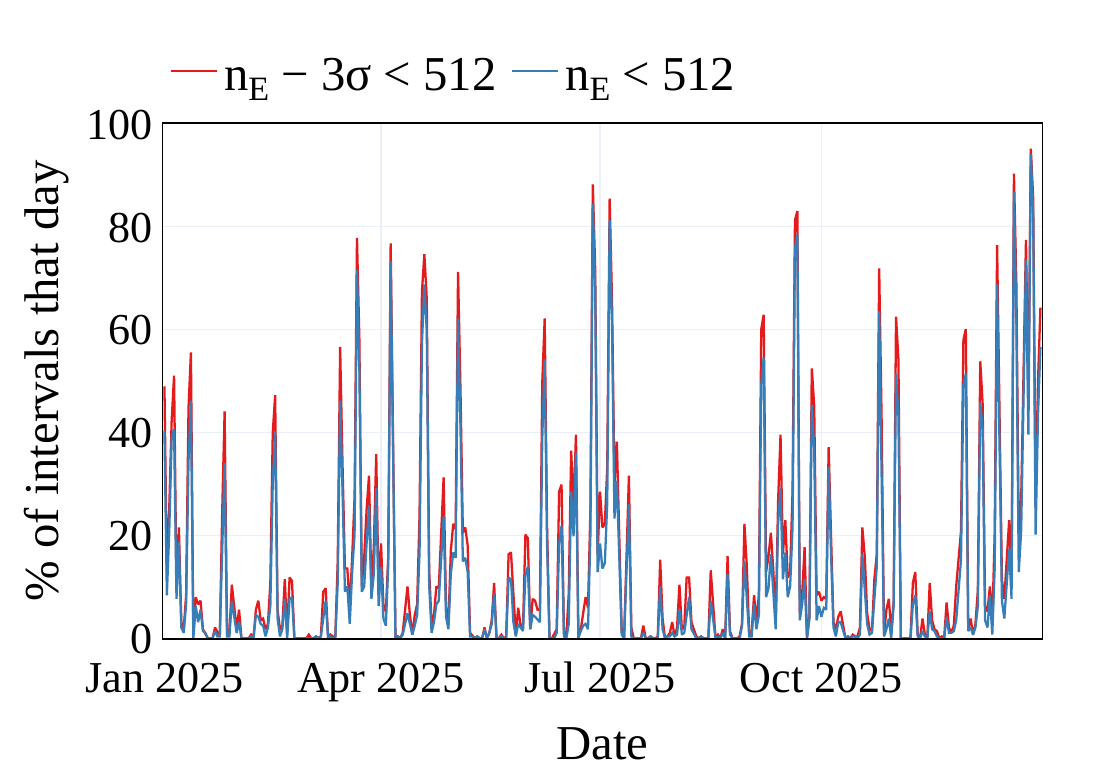}
        \caption{5 minutes, per-day failure rate}
        \label{fig:ETH_lb3_bd_5}
    \end{subfigure}
    \vspace{0.2cm}
    \begin{subfigure}{0.35\textwidth}
        \centering
        \includegraphics[width=\linewidth]{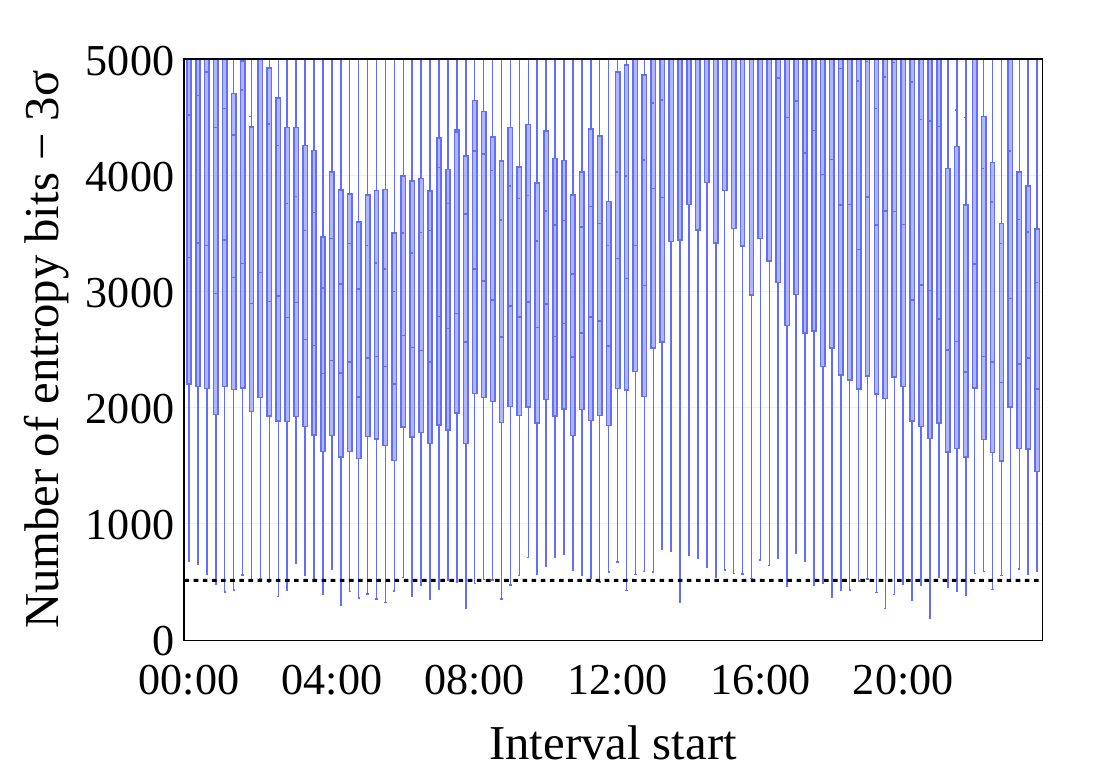}
        \caption{15 minutes}
        \label{fig:ETH_lb3_15}
    \end{subfigure}
    \hfill
    \begin{subfigure}{0.35\textwidth}
        \centering
        \includegraphics[width=\linewidth]{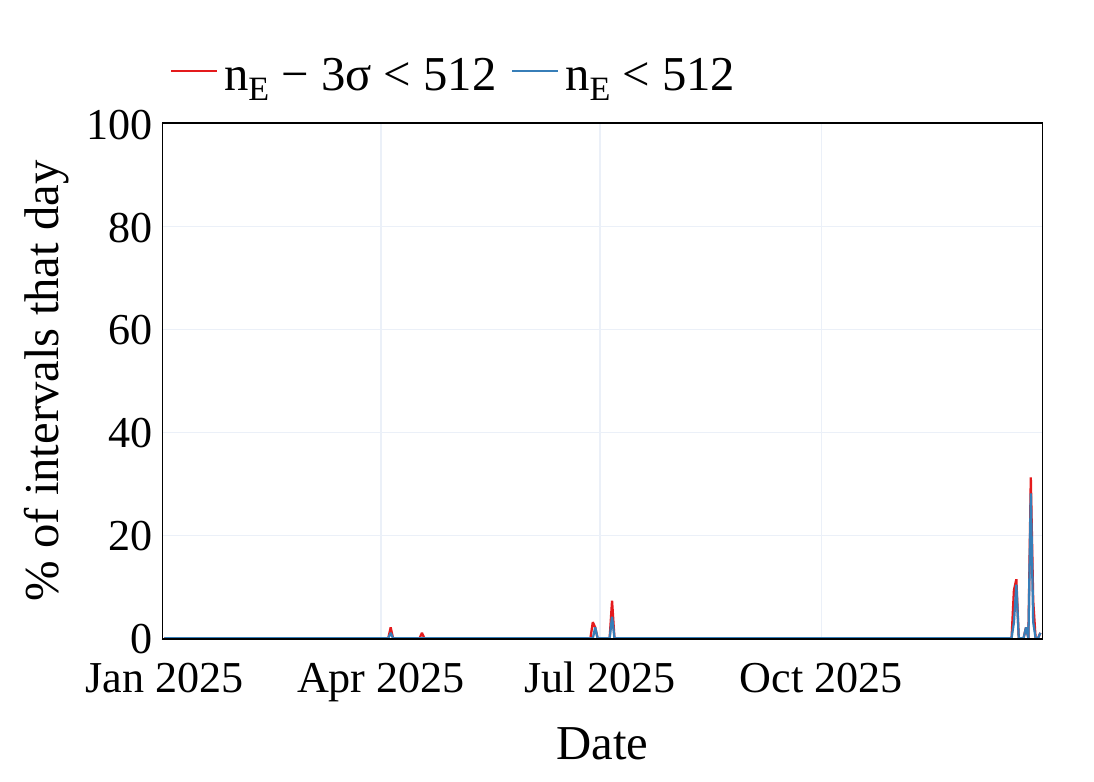}
        \caption{15 minutes, per-day failure rate}
        \label{fig:ETH_lb3_bd_15}
    \end{subfigure}
    \vspace{0.2cm}
    \begin{subfigure}{0.35\textwidth}
        \centering
        \includegraphics[width=\linewidth]{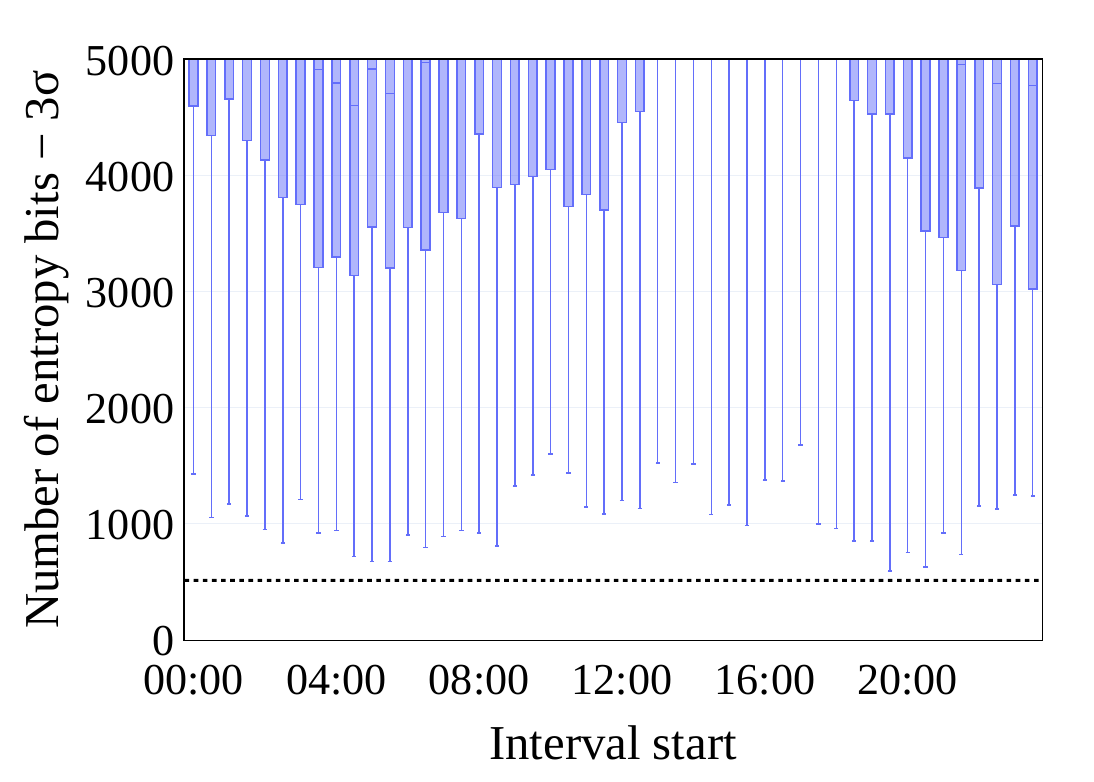}
        \caption{30 minutes}
        \label{fig:ETH_lb3_30}
    \end{subfigure}
    \hfill
    \begin{subfigure}{0.35\textwidth}
        \centering
        \includegraphics[width=\linewidth]{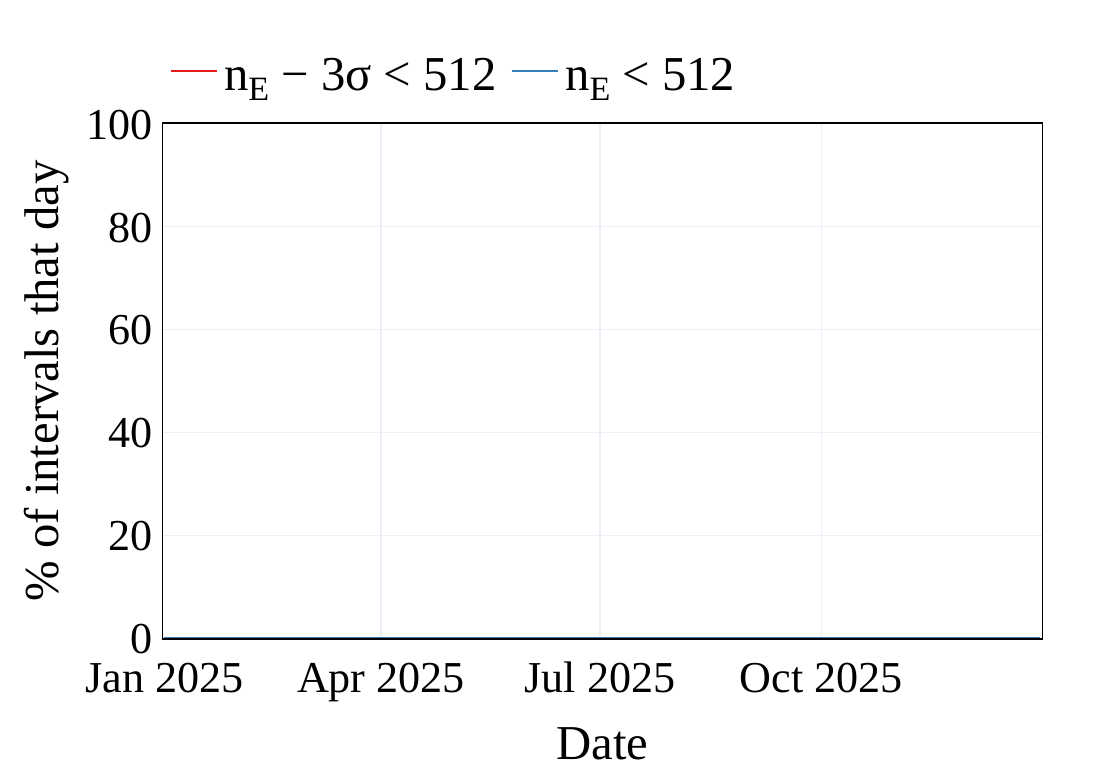}
        \caption{30 minutes, per-day failure rate}
        \label{fig:ETH_lb3_bd_30}
    \end{subfigure}
    \caption{Conservative counterpart of \Cref{fig:ETH_entropy_4x2_combined} for
    ETHUSDT; layout and colours as in \Cref{fig:BTC_lb3}.}
    \label{fig:ETH_lb3}
\end{figure}

\begin{figure}[p]
    \centering
    \begin{subfigure}{0.35\textwidth}
        \centering
        \includegraphics[width=\linewidth]{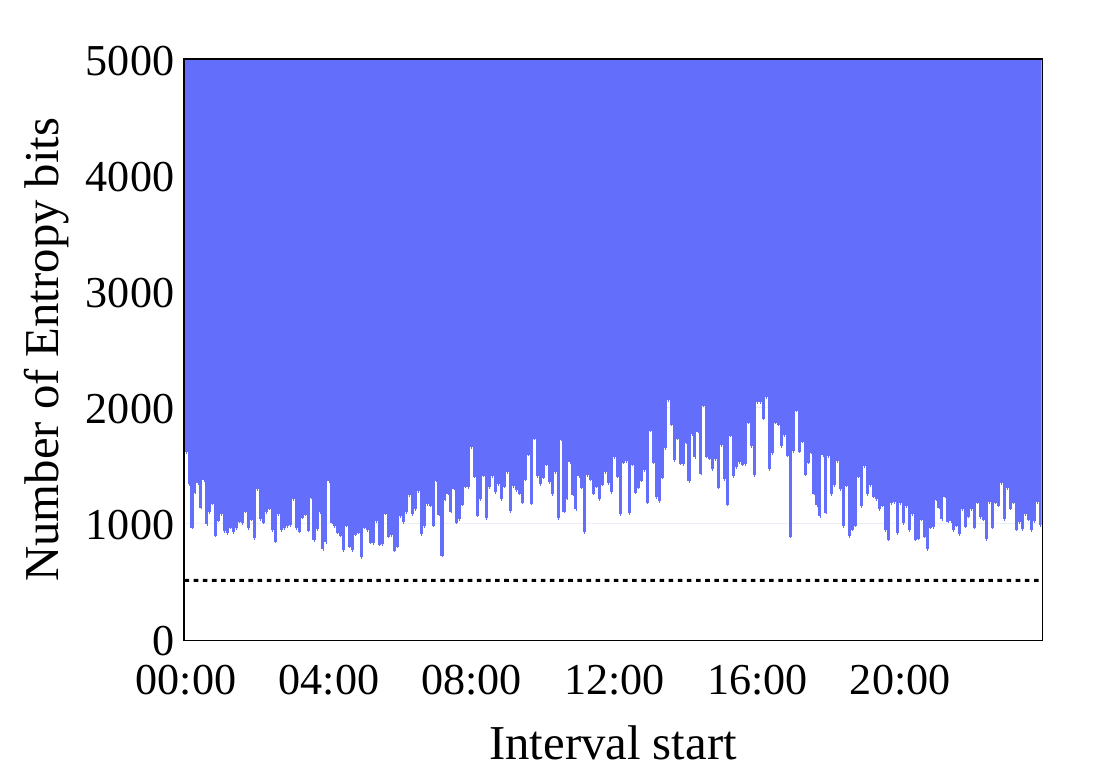}
        \caption{5 minutes}
        \label{fig:MIXED_5_combined}
    \end{subfigure}
    \hfill
    \begin{subfigure}{0.35\textwidth}
        \centering
        \includegraphics[width=\linewidth]{figures/nentrbits_plots/nentrbits_bydate_MIXED_5.pdf}
        \caption{5 minutes, per-day}
        \label{fig:MIXED_bd_5_combined}
    \end{subfigure}
    \vspace{0.2cm}
    \begin{subfigure}{0.35\textwidth}
        \centering
        \includegraphics[width=\linewidth]{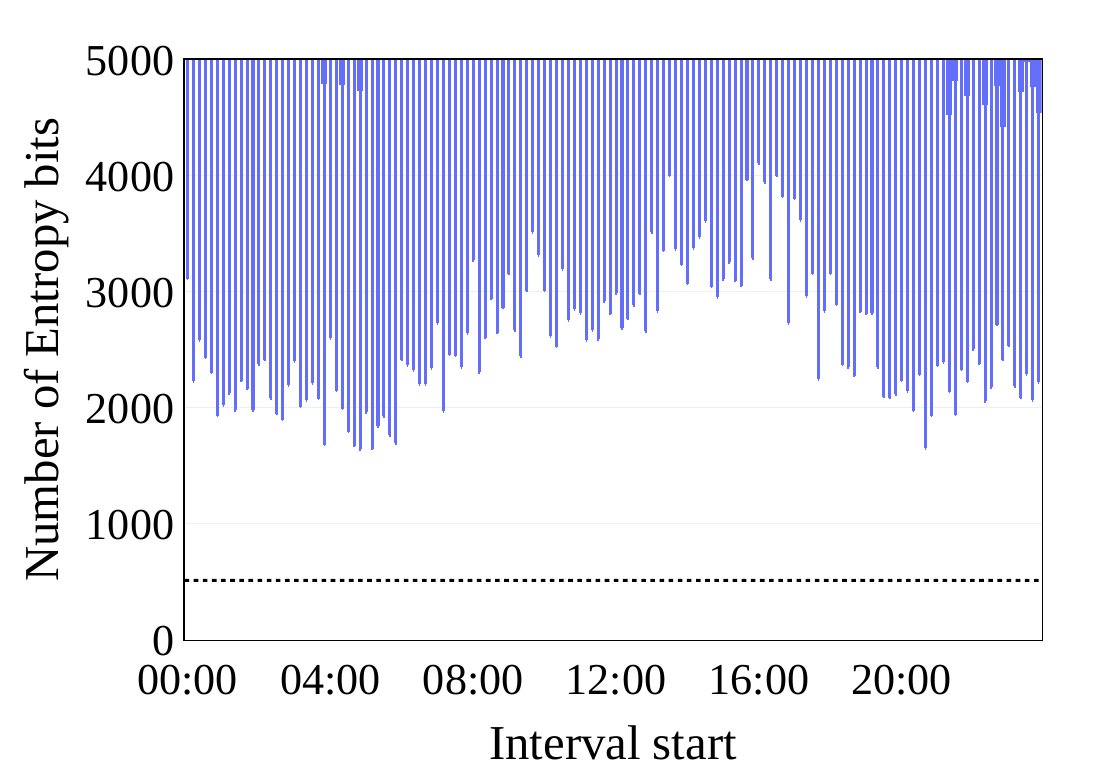}
        \caption{10 minutes}
        \label{fig:MIXED_10_combined}
    \end{subfigure}
    \hfill
    \begin{subfigure}{0.35\textwidth}
        \centering
        \includegraphics[width=\linewidth]{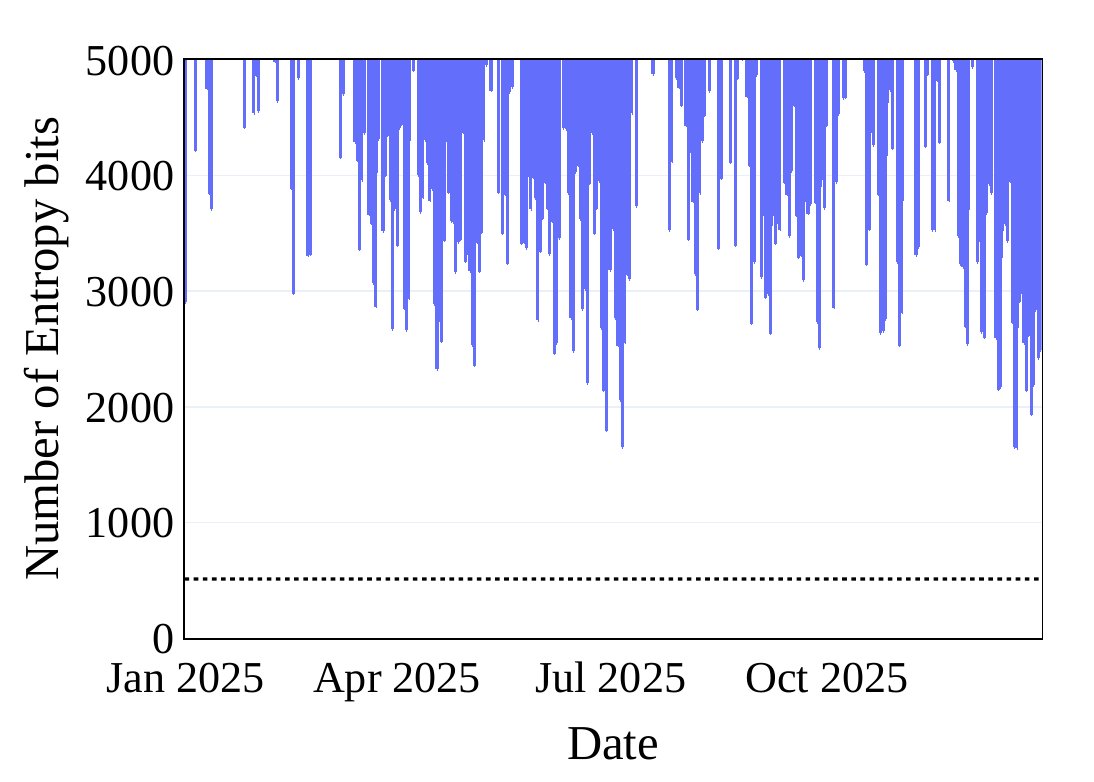}
        \caption{10 minutes, per-day}
        \label{fig:MIXED_bd_10_combined}
    \end{subfigure}
    \vspace{0.2cm}
    \begin{subfigure}{0.35\textwidth}
        \centering
        \includegraphics[width=\linewidth]{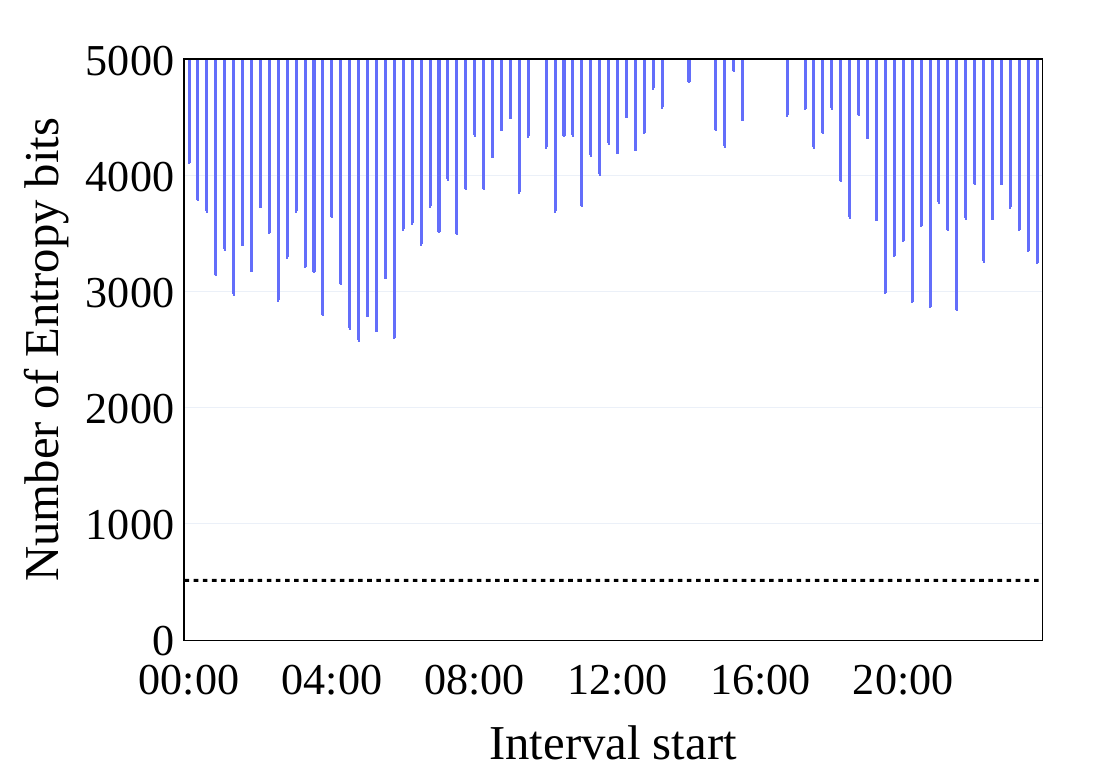}
        \caption{15 minutes}
        \label{fig:MIXED_15_combined}
    \end{subfigure}
    \hfill
    \begin{subfigure}{0.35\textwidth}
        \centering
        \includegraphics[width=\linewidth]{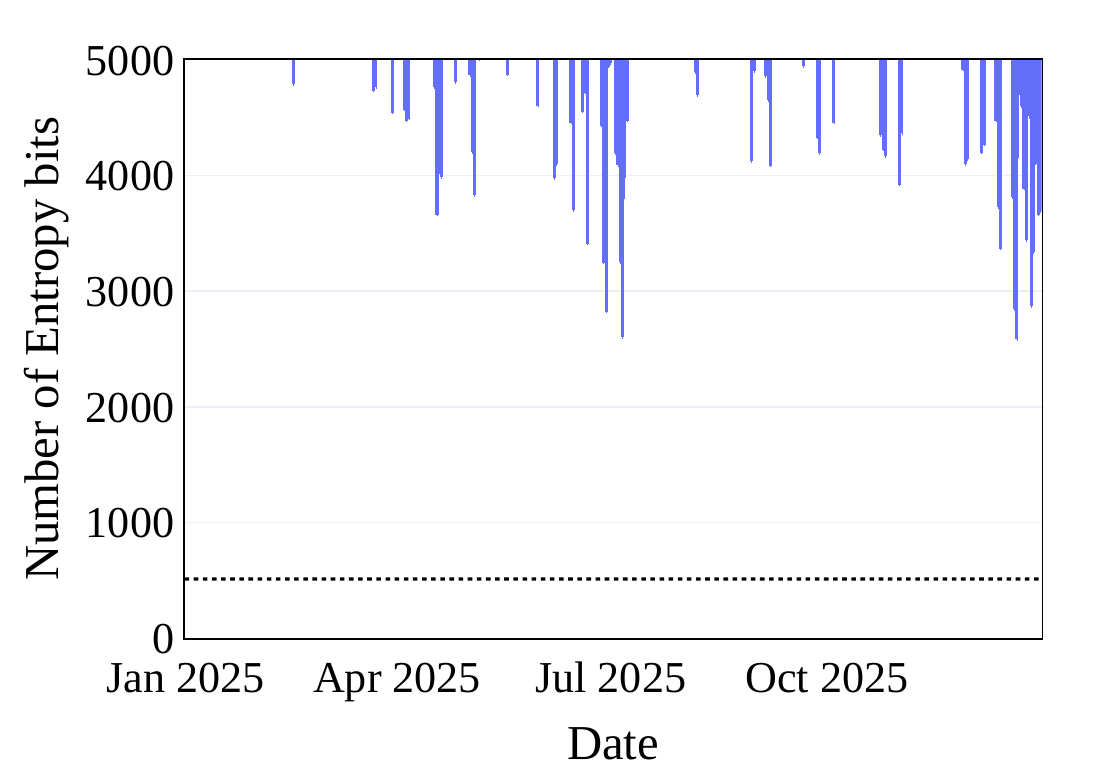}
        \caption{15 minutes, per-day}
        \label{fig:MIXED_bd_15_combined}
    \end{subfigure}
    \vspace{0.2cm}
    \begin{subfigure}{0.35\textwidth}
        \centering
        \includegraphics[width=\linewidth]{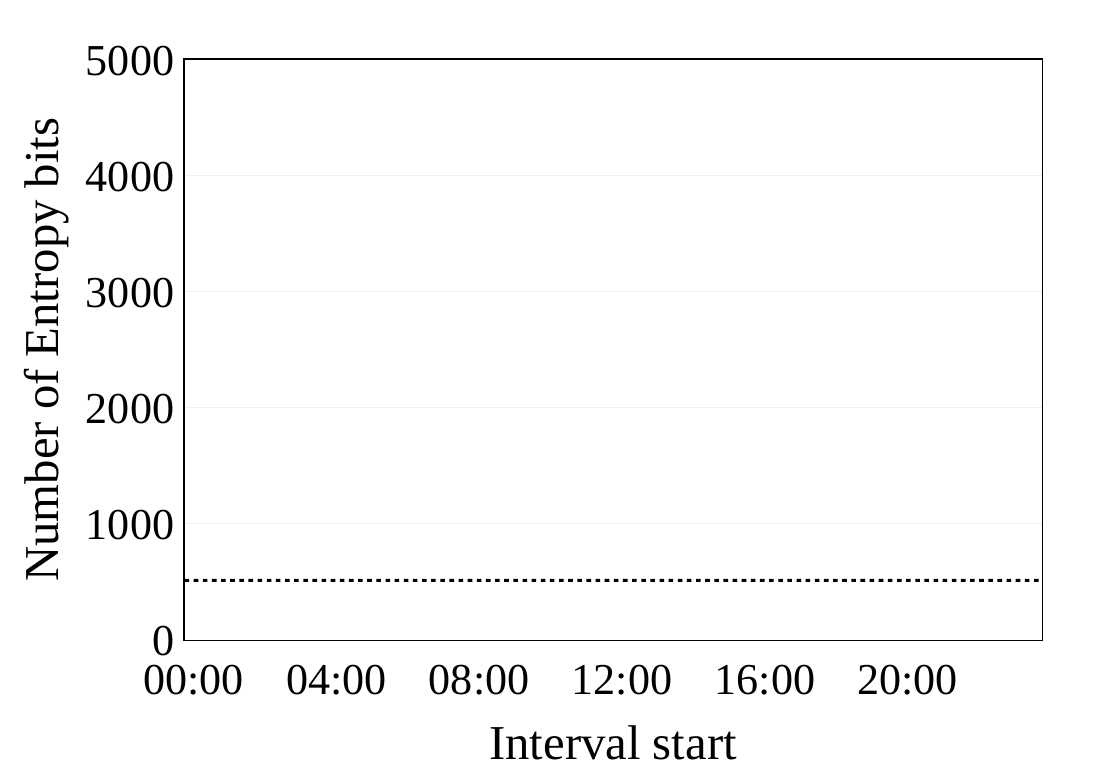}
        \caption{30 minutes}
        \label{fig:MIXED_30_combined}
    \end{subfigure}
    \hfill
    \begin{subfigure}{0.35\textwidth}
        \centering
        \includegraphics[width=\linewidth]{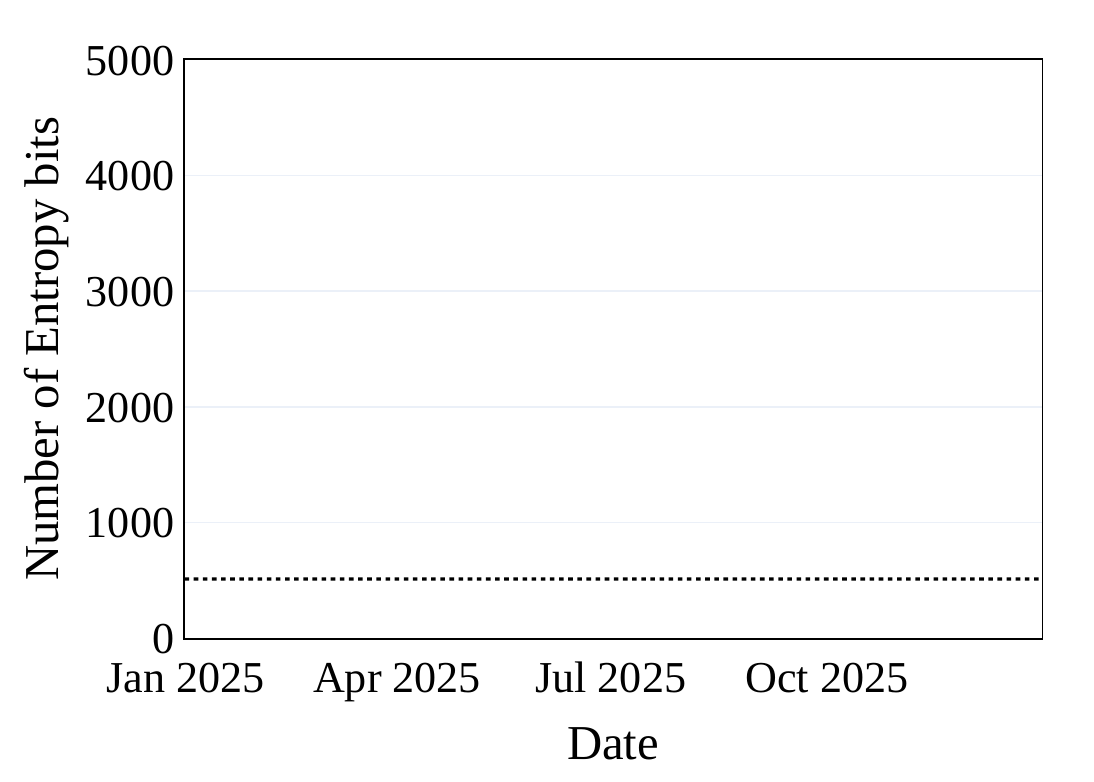}
        \caption{30 minutes, per-day}
        \label{fig:MIXED_bd_30_combined}
    \end{subfigure}
    \caption{Number of entropy bits $n_E$ created by strings of \textsc{Mixed} in
    intervals of $5$, $10$, $15$, $30$ minutes. Left column: distribution by interval
    start; right column: per-day distribution over 2025. The $30$-minute plots look
    empty because the values are always greater than $5000$, that is, above the top of
    the common vertical axis.}
    \label{fig:MIXED_entropy_4x2_combined}
\end{figure}

\begin{figure}[p]
    \centering
    \begin{subfigure}{0.35\textwidth}
        \centering
        \includegraphics[width=\linewidth]{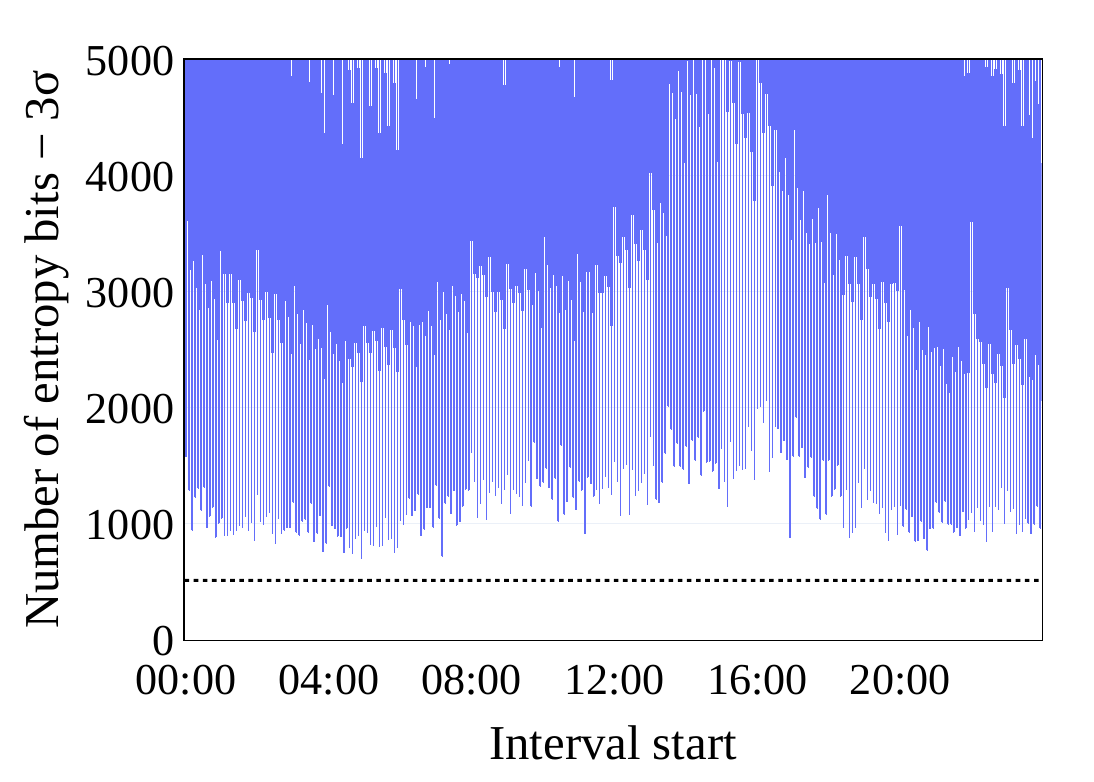}
        \caption{5 minutes}
        \label{fig:MIXED_lb3_5}
    \end{subfigure}
    \hfill
    \begin{subfigure}{0.35\textwidth}
        \centering
        \includegraphics[width=\linewidth]{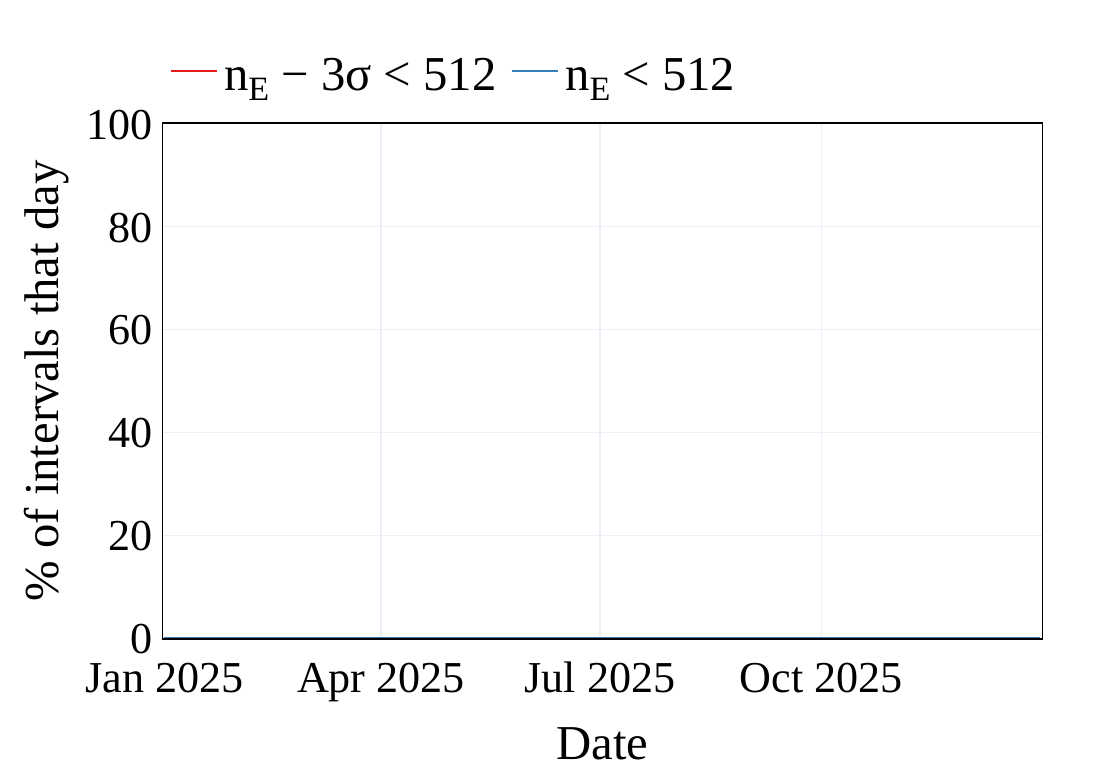}
        \caption{5 minutes, per-day failure rate}
        \label{fig:MIXED_lb3_bd_5}
    \end{subfigure}
    \vspace{0.2cm}
    \begin{subfigure}{0.35\textwidth}
        \centering
        \includegraphics[width=\linewidth]{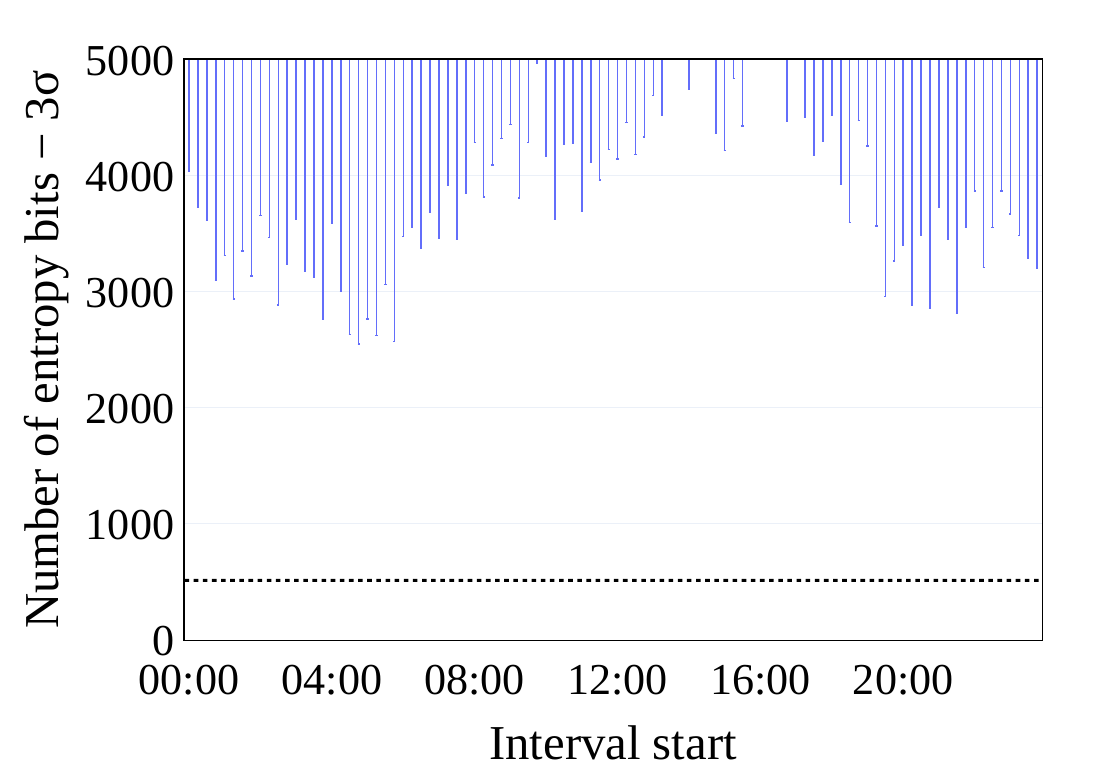}
        \caption{15 minutes}
        \label{fig:MIXED_lb3_15}
    \end{subfigure}
    \hfill
    \begin{subfigure}{0.35\textwidth}
        \centering
        \includegraphics[width=\linewidth]{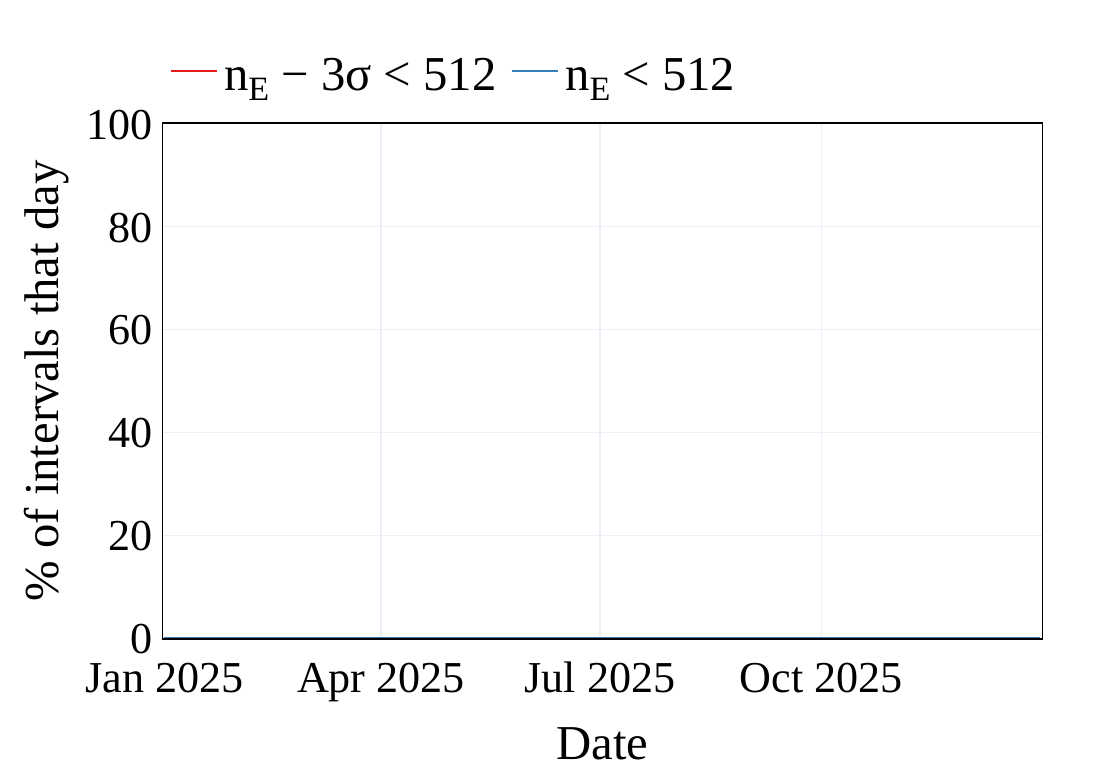}
        \caption{15 minutes, per-day failure rate}
        \label{fig:MIXED_lb3_bd_15}
    \end{subfigure}
    \vspace{0.2cm}
    \begin{subfigure}{0.35\textwidth}
        \centering
        \includegraphics[width=\linewidth]{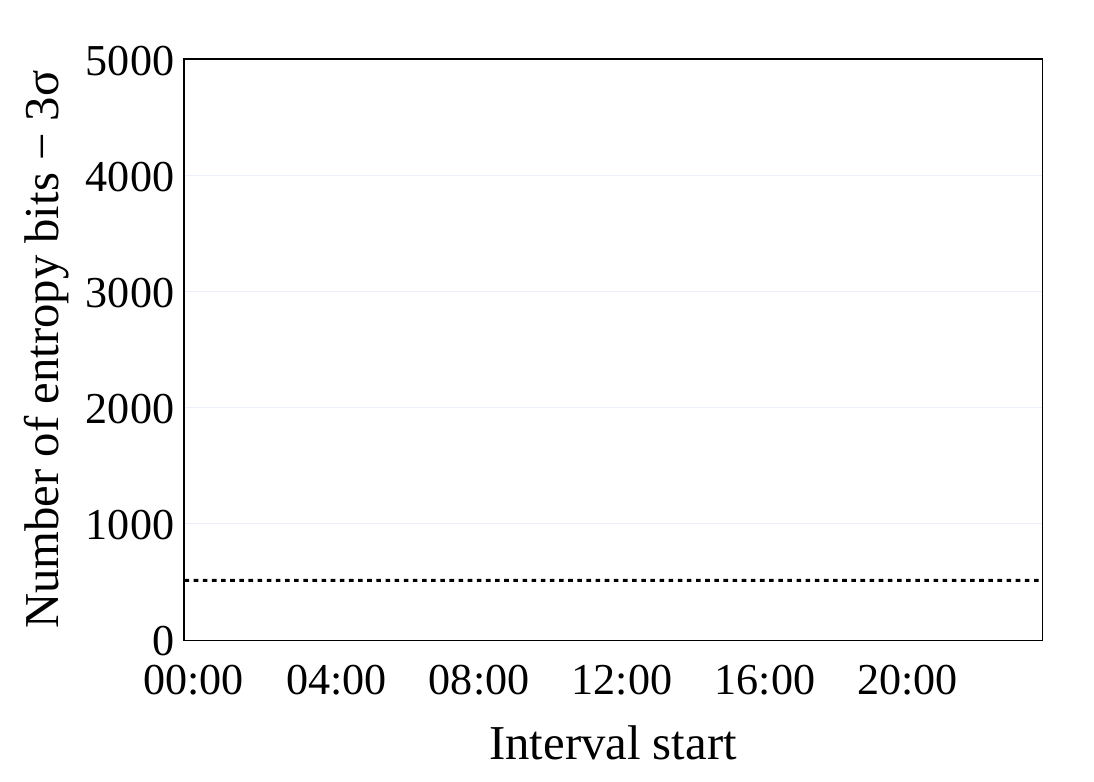}
        \caption{30 minutes}
        \label{fig:MIXED_lb3_30}
    \end{subfigure}
    \hfill
    \begin{subfigure}{0.35\textwidth}
        \centering
        \includegraphics[width=\linewidth]{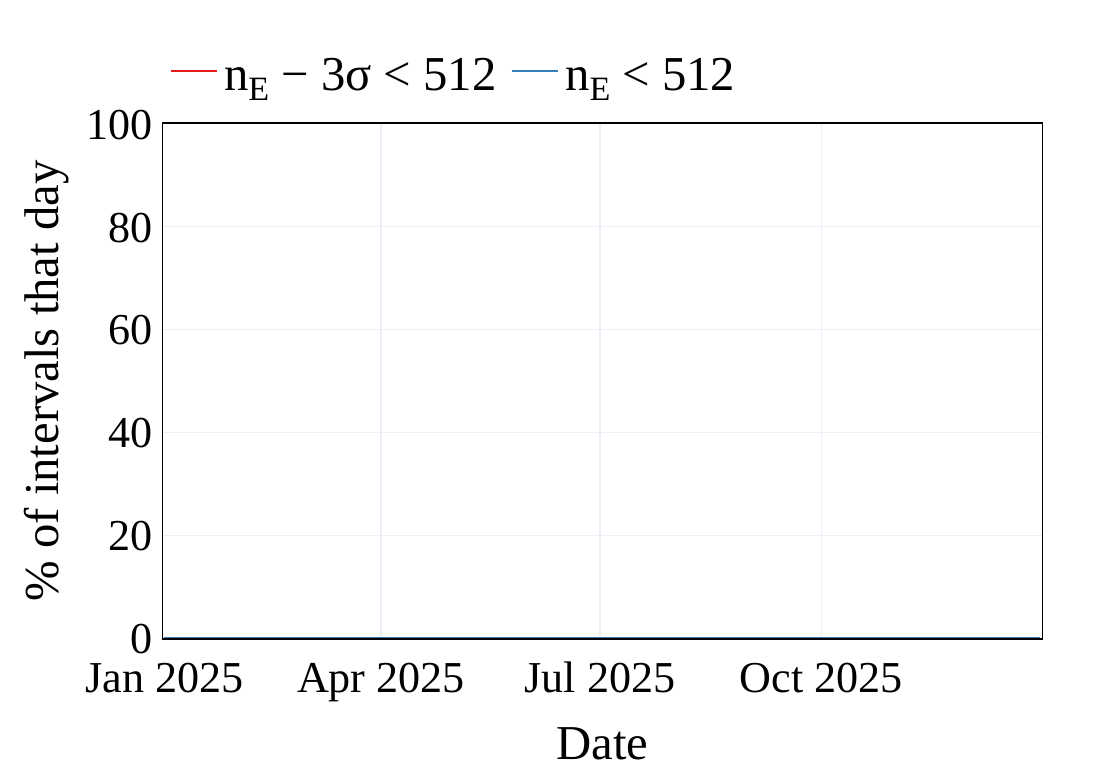}
        \caption{30 minutes, per-day failure rate}
        \label{fig:MIXED_lb3_bd_30}
    \end{subfigure}
    \caption{Conservative counterpart of \Cref{fig:MIXED_entropy_4x2_combined} for
    \textsc{Mixed}; layout and colours as in \Cref{fig:BTC_lb3}. Both failure-rate
    curves are identically zero at every window length, and the $n_E-3\sigma_{n_E}$
    boxplots sit above the top of the vertical axis from $15$ minutes on.}
    \label{fig:MIXED_lb3}
\end{figure}

Figures~\ref{fig:BTC_entropy_4x2_combined}--\ref{fig:MIXED_lb3} show the windows of Sections~\ref{subsec:bound} and~\ref{app:circuitbreaker} for BTCUSDT, ETHUSDT and \textsc{Mixed} at every window length.

\end{appendix}

\end{document}